\documentclass[aps,prd,onecolumn,superscriptaddress,showpacs,nofootinbib,floatfix,onecolumn]{revtex4-2}

\usepackage{dcolumn}
\usepackage{bm}

\usepackage{graphicx}
\usepackage{float}
\usepackage{epsfig,subfigure}
\usepackage{longtable}
\usepackage{booktabs}
\usepackage{rotating}
\usepackage{amsmath}
\usepackage{amsfonts}
\usepackage{amssymb}
\usepackage{accents}
\usepackage{dcolumn}
\usepackage[colorlinks,linkcolor={red!50!black},citecolor={blue!80!black},urlcolor={blue!80!black}]{hyperref}
\usepackage{multirow}
\usepackage[table]{xcolor}
\usepackage{makecell}

\usepackage{color}
\usepackage{txfonts}
\RequirePackage[capitalise]{cleveref}
\RequirePackage{siunitx}
\RequirePackage[status=draft,inline,nomargin,marginclue]{fixme}
\fxusetheme{color}
\FXRegisterAuthor{fh}{afh}{FH}

\usepackage{dashrule}

\usepackage{xspace}

\newcommand{\LQCD}{\Lambda_\text{QCD}}
\newcommand{\alphaS}{\alpha_\mathrm{s}}
\newcommand{\alphasmZ}{\alphaS(m_\mathrm{Z})}
\newcommand{\sqrts}{\sqrt{s}}

\newcommand{\sqrtsNN}{\sqrt{s_{_{\mbox{\rm \tiny{NN}}}}}}

\newcommand{\pT}{p_{_\mathrm{T}}}
\newcommand{\mT}{m_{_\mathrm{T}}}

\newcommand{\xmin}{x_\text{min}}
\newcommand{\xgmin}{x_\text{gridmin}}
\newcommand{\Qmin}{Q_\text{min}}

\newcommand{\mtop}{m_\text{top}}
\newcommand{\mQ}{m_\mathrm{Q}}
\newcommand{\mc}{m_\mathrm{c}}
\newcommand{\mb}{m_\mathrm{b}}

\newcommand{\mhQ}{m_\mathrm{h}}

\newcommand{\muF}{\mu_\mathrm{F}}
\newcommand{\muR}{\mu_\mathrm{R}}
\newcommand{\muFF}{\mu_\mathrm{FF}}

\newcommand{\pp}{\text{p-p}}

\newcommand{\ppbar}{\text{p-\=p}}
\newcommand{\pA}{\text{p-A}}
\newcommand{\pX}{\text{p-X}}

\newcommand{\pPb}{p-Pb}

\newcommand{\epem}{\mathrm{e}^+\mathrm{e}^-}

\newcommand{\pythia}{{\sc pythia}}
\newcommand{\powheg}{{\sc powheg}}
\newcommand{\fonll}{{\sc fonll}}
\newcommand{\mnr}{{\sc mnr}}
\newcommand{\mcatnlo}{{\sc mc@nlo}}
\newcommand{\Matrix}{{\sc Matrix}}

\newcommand{\toppp}{{\sc top++}}
\newcommand{\hathor}{{\sc Hathor}}
\newcommand{\lhapdf}{{\sc LHAPDF}}
\newcommand{\sacotmt}{\textsc{SACOT}-$m_{\mathrm{T}}$}
\newcommand{\MaunaKea}{\texttt{MaunaKea}\@\xspace}
\newcommand{\pineappl}{\texttt{PineAPPL}\@\xspace}

\newcommand{\MSbar}{\overline{\mathrm{MS}}}
\newcommand{\qqbar}{\mathrm{q}\overline{\mathrm{q}}}
\newcommand{\QQbar}{\mathrm{Q}\overline{\mathrm{Q}}}
\newcommand{\ccbar}{\mathrm{c}\overline{\mathrm{c}}}
\newcommand{\bbbar}{\mathrm{b}\overline{\mathrm{b}}}
\newcommand{\ttbar}{\mathrm{t}\overline{\mathrm{t}}}

\newcommand{\hQ}{\mathrm{h}_\mathrm{Q}}
\newcommand{\jpsi}{J/\psi}

\newcommand{\ups}{\Upsilon}

\newcommand{\sigmaQQbar}{\sigma(\QQbar)}
\newcommand{\sigmaccbar}{\sigma(\ccbar)}

\newcommand{\sigmabbbar}{\sigma(\bbbar)}

\newcommand*{\eg}{e.g.\@\xspace}
\newcommand*{\ie}{i.e.\@\xspace}
\newcommand*{\cm}{c.m.\@\xspace}

\newcommand{\orcid}[1]{\href{https://orcid.org/#1}{\hspace*{0.1em}\raisebox{-0.45ex}{\includegraphics[width=1em]{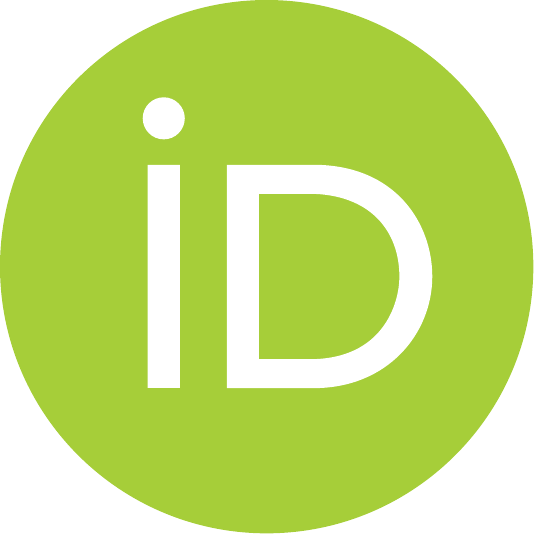}}}}

\newcommand{\mub}{$\mu$b}

\begin{document}

\title{Inclusive charm and bottom quark pair production cross sections\\ at hadron colliders at next-to-next-to-leading-order accuracy}

\author{David~d'Enterria\orcid{0000-0002-5754-4303}}\email{david.d'enterria@cern.ch}
\affiliation{CERN, EP Department, CH-1211 Geneva, Switzerland}
\author{Felix~Hekhorn\orcid{0000-0001-8752-8008}}\email{felix.a.hekhorn@jyu.fi}
\author{Ilkka~Helenius\orcid{0000-0003-1998-038X}}\email{ilkka.m.helenius@jyu.fi}
\affiliation{University of Jyvaskyla, Department of Physics, P.O. Box 35, FI-40014 University of Jyvaskyla, Finland}
\affiliation{Helsinki Institute of Physics, P.O. Box 64, FI-00014 University of Helsinki, Finland}
\author{V\u{a}n~D\~{u}ng~L\^{e}\orcid{0009-0006-2306-1000}}\email{dunglvht@gmail.com}
\affiliation{Helsinki Institute of Physics, P.O. Box 64, FI-00014 University of Helsinki, Finland}
\author{Hannu~Paukkunen\orcid{0000-0001-8815-4255}}\email{hannu.paukkunen@jyu.fi}
\affiliation{University of Jyvaskyla, Department of Physics, P.O. Box 35, FI-40014 University of Jyvaskyla, Finland}
\affiliation{Helsinki Institute of Physics, P.O. Box 64, FI-00014 University of Helsinki, Finland}

\begin{abstract}
\noindent 
The inclusive cross sections for charm ($\ccbar$) and bottom ($\bbbar$) quark-antiquark pair production in proton-proton, proton-antiproton, and proton-nucleus collisions are studied over a wide range of center-of-mass energies, $\sqrts\approx 10$~GeV--400~TeV. All existing data over $\sqrts\approx 10$~GeV--14~TeV are collected and compared to calculations at next-to-next-to-leading-order (NNLO) accuracy using the new fixed-order \MaunaKea open-source code for varying sets of parton distribution functions (PDFs). Relative to next-to-leading-order (NLO) predictions, the NNLO cross sections are enhanced by up to a factor of two, with the associated theoretical scale uncertainties reduced by the same amount, leading to agreement with experimental data over the full range of collision energies. The NNLO results are also compared with NLO predictions obtained within the SACOT-$\mT$ general-mass variable-flavour-number scheme. Despite still sizable theoretical and experimental uncertainties, $\ccbar$ cross section at multi-TeV energies can provide extra constraints on the gluon density at very small-$x$ in global PDF analyses. In the bottom sector, more precise cross section measurements at low energies, $\sqrts\approx 10$--100~GeV, can help constraint the bottom-quark pole mass.

\end{abstract}


\maketitle

\vspace{-0.5cm}

\tableofcontents

\section{Introduction}
\label{sec:intro}

Substantial theoretical progress has been achieved over the past decade in the computation of higher-order corrections for hard-scattering cross sections in hadronic collisions within perturbative quantum chromodynamics (pQCD), in parallel with the steadily improved experimental precision achieved in proton-proton (\pp) measurements at the CERN Large Hadron Collider (LHC). Advances in analytical techniques, numerical methods, and automation have enabled increasingly precise predictions for a wide range of observables~\cite{Huston:2023ofk}. At present, state-of-the-art fixed-order calculations for the cross sections of hard processes in \pp\ collisions are available at next-to-next-to-leading-order (NNLO) accuracy in the expansion of the strong coupling constant, $\alphaS$, often supplemented by resummation of logarithmically enhanced contributions up to next-to-next-to-leading-log (NNLL) accuracy~\cite{Andersen:2024czj,Huss:2025nlt}, with an increasing number of processes being computed at next-to-NNLO (N$^3$LO)~\cite{Caola:2022ayt,Andersen:2024czj,Huss:2025nlt}.

Among the pQCD processes in hadronic collisions, the production of heavy quarks --top, bottom, and charm quarks-- predominantly proceeds  through gluon-gluon fusion processes, $gg\to\QQbar$ (with heavy $\rm Q=c,\,b,\,t$ quarks), and thereby constitute a particularly sensitive probe of the gluon density in the hadrons.
The production of pairs of top ($\ttbar$), bottom ($\bbbar$), and  charm ($\ccbar$) quarks is characterized by intrinsically hard scales, corresponding to at least twice their (pole) quark masses, $\mQ\approx 172.5,\,4.8,\,1.67$~GeV~\cite{ParticleDataGroup:2024cfk} respectively, which are much larger than the nonperturbative QCD scale $\LQCD\approx 0.2$~GeV, thereby enabling the use of the pQCD framework to compute their cross sections. Within the factorization theorem framework~\cite{Collins:1989gx}, these cross sections are obtained from convolutions of partonic subprocess cross sections (``matrix elements''), computed perturbatively as expansions in powers of $\alphaS$, with parton distribution functions (PDFs), evolving according to the DGLAP equations~\cite{Gribov:1972ri,Dokshitzer:1977sg,Altarelli:1977zs}, that encode the longitudinal momentum fraction ($x$) and virtuality ($Q^2$) carried by the constituent partons of the colliding hadrons.
Given the phenomenological significance of the top quark at the LHC, NNLO calculations were first performed for $\ttbar$ production with the \hathor~\cite{Aliev:2010zk} and \toppp~\cite{Czakon:2011xx,Czakon:2013goa,Czakon:2012pz,Czakon:2013goa,Barnreuther:2012wtj} codes, achieving very good agreement with experimental data~\cite{ATLAS:2024kxj,CMS:2024gzs}.
The high precision of both $\ttbar$ data and theory has enabled not only to constrain the high-$x$ region of the gluon PDF~\cite{NNPDF:2021njg, Bailey:2020ooq, Hou:2019efy, Alekhin:2024bhs}, but also accurately measure $\mtop$~\cite{ATLAS:2024kxj,CMS:2024irj}, as well as precisely extract $\alphaS$~\cite{Klijnsma:2017eqp,CMS:2018fks,dEnterria:2022hzv}.

For the production of the lighter charm and bottom quarks, the theoretical state-of-the-art has been for many years the next-to-leading order (NLO) accuracy for their differential cross sections (calculated \eg\ with the \mnr\ code~\cite{Mangano:1991jk}), complemented with next-to-leading-log (NLL) transverse momentum resummation often obtained in a general-mass variable-flavour-number scheme (GM-VFNS)~\cite{Cacciari:1998it,Kniehl:2004fy,Kniehl:2005mk,Kniehl:2007erq,Kniehl:2011bk,Helenius:2018uul,Xie:2021ycd,Helenius:2023wkn} (such as in the \fonll\ code~\cite{Cacciari:2012ny,Cacciari:2015fta}).
Matching fixed-order calculations, \eg\ from \powheg~\cite{Frixione:2007vw,Alioli:2010xd} or \mcatnlo~\cite{Frixione:2007vw}, with a parton-shower (PS) algorithm from general-purpose event generators such as \pythia~\cite{Sjostrand:2006za,Bierlich:2022pfr}, provides an NLO-accurate description of the complete collision event that can be used to compare with experimental measurements applying arbitrary kinematical cuts. More recent developments include differential NNLO predictions~\cite{Catani:2020kkl,Czakon:2021ohs,Mazzitelli:2023znt,Czakon:2024tjr} for $\bbbar$ production (such as in the \Matrix\ framework~\cite{Grazzini:2017mhc}), 
or transverse-momentum-dependent approaches~\cite{Shao:2025dzw}. Whereas inclusive charm-anticharm and bottom-antibottom production cross sections can in principle be computed at NNLO accuracy using the same computational frameworks employed for $\ttbar$ production, as done in~\cite{dEnterria:2016ids,dEnterria:2016yhy,Accardi:2016ndt}, no systematic study has to date compared the full set of available inclusive $\ccbar$ and $\bbbar$ measurements with NNLO predictions including a comprehensive assessment of the associated theoretical uncertainties. Several factors explain this situation. First, the relatively low energy scales involved in inclusive charm production lead to sizable theoretical uncertainties --particularly from missing higher-order pQCD corrections, PDFs, and charm quark mass-- thereby reducing the predictive power of the calculations.
Second, LHC measurements have indicated~\cite{LHCb:2013xam,CMS:2019uws,ALICE:2021dhb,CMS:2023frs} that charmed hadrons are produced at the LHC in different mutual proportions than observed at $\epem$ collisions or in deep-inelastic scattering (DIS) at HERA. The observed relative reduction (increase) of charmed meson (baryon) yields in \pp\ compared with $\epem$ collisions has complicated the proper extraction of inclusive charm production cross sections from the experimentally measured hadronic final states. Similar observations have been made also 
in the case of bottom quarks~\cite{LHCb:2023wbo}, though the available data are currently more limited.

The enhanced heavy-quark baryon-to-meson production ratio observed in \pp\ collisions at the LHC has been interpreted as evidence for environment-dependent heavy-quark hadronization, driven by the high density of colour charges produced at TeV energies, in a situation conceptually similar to that first proposed to account for analogous effects originally seen at heavy-ion colliders~\cite{Sorensen:2005sm,Martinez-Garcia:2007mzr,Oh:2009zj,Andronic:2015wma}. In \pp\ events at the LHC, the presence of large number of semihard multiparton interactions leads to heavy quarks hadronizing within a dense system of surrounding partons and beam remnants, where effects such as colour reconnection~\cite{Altmann:2024kwx}, coalescence (recombination) with nearby light quarks~\cite{Alberico:2011zy,Song:2018tpv,Minissale:2020bif}, and large diquark formation may enhance the probability to form baryons at the expense of mesons compared with the ``vacuum'' fragmentation measured in $\epem$ collisions. Interestingly, the observed baryon-to-meson enhancement has been found to be nearly equal in \pp\ and \pPb\ collisions at the LHC~\cite{ALICE:2024ocs,LHCb:2023wbo}, whereas the surrounding hadronic activity is twice larger in the latter than in the former system~\cite{ALICE:2022kol}, indicating that the system-size dependence of the baryon-enhancement appears to be relatively weak. The LHC data indicate that the enhanced baryon-to-meson production fractions appear at low transverse momentum $\pT$, whereas for increasing $\pT$ values the production fractions become roughly consistent with those from $\epem$ collisions and DIS. From the point of view of collinear factorization, this suggests the presence of significant higher-twist (HT) contributions not describable with universal parton-to-hadron fragmentation functions (FFs) and/or PDFs without some extra modeling. Evidence for the presence of such effects in \pp\ collisions at the LHC has been also suggested even in high-$\pT$ light-hadron production~\cite{dEnterria:2013sgr}. Whether the enhanced baryon-to-meson production ratio can be uniquely classified as a final-state effect, and/or as some other HT dynamics, remains an open question under study.

The main goal of this work is to confront fixed-order NNLO predictions for inclusive charm- and bottom-quark production cross sections, including a systematic assessment of all sources of theoretical uncertainties, with the available experimental data from hadronic collisions over a broad range of center-of-mass (\cm) energies, spanning $\sqrts\approx 10$~GeV--13~TeV, and extending up to the $\sqrts\approx 100$~TeV regime foreseen at the CERN Future Circular Collider (FCC)~\cite{Mangano:2016jyj, FCC:2018vvp}, and even $\sqrts \approx 400$~TeV reached in ultrahigh-energy collisions of cosmic protons with the upper atmosphere~\cite{dEnterria:2018kcz}. To this end, we first discuss the measurements of the charm- and bottom-quark fragmentation fractions used to derive the $\sigmaQQbar$ cross sections from hadronic final states, and compile all existing measurements of inclusive $\ccbar$ and $\bbbar$ production cross sections in \pp, proton-antiproton (\ppbar), and proton-nucleus (\pA) collisions (\cref{sec:exp}). We then introduce a new numerical NNLO implementation code, called \MaunaKea~\cite{felix_hekhorn_2024_14185847}, based on the \toppp\ matrix elements interfaced with the \pineappl\ fast interpolation grids~\cite{Carrazza:2020gss,christopher_schwan_2025_15635174} (\cref{sec:th_MaunaKea}). All sources of theoretical uncertainties, including missing higher-order corrections, PDFs, heavy-quark masses, and $\alphaS$ are examined. The \MaunaKea\ predictions are subsequently compared with those obtained from resummed NLO calculations using the SACOT-$\mT$ GM-VFNS framework (\cref{subsec:resumm}). A detailed assessment of the level of theory and data agreement is presented in \cref{sec:compare}. Finally, the main results are summarized in \cref{sec:summ}. The appendices~\ref{sec:app} and~\ref{sec:app2} provide the charm and bottom NNLO cross sections computed in this work in a tabulated form.

\section{Experimental data}
\label{sec:exp}

The production yields of charm and bottom quarks in hadronic collisions increase significantly with \cm\ energy. At the top LHC energies, they represent about 20\% (charm) and 1\% (bottom) of the total inelastic \pp\ cross section of $\sigma(\pp\to \rm X)\approx 75$~mb~\cite{dEnterria:2016ids}. Virtually all heavy quarks are produced in pairs via QCD processes because single heavy-quark production is possible only through quark-flavour-violating weak interactions that are relatively suppressed by many orders-of-magnitude. Experimentally, one measures the fragmentation products of c- and b-quarks in the form of D, B mesons (including their excited $\rm D^*, B^*$ states), $\Lambda_\mathrm{c,b}$, $\textstyle\sum_\mathrm{c,b}$, $\Omega_\mathrm{c,b}$ baryons, plus quarkonia bound states~\cite{Brambilla:2010cs,Lansberg:2019adr,Chapon:2020heu} of the charmonium ($\rm \jpsi$) and bottomonium ($\ups$) families. Both charm and bottom hadrons decay via the weak interaction and therefore have relatively long lifetimes, leading to typical average decay lengths of $c\tau\approx \SI{150}{\mu\m}$ for charm and $c\tau\approx \SI{470}{\mu\m}$ for bottom hadrons~\cite{ParticleDataGroup:2024cfk}. Since bottom hadrons also decay into charm hadrons, their successive decay chain results in final states with significantly displaced vertices. Collider experiments heavily exploit this property to measure heavy-quark production by identifying the displaced secondary and tertiary vertices of their fragmentation hadron decays with high-resolution silicon vertex detectors. 
We discuss first the experimental charm- and bottom-quark fragmentation fractions, and then provide a systematic compilation of all charm-anticharm and bottom-antibottom measurements performed to date at hadronic colliders.

\subsection{Heavy-quark fragmentation fractions}

Most of the inclusive $\ccbar$ and $\bbbar$ cross sections experimentally determined so far are based on measurements of a fraction of the decay channels into heavy-quark hadrons plus, where needed, model-dependent extrapolations from the visible transverse-momentum and rapidity $(\pT,y)$ range to the full phase space. The final extraction of the inclusive $\sigma(\ccbar)$ and $\sigma(\bbbar)$ cross sections is performed by dividing the measured inclusive heavy-quark hadron cross section, $\sigma(\hQ)$, by the corresponding total charm or bottom decay fragmentation fractions\footnote{
 Hereafter, expressions such as $f(\rm Q \to h_Q)$ refer to heavy-quark or antiquark hadron production that, by charge conjugation, satisfy $f({\rm Q \to h_Q})\equiv f(\rm \overline{Q} \to h_{\overline{Q}})$.}, $f(\rm Q \to h_Q)$:
\begin{align}
\sigma(\QQbar) & =\sigma(\textstyle\sum\hQ)= \frac{\sigma(\hQ)}{\ f(\rm Q \to h_Q)},\, \mbox{ for Q = c, b}. 
\label{eq:FF}
\end{align}
Whenever different heavy-quark hadron species $\mathrm{h_Q}$ are measured simultaneously in the same colliding system, their cross sections and fragmentation fractions are conveniently added up or combined (\eg\ in the case of measurements of ground and excited hadronic states, where the latter feed the yields of the former) to obtain the corresponding $\sigma(\QQbar)$ value. In addition, several $\sigma(\ccbar)$ extractions exist also based on measurements of high-$\pT$ displaced single leptons ($\ell^\pm =$~e$^\pm,\,\mu^\pm$) produced in the charmed hadron decays, which use instead the inclusive fragmentation fraction $f(\rm c \to \ell^+) = (9.71\pm 0.32)\%$~\cite{ALEPH:2005ab} to extrapolate to a total heavy-Q cross section.

\begin{figure}[htpb!]
\centering
\hspace*{0.5cm}\includegraphics[width=0.8\textwidth]{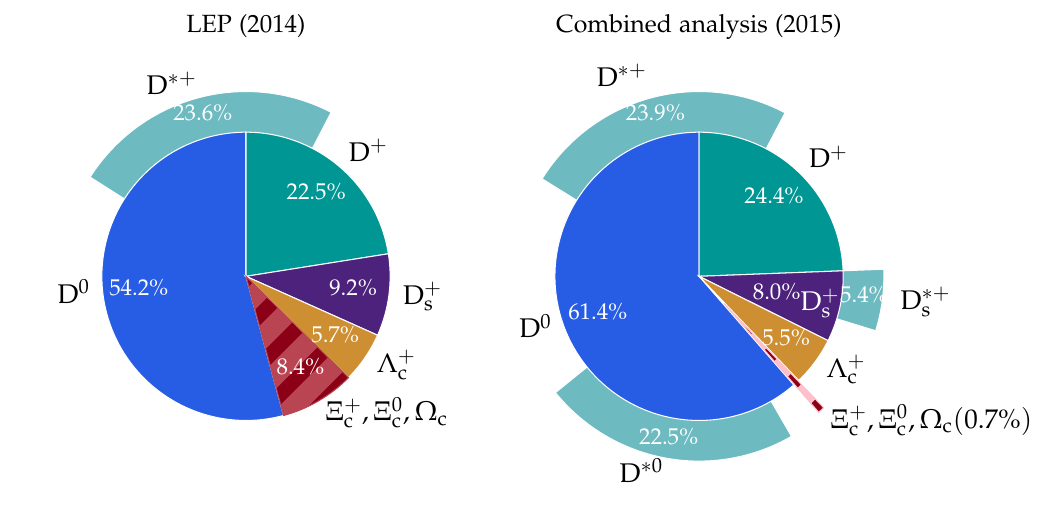}
\includegraphics[width=0.8\textwidth]{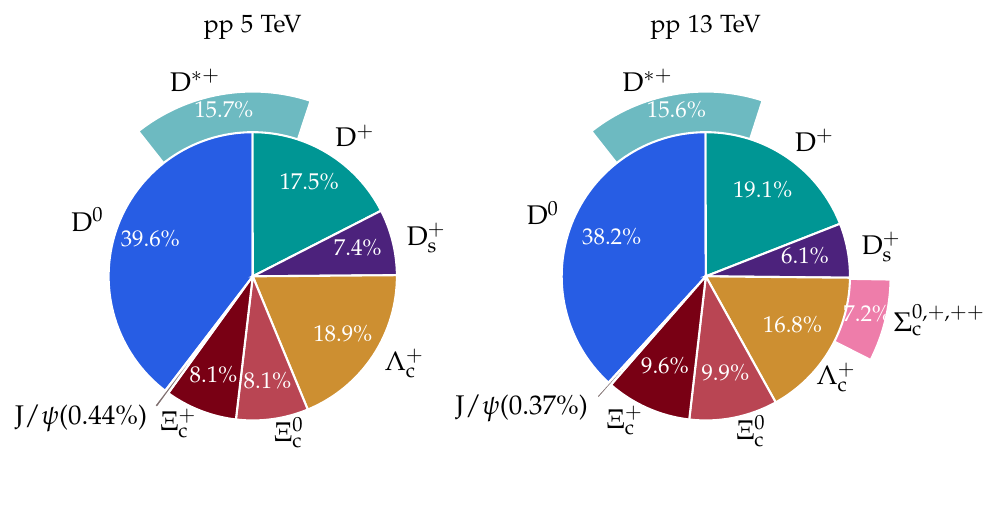}
\caption{Charm-quark fragmentation fractions $f(\rm c\to h_c)$ into specific charmed hadrons as obtained from different analyses using data from $\epem$ at LEP (upper left)~\cite{Gladilin:2014tba}, a combination of $\epem$, DIS and \pp\ collisions (upper right)~\cite{Lisovyi:2015uqa}, and ALICE in \pp\ at 5 TeV (lower left)~\cite{ALICE:2021dhb} and 13 TeV (lower right)~\cite{ALICE:2023sgl}. 
\label{fig:fragfractions_charm}}
\end{figure}

The heavy-quark fragmentation fraction into a specific hadron $f(\rm Q \to h_Q)$ is a soft nonperturbative process, which cannot be calculated with pQCD techniques, and instead one uses experimental data for their determination. The charm-quark fragmentation fractions are shown in Fig.~\ref{fig:fragfractions_charm} as measured in $\epem$ collisions at LEP~\cite{Gladilin:2014tba} (upper left), derived from a combination of selected $\epem$, DIS, and \pp\ collisions~\cite{Lisovyi:2015uqa} (upper right), and determined by ALICE measurements in \pp\ collisions at 5~TeV~\cite{ALICE:2021dhb} (lower left) and 13~TeV~\cite{ALICE:2023sgl} (lower right).

\begin{figure}[htpb!]
\centering
\includegraphics[width=0.95\textwidth]{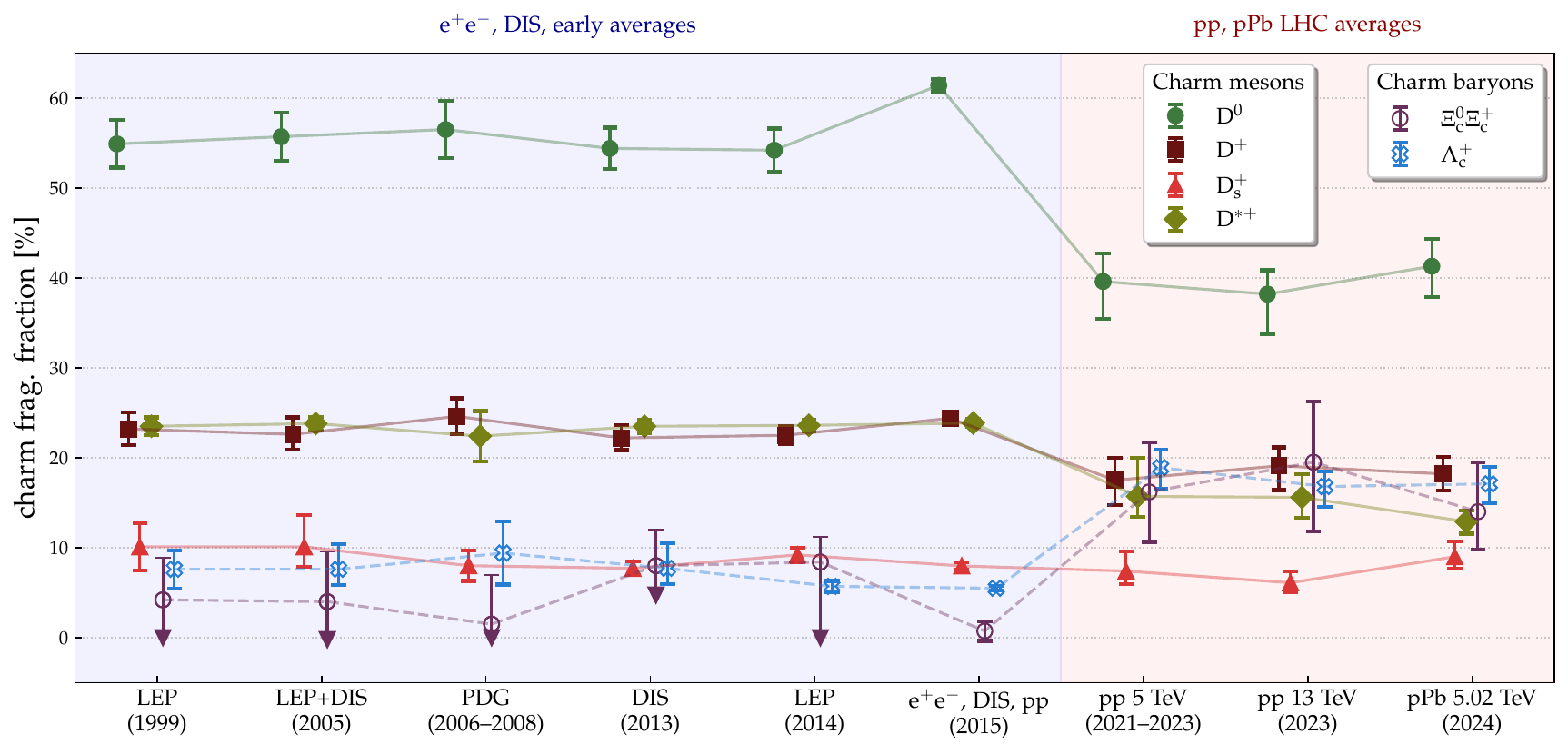}
\caption{Time evolution of charm-quark fragmentation fractions into charm mesons and baryons for different colliding systems as determined from LEP data (1999)~\cite{Gladilin:1999pj}, combined LEP and DIS analysis (2005)~\cite{ZEUS:2005pvv}, reported in the PDG over 2006--2008~\cite{ParticleDataGroup:2008zun}, DIS data update (2013)~\cite{ZEUS:2013fws}, LEP data update (2014)~\cite{Gladilin:2014tba}, combined $\epem$, DIS, and \pp\ analysis (2015)~\cite{Lisovyi:2015uqa}, ALICE \pp\ 5 TeV (2021--23)~\cite{ALICE:2021dhb,ALICE:2023sgl}, ALICE \pp\ 13 TeV (2023)~\cite{ALICE:2023sgl}, and ALICE \pPb\ 5.02 TeV (2024)~\cite{ALICE:2024ocs}. These different fragmentation fractions have been used throughout different time periods to extract $\sigmaccbar$ in hadronic collisions via Eq.~(\ref{eq:FF}) as discussed in the text.
\label{fig:fragfractions_collection}}
\end{figure}

Up until about 2015, the data indicated that charm meson and baryon production represented about 85--90\% and 15--10\%, respectively, of the total charm-quark fragmentation products. However, recent \pp\ measurements at the LHC~\cite{LHCb:2013xam,CMS:2019uws,ALICE:2021dhb,CMS:2023frs, ALICE:2023sgl,ALICE:2024ocs} show an increased proportion of baryons at the expense of mesons, with the mesons (baryons) contributing about 65\% (35\%) of the total charm production yields. This enhanced charm baryon over meson production observed at the LHC, compared with that measured in $\epem$ collisions, is indicative of a violation of charm-quark fragmentation universality. Whether this is due to final-state hadronization effects, as aforementioned, or indicative of other HT phenomena (\eg\ enhanced heavy-quark production among the beam remnants outside of the experiment acceptances, or other possibilities) remains an open question. In any case, extrapolations from charm meson to total $\ccbar$ cross sections using Eq.~(\ref{eq:FF}) at the LHC have to take into account such a non-universality or, otherwise, by using the ``vacuum'' $\epem$ fragmentation fractions, one would extract lower $\sigma(\ccbar)$ values. The evolution over time of the values of the charm hadron-to-quark fragmentation fractions measured across various collision systems is shown in Fig.~\ref{fig:fragfractions_collection}. This compilation plot helps clarify some of the differences in charm cross section values inferred from measurements of specific charmed hadrons at different periods, as discussed below. 

\begin{figure}[htpb!]
\centering
\includegraphics[height=0.28\textheight]{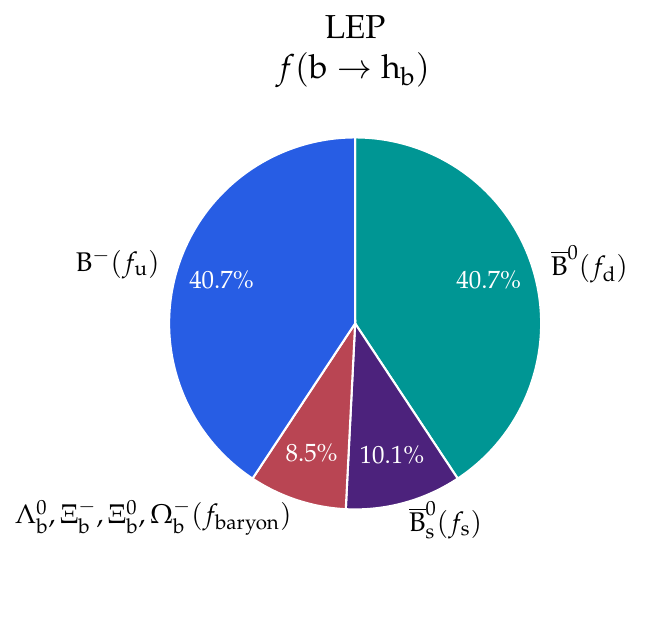}
\includegraphics[height=0.28\textheight]{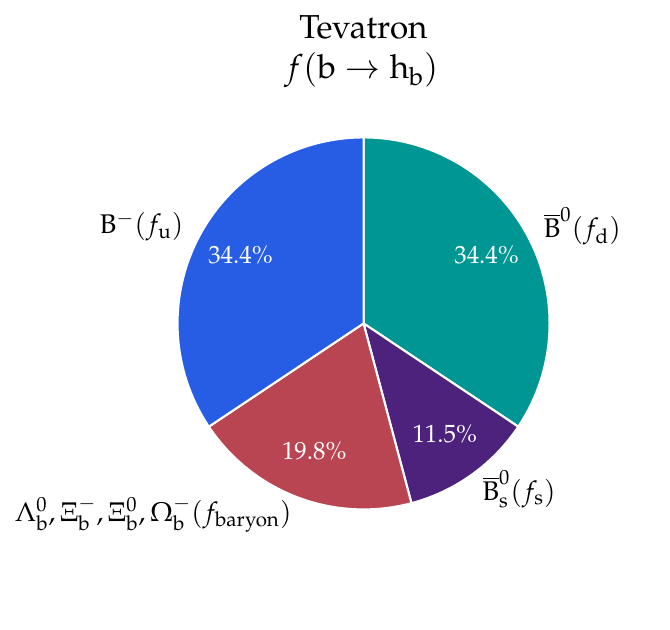}
\includegraphics[height=0.28\textheight]{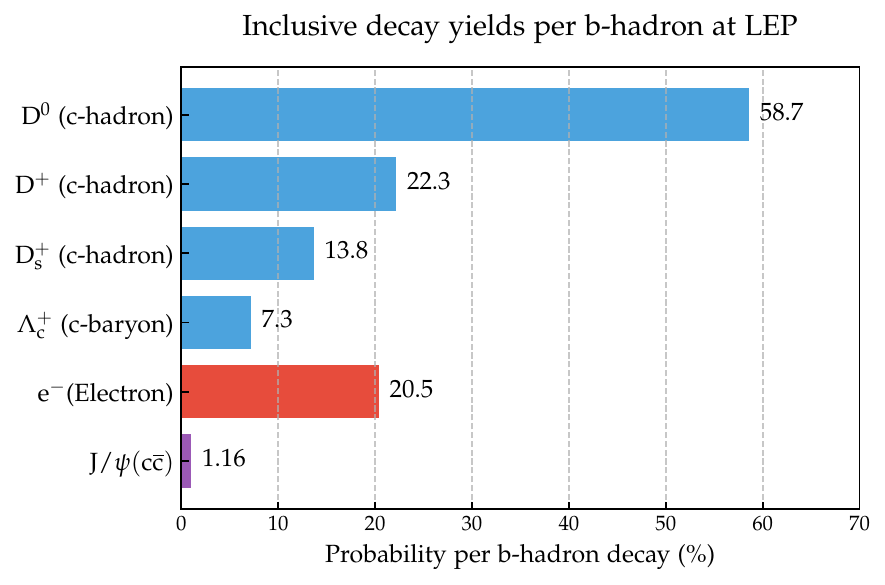}
\caption{Bottom-quark fragmentation fractions into different b-hadrons measured at LEP~\cite{ALEPH:2005ab} (upper left) and Tevatron~\cite{HFLAV:2019otj} (upper right), and inclusive yields to c-hadrons and electrons measured at LEP~\cite{Gladilin:2014tba} (lower panel). The symbols $f_{\rm u}, f_{\rm d}, f_{\rm s}$, and $f_{\rm baryon}$ are typically used to indicate, respectively, the fractions of $\rm B^+$, $\rm B^0$, $\rm B_s^0$, and weakly-decaying bottom baryons in b-quark fragmentation.
\label{fig:fragfractions_bottom}}
\end{figure}

The measurement of $\bbbar$ cross sections relies on the direct reconstruction of B mesons or on the measurement of ``nonprompt'' (n.p.) charmed hadrons or leptons, which themselves originate from the weak decays of bottom-quark hadrons through the decay chain 
$\rm b \to h_b \to h_c (\to \ell^+)$. Such n.p.\ particles are reconstructed farther away from the interaction point, in tertiary charm-hadron decay vertices (where the primary and secondary vertices correspond to the \pp\ collision point, and the B-hadron decay point, respectively). The fragmentation fractions of bottom quarks into different bottom hadrons are shown in Fig.~\ref{fig:fragfractions_bottom} as extracted from $\epem$ collisions at LEP~\cite{ALEPH:2005ab} (upper left) and from \ppbar\ collisions at Tevatron (upper right). The lower histogram shows the b-quark fragmentation fractions into c-hadrons and displaced electrons measured at LEP~\cite{Gladilin:2014tba}. Some evidence of increased bottom baryon-to-meson ratios in hadron-hadron compared with $\epem$ collisions has been observed at Tevatron~\cite{HFLAV:2019otj} (Fig.~\ref{fig:fragfractions_bottom}, upper right chart) and at the LHC at forward rapidities~\cite{LHCb:2023wbo}.

About a half of the experimental $\sigmabbbar$ cross sections are derived based on the measurement of very displaced n.p.\ $\rm \jpsi$ mesons and/or charged leptons $\ell^\pm$ from intermediate charmed-hadron decays. The general method to obtain the total bottom cross sections from such n.p.\ decay products is through the expression,
\begin{align}\label{eq:sigma_tot_hb}
    \sigma(\bbbar) &= \frac{\sigma(\text{n.p. } \rm \jpsi \text{ or } \rm \ell^\pm )}{2\ \mathcal{B}{(\rm \rm b \to \jpsi \text{ or } \rm \ell^+)}} , 
\end{align}
which relies on the well-measured inclusive decay branching fractions $\mathcal{B}{(\rm b \to \jpsi\,X)} = (1.16\pm0.10)\%$ and $\mathcal{B}{(\rm b \to e^\pm\,X)} = (20.5\pm0.7)\%$~\cite{ParticleDataGroup:2024cfk}, and the factor of two in the denominator accounts for the contributions from both b and $\rm \overline{b}$ hadron decays. The $\mathcal{B}({\rm b \to \jpsi\, X})$ value is an average from LEP measurements around the Z peak~\cite{ALEPH:1992oah,L3:1993jrt,DELPHI:1994udj}. These experiments first measure the inclusive branching fraction $\mathcal{B}({\rm Z\to \jpsi\, X}) = (0.351 \pm 0.024)\%$, from which a small fraction of prompt-$\rm \jpsi$ events, $f^\text{prompt}_\mathrm{\jpsi} \approx 2$--8\% is subtracted, to isolate the inclusive $\mathcal{B}({\rm Z\to b \to \jpsi \,X})$ component. Finally, this value is divided by $\mathcal{B}({\rm Z \to \bbbar})=15.12\%$ to obtain the branching fraction:
$\mathcal{B}({\rm h_b \to \jpsi\,X}) = \mathcal{B}({\rm Z \to b \to \jpsi\,X})/(2\,\mathcal{B}({\rm Z\to \bbbar}))$.
The factor of two in the denominator accounts for the fact that each  ${\rm Z \to \bbbar}$ event provides two independent opportunities to produce a $\rm \jpsi$ meson, one from the b quark and one from the $\rm \overline{b}$ antiquark.
The total inclusive electron/positron yield from b-hadrons, $\mathcal{B}({\rm b \,(\to \accentset{(-)}{c}) \to e^\pm\,X}) = (20.5 \pm 0.7)\%$, 
is the sum of three distinct decay components~\cite{DELPHI:1992yxh,L3:1996zkv,OPAL:1999jit,ALEPH:2001skl}: the primary semileptonic decay $\mathcal{B}({\rm b \to e^- \overline{\nu}~ X}) = (10.86\pm 0.35)\%$, the dominant cascade decay 
$\mathcal{B}({\rm b(\nrightarrow e^-\overline{\nu}) \to c \to e^+ \nu X}) = (8.02\pm0.19)\%$,
and the secondary cascade decay 
$\mathcal{B}({\rm b \to \overline{c} \to e^- \overline{\nu}~ X}) = \left(1.6^{+0.5}_{-0.4}\right)\%$
through the charm quark pair production channel ($\rm b\to \ccbar s$).
Because other hadronic Z decays are a copious source of non-bottom leptonic backgrounds, these measurements do not rely on an intermediate inclusive Z branching fraction. Instead, events containing b-hadrons are selected using lifetime tagging. The analysis then exploits the fact that electrons from primary b-hadron decays ($\rm b \to e$) exhibit characteristically large momentum and transverse momentum relative to the jet axis. This allows the extraction of $\mathcal{B}({\rm b \to e^- \overline{\nu} X})$ and the cascade fractions directly by performing multi-parameter fits to various distributions, such as electron momentum, transverse momentum, and dilepton correlations. This fitting procedure statistically separates the primary signal from the cascade decays ($\rm b \to c \to e^+$, and $\rm b \to \overline{c} \to e^-$), primary charm decays (${\rm Z \to c\overline{c} \to e^\pm\,X}$), and non-prompt sources such as photon conversions and misidentified hadrons. Ultimately, these individual experimental measurements were combined by the LEP Electroweak Heavy Flavour Working Group to perform a global fit of multiple electroweak parameters~\cite{ALEPH:1996aa,ALEPH:2005ab}.\\

As a last comment regarding the different heavy-quark fragmentation fractions observed at LEP and the LHC, it is important to keep in mind that the experimental geometric acceptances and kinematic selections for final states in $\epem$ and hadron-hadron collisions differ significantly (and, to some extent, also between the Tevatron and the LHC). Whereas LEP experiments detect almost the full phase space for hadron production, this is not true for hadron-collider experiments where a fraction of the particle activity goes undetected below $\pT$ thresholds and/or beyond the detector rapidity coverage (\eg\ along the beam direction). Consequently, simple combinations of the LEP and LHC fragmentation fractions into a global average are unsuitable. 
In this context, LHC heavy-quark hadron measurements can be more consistently extrapolated by either using the fragmentation fractions measured in situ, or alternatively by employing the LEP fractions as a vacuum fragmentation baseline, but stating clearly the underlying assumptions of either choice.

\subsection{Measurements of $\ccbar$ cross sections} 
\label{sec:charm}

Table~\ref{tab:charmmeasurements} collects more than 50 measurements of charm cross sections in hadronic collisions from $\sqrtsNN \approx 11$~GeV (fixed target) up to 13 TeV (LHC). For each result, the first three columns list, respectively, the experiment and colliding system\footnote{For entries with multiple citations, the first one typically provides the numerical values listed, while the subsequent references describe similar previous (often superseded) measurements performed by the same experiment for the same colliding system and final states.}, the detected charm-hadron species (or their decay particles), and their fiducial kinematic range. The latter include the $\pT$ and $y$ (or pseudorapidity $\eta$), and/or invariant mass ($m_\mathrm{ee}$), and/or longitudinal momentum fraction ($x_\mathrm{F}$) selection criteria applied by each experimental analysis. The three next columns list, respectively, the fragmentation fraction $f(\rm c \to h_c)$ used to extrapolate the hadron cross section $\sigma(\hQ)$ to a $\sigmaccbar$ value via Eq.~\eqref{eq:FF}, the measured fiducial cross section, and the typical extrapolation factor used to convert the fiducial into a total 
cross section, usually obtained with MC simulations via
\begin{align}
    \sigma_\text{tot}(\textstyle\sum \hQ) &= \left(\frac{\sigma_\text{tot}^\text{MC}(\textstyle\sum \hQ)}{\sigma_\text{fid}^\text{MC}(\textstyle\sum \hQ)}\right) \cdot \sigma_\text{fid}(\textstyle\sum\hQ) \,\mbox{ with MC = \fonll, \powheg, \pythia\,{\small 6}, \pythia\,{\small 8},...
    }\label{eq:total_phase_extrapolation}
\end{align}
Typical fiducial extrapolation factors amount to 5--10, with relatively large uncertainties due to the lack of experimental data in the detector ``blind spots'', at low $\pT$ and/or at very forward-backward rapidities, which limit the validation of MC model predictions. The last column lists the final derived charm-anticharm production cross section. Whenever two (three) uncertainties are quoted, the first one corresponds to the statistical uncertainty, and the second to the combined systematic and extrapolation uncertainty (or the third to the extrapolation uncertainty alone). In case a given publication gives several $\ccbar$ cross sections obtained from different MC generator extrapolations via Eq.~(\ref{eq:total_phase_extrapolation}), we quote their average and add in quadrature the standard deviation of the different MC central values as an additional uncertainty.
All \pA\ (and the few d-A) cross sections are quoted normalized per nucleon-nucleon collision, \ie\ divided by nuclear mass number $A$ (or by $2A$ for the deuteron-nucleus collisions). 

\begin{table}[htbp!]
 \caption{Experimental cross sections for charm-quark pair production in \pp, \ppbar, and \pA\ collisions from $\sqrts \approx 11$~GeV to 13~TeV. For each colliding system, we list the measured charmed hadrons (or their decay products), the fiducial  $(\pT,\, y,\,\eta,\,x_\mathrm{F},\,m_{\ell\ell})$ range, the hadron-to-charm-quark fragmentation fraction, the measured fiducial $\sigmaccbar$ value, the extrapolation factor to the total phase space, and the final inclusive $\sigmaccbar$ cross section. Whenever two (three) uncertainties are given, the first refers to statistical, the second to systematics plus extrapolation (the third to extrapolation alone) sources. Results indicated as `(t.w.)' are obtained as described in the text, with only one total uncertainty quoted (quadratic sum of all uncertainties). Rows in gray indicate cross section values compared with theoretical ones in Fig.~\ref{fig:ccbar_exp_xsecs}.
\label{tab:charmmeasurements}}
\centering
\resizebox{\textwidth}{!}{
\begin{tabular}{llccccclllllcc}
\toprule
 & \makecell[c]{detected \\particles} & \makecell[c]{fiducial kinematic \\phase space} & \makecell[c]{$f(\rm c\to h_c)$ \\fragmentation fraction} & \makecell[c]{fiducial cross section (\mub) \\$\sigma \pm \Delta\sigma_\mathrm{stat} \pm \Delta\sigma_\mathrm{sys} \pm \Delta\sigma_\mathrm{extr}$ } & \makecell[c]{full phase space \\extrapolation factor} & \makecell[c]{total cross section (pp equiv.) (\mub) \\$\sigma \pm \Delta\sigma_\mathrm{stat} \pm \Delta\sigma_\mathrm{sys} \pm \Delta\sigma_\mathrm{extr}$  \mub} \\
experiment ($\sqrts$) &  &  &  &  &  &  \\
\midrule

 ALICE (13 TeV pp)~\cite{ALICE:2023sgl} & $ \rm D^0,D^+,D_s^+,D^{*+},\Lambda_c^+,\Xi_c^+$ & all $\pT$, $|y|<0.5$ & none ( $\sum\rm h_c$) & $2031\pm 61\ ^{+135}_{-141}\ ^{+230}_{-140}$ & $10^{+50\%}_{-10\%}$~(t.w.) & $20000^{+11000}_{-3000}$ ~~(t.w.) \\

 ALICE (13 TeV pp)~\cite{ALICE:2018gev} & $ \rm \rm e^\pm$ & all $\pT$, $|y|<0.5$ & PDG (2016) & $1200 \pm 400$~(t.w.) & $10^{+50\%}_{-10\%}$~(t.w.) & $12000^{+7000}_{-4000}$ ~~(t.w.) \\

 LHCb (13 TeV pp)~\cite{LHCb:2015swx} & $ \rm D^0,D^+,D_s^+,D^{*+}$ & $\pT$=0--8 GeV, $2<y<4.5$ & PDG (2008) & $2369 \pm 3 \pm 192$  & $5^{+50\%}_{-50\%}$~(t.w.) & $12000 \pm 6000$ ~(t.w.) \\

 CMS (13 TeV pp)~\cite{CMS:2021lab} & $ \rm D^0,D^+,D^{*+}$ & $\pT$=4--100 GeV, $|\eta|<2.1$ & 13 TeV pp & $1200 \pm 200$~(t.w.) & $9.9^{+50\%}_{-10\%}$~(t.w.) & $12000^{+7000}_{-3000}$ ~(t.w.) \\

 \rowcolor{lightgray} ALICE+LHCb comb. (13 TeV pp)~\cite{Yang:2025qzo} & $ \rm D^0$ & (combined) & 13 TeV pp & (combined) & -- & $17430^{+2700}_{-1960} $ \\
\hline
 LHCb (8.16 TeV pPb)~\cite{LHCb:2023rpm} & $ \rm D^+,D_s^+$ & \Gape[0pt][2pt]{\makecell{ $\pT$=1--13 GeV\\, $1.5<y<4$ , $-5<y<-2.5$}} & 5.02 TeV pp & $5100 \pm 500$ ~(t.w.) & $3^{+50\%}_{-10\%}$~(t.w.) & $15000^{+8000}_{-2000}$ ~(t.w.) \\
\hline
 ALICE (7 TeV pp)~\cite{ALICE:2021dhb} & $ \rm D^0,D^+,D_s^+,\Lambda_c^+,\Xi_c$ & all $\pT$, $|y|<0.5$ & 5.02 TeV pp & $1347\pm 97\pm 104\ ^{+163}_{-131}$ & $8.6^{+50\%}_{-10\%}$~(t.w.) & $12000^{+6000}_{-2000}$~(t.w.) \\

 ALICE (7 TeV pp)~\cite{ALICE:2012acz} & $ \rm \rm e^\pm$ & all $\pT$, $|y|<0.5$ & LEP+DIS (2005) & $1100\pm 200\ ^{+600}_{-700}\ ^{+200}_{-100}$ & $8.8^{+35\%}_{-10\%}$ & $9700 \pm 1700^{+5200}_{-5600}\ ^{+3400}_{-500}$ \\

 ALICE (7 TeV pp)~\cite{ALICE:2018fvj} & $ \rm \rm e^\pm$ & \Gape[0pt][2pt]{\makecell{$\pT>0.4$ GeV, $|y|<0.8$,\\ $m_{\rm ee}$=1.1--2.7 GeV}} & PDG (2008) & -- & -- & $9700 \pm 3000$  \\

 ALICE (7 TeV pp)~\cite{ALICE:2012mhc},\cite{ALICE:2012inj} & $ \rm D^0,D^+,D^{*+}$ & all $\pT$, $|y|<0.5$ & LEP+DIS (2005) & $1000^{+300}_{-400}$~(t.w.) & $8.5^{+60\%}_{-5\%}$ & $8500 \pm 500^{+1100}_{-2400}\ ^{+5000}_{-400}$ \\

 ALICE (7 TeV pp)~\cite{ALICE:2017thy} & $ \rm \Lambda_c^+$ & $\pT$=1--8 GeV, $|y|<0.5$ & 5.02 TeV pp & $1200^{+400}_{-400}$~(t.w.) & $10^{+50\%}_{-10\%}$~(t.w.) & $12000^{+7000}_{-4000}$ ~(t.w.) \\

 ALICE (7 TeV pp)~\cite{ALICE:2016yta} & $ \rm D^0,D^+,D_s^+,D^{*+}$ & all $\pT$, $|y|<0.5$ & LEP (2014) & $988 \pm 81^{+126}_{-206}$  & $(8.57^{+2.52}_{-0.38})/1.034$ & $8180 \pm 670^{+1010}_{-1680}\ ^{+2400}_{-360}$ \\

 ALICE (7 TeV pp)~\cite{ALICE:2017olh} & $ \rm D^0,D^+,D_s^+,D^{*+}$ & all $\pT$, $|y|<0.5$ & LEP (2014) & $954\pm 69\pm 97$ & $(8.56^{+2.51}_{-0.42})/1.034$ & $7890 \pm 570 \pm 760^{+2320}_{-450}$ \\

 LHCb (7 TeV pp)~\cite{LHCb:2013xam} & $ \rm D^0,D^+,D_s^+,D^{*+},\Lambda_c$ & $\pT<$ 8 GeV, $2<y<4.5$ & PDG (2008) & $1419 \pm 12 \pm 133$  & -- & -- \\

 \rowcolor{lightgray} ALICE+LHCb (7 TeV pp)~\cite{ALICE:2017olh,LHCb:2013xam} & $ \rm D^0,D^+,D_s^+,D^{*+},\Lambda_c$ & (combined) & LEP (2014), PDG (2008) & (combined) & -- & $7440 \pm 140 \pm 570^{+130}_{-70}$ \\

 ATLAS (7 TeV pp)~\cite{ATLAS:2015igt} & $ \rm D^+,D_s^+,D^{*+}$ & $\pT$=3.5--100 GeV, $|\eta|<2.1$ & LEP (2014) & $720 \pm 70$~(t.w.) & (12--14) $\ ^{+44\%}_{-40\%}$ & $8600 \pm 300 \pm 800^{+3800}_{-3400}$ \\

 \rowcolor{lightgray} CMS (7 TeV pp)~\cite{CMS:2024rfu} & $ \rm D^{*+}$ & $\pT>$  GeV 1 GeV, $|\eta|<2.5$ & 5.02 TeV pp & -- & -- & $9390^{+1350}_{-1490}$   \\

 CMS (7 TeV pp)~\cite{CMS:2011rxs} & $ \rm \jpsi$, $\psi(2S)$ & $\pT^\mathrm{\jpsi}$=8--30 GeV, $|y|<2.4$ & 5.02 TeV pp & $85 \pm 15$~(t.w.) & -- & -- \\

 CMS (7 TeV pp)~\cite{CMS:2010nis} & $ \rm \jpsi$ & $\pT^\mathrm{\jpsi}$=6.5--30 GeV, $|y|<2.4$ & 5.02 TeV pp & $140 \pm 30$~(t.w.) & -- & -- \\
\hline
 ALICE (5.02 TeV pPb)~\cite{ALICE:2024ocs} & $ \rm D^0,D^+,D_s^+,D^{*+},\Lambda_c^+,\Xi_c^0$ & all $\pT$, $-0.96<y<0.04$ & 5.02 TeV pPb & $1060 \pm 30\ ^{+50}_{-60}\pm 150$  & $16^{+50\%}_{-10\%}$ & $8700^{+4600}_{-1600}$ ~(t.w.) \\

 ALICE (5.02 TeV pp)~\cite{ALICE:2023sgl}~\cite{ALICE:2021mgk}\cite{ALICE:2019nxm}\cite{ALICE:2020wla}\cite{ALICE:2021dhb} & $ \rm D^0,D^+,D_s^+,D^{*+},\Lambda_c^+,\Xi_c^+$ & all $\pT$, $|y|<0.5$ & none ( $\sum\rm h_c$) & $1148\pm 43^{+62}_{-65}\ ^{+124}_{-84} $ & $8.1^{+50\%}_{-10\%}$~(t.w.) & $9300^{+4800}_{-1300}$ ~(t.w.) \\

 LHCb (5.02 TeV pp)~\cite{LHCb:2016ikn} & $ \rm D^0,D^+,D_s^+,D^{*+}$ & $\pT<$ 8 GeV, $2<y<4.5$ & PDG (2008) & $1193 \pm 3 \pm 89$  & $4.8^{+50\%}_{-10\%}$~(t.w.) & $5700^{+2900}_{-700}$ ~(t.w.) \\

 LHCb (5.02 TeV Pbp)~\cite{LHCb:2017yua} & $ \rm D^0$ & \Gape[0pt][2pt]{\makecell{ $\pT<$ 10 GeV\\, $1.5<y<4$ , $-5<y<-2.5$}} & 5.02 TeV pPb & $2800 \pm 200$ ~(t.w.) & $2.4^{+50\%}_{-10\%}$ & $6800^{+3400}_{-800}$ ~(t.w.) \\

 CMS (5.02 TeV pp)~\cite{CMS:2017uuv} & $ \rm \jpsi$, $\psi(2S)$ & $\pT^\mathrm{\mu\mu}$=6.5--50 GeV, $|y|<2.4$ & 5.02 TeV pp & $43^{+6}_{-6}$~(t.w.) & -- & -- \\

 CMS (5.02 TeV pp)~\cite{CMS:2017qjw} & $ \rm D^0$ & $\pT$=2--100 GeV, $|y|<1$ & 5.02 TeV pp & $880 \pm 170$~(t.w.) & $7^{+50\%}_{-10\%}$~(t.w.) & $6100^{+3300}_{-1300}$ ~(t.w.) \\

 CMS (5.02 TeV pp)~\cite{CMS:2019uws} & $ \rm \Lambda_c^+$ & $\pT$=5--20 GeV, $|y|<1$ & 5.02 TeV pp & $60 \pm 20$~(t.w.) & $40^{+50\%}_{-10\%}$~(t.w.) & $2300^{+1500}_{-900}$ ~(t.w.) \\

 \rowcolor{lightgray} ALICE+LHCb+CMS (5.02 TeV pp)~\cite{Bierlich:2023ewv} & $ \rm D^0,D^{*+},D^+,D_s^+,\Lambda_c^+,\Xi_c$ & (combined) & dynamical hadronisation & (combined) & -- & $8340\ ^{+220}_{-220}\ ^{+370}_{-370}\ ^{+769}_{-460}$  \\
\hline
 \rowcolor{lightgray} ALICE (2.76 TeV pp)~\cite{ALICE:2012inj} & $ \rm D^0,D^+,D^{*+}$ & all $\pT$, $|y|<0.5$ & LEP+DIS (2005) & $660^{+280}_{-240}$~(t.w.) & $7.3^{+60\%}_{-10\%}$ & $4800 \pm 800^{+1000}_{-1300}\ ^{+2600}_{-400}$ \\

 ALICE (2.76 TeV pp)~\cite{ALICE:2021dhb} & $ \rm D^0,D^+,D_s^+,\Lambda_c^+,\Xi_c$ & all $\pT$, $|y|<0.5$ & 5.02 TeV pp & $1126\pm 303\ ^{+258}_{-429}\ ^{+420}_{-140}$ & $7.3^{+50\%}_{-10\%}$~(t.w.) & $8200^{+5900}_{-4100}$~(t.w.) \\
\hline
 \rowcolor{lightgray} CDF (1.96 TeV p\=p)~\cite{CDF:2016lht} & $ \rm D^+$ & $\pT=1.5\mbox{--}14.5$ GeV, $|y|<1$ & LEP (2014) & $320 \pm 50$~(t.w.) & $5.2^{+50\%}_{-10\%}$~(t.w.) & $1700^{+900}_{-300}$ ~(t.w.) \\
\hline
 PHENIX (0.2 TeV pp)~\cite{PHENIX:2013klq} & $ \rm \rm e^\pm,\mu^\pm$ & -- & PDG (2006) & -- & -- & $538 \pm 46 \pm 197 \pm 174$  \\

 PHENIX (0.2 TeV pp)~\cite{PHENIX:2008qav} &  $\rm e^\pm$ & all $\pT$, $|y|<0.5$ & PDG (2006) & $118 \pm 8 \pm 50 $  & $4.6$ & $544 \pm 39 \pm 142 \pm 200$  \\

 \rowcolor{lightgray} PHENIX (0.2 TeV pp)~\cite{PHENIX:2017ztp} & $ \rm \rm e^\pm$ & all $\pT$, $|y|<0.5$ & PDG (2006) & $130 \pm 70$ ~(t.w.) & $3.4\pm 0.2$ & $440 \pm 230$ \\

 PHENIX (0.2 TeV dAu)~\cite{PHENIX:2017ztp} & $ \rm \rm e^\pm$ & -- & PDG (2006) & -- & -- & $490 \pm 280$ \\

 PHENIX (0.2 TeV pp)~\cite{PHENIX:2005iez} & $ \rm \rm e^\pm$ & all $\pT$, $|y|<0.5$ & PDG (2006) & $200\pm 30\pm 110$ & $4.6$ & $920 \pm 150 \pm 540$  \\

 PHENIX (0.2 TeV pp)~\cite{PHENIX:2006tli} & $ \rm \rm e^\pm$ & all $\pT$, $|y|<0.5$ & PDG (2006) & $123 \pm 12 \pm 37$  & $4.6$ & $567 \pm 57 \pm 193$  \\

 PHENIX (0.2 TeV pp)~\cite{PHENIX:2010xji} & $ \rm \rm e^\pm$ & all $\pT$, $|y|<0.5$ & PDG (2006) & $119 \pm 12 \pm 38$  & $4.6$ & $551 \pm 57 \pm 195$  \\

 \rowcolor{lightgray} STAR (0.2 TeV pp)~\cite{STAR:2012nbd} & $ \rm D^0,D^*$ & all $\pT$, $|y|<0.5$ & PDG (2008) & $170 \pm 45^{+38}_{-59}$  & $4.7 \pm 0.7$ & $797 \pm 210^{+171}_{-270} \pm 119$  \\
\hline
 \rowcolor{lightgray} LHCb (86.6 GeV pHe)~\cite{LHCb:2018jry} & $ \rm \jpsi, D^0$ & all $\pT$, $2<y< 4.6$ & LEP (2014) & $75 \pm 10$~(t.w.) & $1.93$ & $144\pm 12\pm 4$ \\

 \rowcolor{lightgray} LHCb (68.5~GeV pNe)~\cite{LHCb:2022cul} & $ \rm D^0$ & all $\pT$, $-2.29<y<0$ & LEP (2014) & $45 \pm 5$~(t.w.) & $2.02$ & $90 \pm 10$~(t.w.) \\

 \rowcolor{lightgray} HERA-B (41.6~GeV pA, A=C,Ti,W)~\cite{HERA-B:2007rfd} & $ \rm D^0,D^+,D_s^+, D^{*+}$ & all $\pT$, $ x_{\rm F}\in [-0.15,0.05]$ & PDG (2006) & $27 \pm 5$~(t.w.) & $1.82\pm 0.17$  & $49 \pm 5 \pm 7$  \\

 \rowcolor{lightgray} E653 (38.9~GeV pA, A=plastic)~\cite{FermilabE653:1991vmo} & $ \rm \mu^\pm$ & -- & LEP (1999) & -- & -- & $48 \pm 6 \pm 11$  \\

 \rowcolor{lightgray} E789 (38.8 GeV pAu)~\cite{Lourenco:2006vw}\cite{E789:1994nhc} & $ \rm D^0,\bar{D^0}$ & -- & LEP (2014) & -- & -- & $14 \pm 5$ ~(t.w.) \\

 \rowcolor{lightgray} E743 (38.2~GeV pp)~\cite{Lourenco:2006vw}\cite{Ammar:1988ta} & $ \rm D^0,D^+$ & -- & LEP (1999) & -- & -- & $29^{+6}_{-5} \pm 5$  \\

 \rowcolor{lightgray} NA50 (28.9 GeV pA)~\cite{Lourenco:2006vw}\cite{NA38:2000wlp} & $ \rm \jpsi$ & -- & LEP (1999) & -- & -- & $36 \pm 9$   \\

 \rowcolor{lightgray} NA27 (27.4~GeV pp)~\cite{Lourenco:2006vw}\cite{LEBC-EHS:1988oic} & $ \rm D^0,D^+,D^*$ & -- & LEP (1999) & -- & -- & $18 \pm 2$  \\

 \rowcolor{lightgray} NA16 (26.0~GeV pp)~\cite{Lourenco:2006vw}\cite{LEBC-EHS:1983llu} & $ \rm D^0,D^+$ & -- & LEP (1999) & -- & -- & $19^{+10}_{-6}$  \\

 \rowcolor{lightgray} E769 (21.7~GeV pA, A=mixed)~\cite{Lourenco:2006vw}\cite{E769:1996jqf} & $ \rm D^0,D^+,D_s^+,D^*,\Lambda_c$ & -- & LEP (1999) & -- & -- & $11 \pm 2 \pm 0.8$  \\

 \rowcolor{lightgray} NA32 (19.5 GeV pA) \cite{Lourenco:2006vw}\cite{ACCMOR:1988pxc,ACCMOR:1990qxc} & $ \rm D^0,D^+,D_s$ & -- & LEP (1999) & -- & -- & $3 \pm 1$  \\

 SVD-1 (11.6~GeV pp)~\cite{SVD:2001lek} & $ \rm D^0,D^+$ & -- & LEP (1999) & -- & -- & $2^{+1}_{-0.7} \pm 0.3$ \\

 \rowcolor{lightgray} SVD-2 (11.6~GeV pA, A=C,Si,Pb)~\cite{SVD-2:2017ovj} & $ \rm D^0,\bar{D}^0,D^\pm,\Lambda_c^+$ & -- & none ( $\sum\rm h_c$) & -- & -- & $7 \pm 2 \pm 1$  \\
\bottomrule
\end{tabular}

}
\end{table}

Most listed values in Table~\ref{tab:charmmeasurements} are taken directly from the bibliographical references indicated in the first column, unless marked as `(t.w.)' (standing for `this work') in which case they have been obtained from our studies here by adding up the measured hadronic fiducial cross sections, applying the quoted fragmentation fractions to obtain fiducial charm-level cross section, and extrapolating them to the full phase space with the listed factors to obtain the final $\sigmaccbar$ values. All results estimated here this way have a single uncertainty obtained by adding all sources in quadrature and rounding it up conservatively. In the before-last column, the phase-space extrapolation factors are either taken directly from the original papers, or estimated here from \fonll\ predictions for each colliding system, using the central \verb|CTEQ6.6| PDFs~\cite{Nadolsky:2008zw} (except for 8.16 TeV where we used \verb|NNPDF30_nlo_as_0118|~\cite{NNPDF:2014otw}) and conservatively assigning $+50\%$ and $-10\%$ uncertainties based on the maximum envelope of differences found between our estimate and other factors used in the literature.

Most final cross sections obtained in this way, for a given colliding system and \cm\ energy, are mutually consistent within their (often large) uncertainties. A few exceptions to this rule are results obtained from measurements of rare hadronic final states (\eg $\Lambda_\mathrm{c}$ mesons) and/or within a reduced fiducial phase space, which have very large extrapolation (fragmentation and phase space) factors, and feature significant lower cross section values. One motivation to provide such a long list of (even superseded) measurements is to observe the impact of several ingredients used in the cross sections derivation. One immediate observation of comparing the $\ccbar$ cross sections for the same colliding system and \cm\ energy is that the choice of ``vacuum'' (``medium'') charm-quark fragmentation fractions can reduce (enhance) the final quoted values by up to 30\%. The rows highlighted in gray in Table~\ref{tab:charmmeasurements} denote those that we consider to be the most accurate and/or precise $\sigmaccbar$ determinations, which are then compared with the theoretical NNLO predictions in Fig.~\ref{fig:ccbar_exp_xsecs} later in this work. This selection is based on several criteria. First, in general we exclude the oldest results that have been superseded by, or combined into, a more recent measurement by the same collaboration(s). The selected results typically cover similar kinematics as the older extractions, but feature smaller uncertainties and/or more detected particles. Second, in general LHC cross sections derived from the sum of ``all" charmed hadrons (indicated as `$\textstyle\sum \rm h_c$' in the table), are preferred over those obtained from fragmentation fractions, $f(\rm c\to h_c)$. Third, \pp\ collisions are prioritized over proton-nucleus collisions to minimize uncertainties due to nuclear PDFs, unless no \pp\ data are available at a given $\sqrts$ value. We note that a few of the values in Table~\ref{tab:charmmeasurements} that combine measurements from independent LHC collaborations, such as~\cite{Bierlich:2023ewv,Yang:2025pmq}, have not been officially obtained by the quoted LHC collaborations themselves.

\subsection{Measurements of $\bbbar$ cross sections} 
\label{sec:bottom}

Table~\ref{tab:beautymeasurements} collects about 50 measurements of bottom cross sections in hadronic collisions from $\sqrts \approx 40$~GeV (fixed target) to 13 TeV (LHC). 
The descriptions of the columns of Table~\ref{tab:beautymeasurements} follow those for the corresponding charm-quark Table~\ref{tab:charmmeasurements}, except that now the bottom fragmentation fraction is not listed as the analyses use different $(\rm b \to h_b)$, $(\rm b \to h_b \to h_c)$, or $(\rm b \to h_b \to h_c \to \ell^\pm)$ decay chains, as discussed previously, and in general the bottom-quark fragmentation fractions implemented in the MC generators employed to correct the experimental measurements follow closely the LEP percentages shown in Fig.~\ref{fig:fragfractions_bottom}.

About half of the values of the table, marked as `(this work)', have been obtained here as done for the charm-quark case, by adding up the b-hadrons fiducial cross sections, and then converting those to quark-level cross section via LEP bottom fragmentation fractions (Fig.~\ref{fig:fragfractions_bottom}).
As done for the similar $\sigmaccbar$ cross sections derived in the previous section, the fiducial phase-space extrapolation factors are either taken directly from the original papers, or estimated here from \fonll\ predictions for each colliding system, using the central \verb|CTEQ6.6| PDFs (except for 8.16 TeV where we used \verb|NNPDF30_nlo_as_0118|) and conservatively assigning uncertainties of $\pm20\%$ (for the b-hadrons) or $\pm50\%$ (for the displaced leptons) based on the maximum envelope of differences found between our estimates and other factors used in the literature. All our derived `(t.w.)' results are consistent within uncertainties with the extractions obtained directly by the experimental collaborations. For \pp\ collisions at $\sqrts = 7$~TeV, we have extrapolated six fiducial cross-section measurements from CMS, corresponding to various different b-quark final states, which all yield mutually consistent $\sigmabbbar$ cross section values. We thus provide their combination, obtained as a standard average of the individual measurements. The resulting $\sigmabbbar = 300\pm 100~\mu$b value, with a conservative uncertainty assigned, has been also added to the table.
The experimental results listed in gray are considered the most accurate and/or precise determinations (often based on the most recent measurements, and following similar criteria as discussed previously for the charm cross sections), and are selected for comparison with theoretical NNLO predictions in Fig.~\ref{fig:bbbar_exp_xsecs} later on. We note that the entry of Ref.~\cite{Bai:2024pxk}, which combines ALICE and LHCb data, does not correspond to an official combination of measurements by these LHC collaborations.

\begin{table}[htbp!]
\caption{Experimental cross sections for bottom-quark pair production in \pp, \ppbar, and \pA\ collisions over $\sqrts \approx 40$~GeV--13~TeV. For each colliding system, we list the measured bottom hadrons or their nonprompt (`n.p.') decay products, the fiducial $(\pT,\, y,\,\eta,\,x_\mathrm{F},\,m_{\ell\ell})$ range, the measured fiducial $\sigmabbbar$ value, and the final inclusive $\sigmabbbar$ cross section derived. Whenever two (three) uncertainties are given, the first refers to statistical, the second to systematics plus extrapolation (and the third to extrapolation) sources. Results indicated as `(t.w.)' (`this work') are obtained as described in the text, with only one total uncertainty quoted (quadratic sum of all uncertainties). Rows in gray indicate cross section values compared with theoretical ones in Fig.~\ref{fig:bbbar_exp_xsecs}.
}
\label{tab:beautymeasurements}
\centering
\resizebox{\textwidth}{!}{
\begin{tabular}{llccccclllllcc}
\toprule
 & \makecell[c]{detected \\particles} & \makecell[c]{fiducial kinematic \\phase space} & \makecell[c]{fiducial cross section (nb) \\$\sigma \pm \Delta\sigma_\mathrm{stat} \pm \Delta\sigma_\mathrm{sys} \pm \Delta\sigma_\mathrm{extr}$ } & \makecell[c]{full phase space \\extrapolation factor} & \makecell[c]{total cross section (pp equiv.) (nb) \\$\sigma \pm \Delta\sigma_\mathrm{stat} \pm \Delta\sigma_\mathrm{sys} \pm \Delta\sigma_\mathrm{extr}$} \\
experiment ($\sqrts$) &  &  &  &  &  \\
\midrule

 ALICE (13.6 TeV pp)~\cite{ALICE:2026xuh} & $ \rm B^0$  & all $\pT$, $|y|<0.5$ & $\left(59^{+35}_{-35}\right) \times 10^{3}$ & $7.6\pm 20\%$~(t.w.) & $\left(450 \pm 280\right) \times 10^{3}$~(t.w.) \\

 ALICE (13 TeV pp)~\cite{ALICE:2024xln} & $ \rm D^0, D^+, D_s^+, \Lambda^+_c$ (n.p.) & all $\pT$, $|y|<0.5$ & $\left(75.2 \pm 3.2 \pm 5.2 ^{+12.3} _{-3.2}\right) \times 10^{3}$ & $7.3\pm 20\%$~(t.w.) & $\left(550^{+150}_{-120}\right) \times 10^{3}$~(t.w.) \\

 ALICE (13 TeV pp)~\cite{ALICE:2018gev} & $ \rm \rm e^\pm$ (from b) & all $\pT$, $|y|<0.5$ & $\left(63.5^{+14}_{-14}\ ^{+9}_{-9}\ ^{+15.5}_{-15.5}\right) \times 10^{3}$ & $7.3\pm 20\%$~(t.w.) & $\left(460 \pm 190\right) \times 10^{3}$~(t.w.) \\

 \rowcolor{lightgray} LHCb (13 TeV pp)~\cite{LHCb:2016qpe} & all $\rm h_b$  & all $\pT$, $2<\eta<5$ & $\left(144 \pm 1 \pm 21\right) \times 10^{3}$ & $3.9$ & $\left(560 \pm 4 \pm 82\right) \times 10^{3}$ \\

 LHCb (13 TeV pp)~\cite{LHCb:2017vec} & $ \rm B^\pm$  & $\pT$=0--40 GeV, $2<y<4.5$ & $\left(110 \pm 10\right) \times 10^{3}$~(t.w.) & $4.9\pm 20\%$ & $\left(520 \pm 110\right) \times 10^{3}$~(t.w.) \\

 LHCb (13 TeV pp)~\cite{LHCb:2015foc} & $ \rm \jpsi$ (n.p.) & $\pT$=0--14 GeV, $2<y<4.5$ & $\left(970 \pm 5 \pm 60\right) \times 10^{3}$ & -- & $\left(495 \pm 2 \pm 52\right) \times 10^{3}$ \\

 \rowcolor{lightgray} ALICE+LHCb comb. (13 TeV pp)~\cite{Bai:2024pxk} & $ \rm D^0, \jpsi$ (n.p.) & -- & -- & -- & $\left(500\pm 40\right) \times 10^{3}$ \\

 CMS (13 TeV pp)~\cite{CMS:2016plw} & $ \rm B^+$  & $\pT$=10--100 GeV, $|y|<1.45$ & $\left(71 \pm 5\right) \times 10^{3}$~(t.w.) & $9.7\pm 20\%$~(t.w.) & $\left(680 \pm 150\right) \times 10^{3}$~(t.w.) \\
\hline
 LHCb (8.16 TeV pPb)~\cite{LHCb:2019avm} & $ \rm B^+,B^0,\Lambda_b$  & \Gape[0pt][2pt]{\makecell{$\pT$=2--20 GeV,\\ $1.5<y<3.5$, $-4.5<y<2.5$}} & $\left(96 \pm 12\right) \times 10^{3}$~(t.w.) & $3.5\pm 20\%$~(t.w.) & $\left(340 \pm 80\right) \times 10^{3}$~(t.w.) \\
\hline
 ALICE (7 TeV pp)~\cite{ALICE:2012vpz} & $ \rm \jpsi$ (n.p.) & all $\pT$, $|y|<0.5$ & $\left(43^{+11}_{-11}\ ^{+9}_{-10}\ ^{+0.6}_{-1.5}\right) \times 10^{3}$ & $4.49^{+0.12}_{-0.10}$ & $\left(282 \pm 74 \pm^{+59}_{-68}\right) \times 10^{3}$ \\

 ALICE (7 TeV pp)~\cite{ALICE:2012acz} & $ \rm \rm e^\pm$ (from b) & all $\pT$, $|y|<0.5$ & $\left(57.7 \pm 3.2 \pm 17.6\right) \times 10^{3}$ & -- & $\left(383 \pm 21 \pm 118\right) \times 10^{3}$ \\

 \rowcolor{lightgray} ALICE (7 TeV pp)~\cite{ALICE:2012acz}+\cite{ALICE:2012vpz} & $ \rm e^\pm ,\jpsi$ (n.p.) & -- & -- & -- & $\left(322 \pm 45 \pm^{+59}_{-63}\right) \times 10^{3}$ \\

 ALICE (7 TeV pp)~\cite{ALICE:2018fvj} & $ \rm \rm e^\pm$ (from b) & \Gape[0pt][2pt]{\makecell{$p_{\rm T}^\text{ee}<8$ GeV,\\ $m_\text{ee}<3.3$ GeV,\ $|y|<0.8$}} & -- & -- & $\left(170 \pm 100\right) \times 10^{3}$ \\

 \rowcolor{lightgray} LHCb (7 TeV pp)~\cite{LHCb:2016qpe} & all $\rm h_b$  & all $\pT$, $2<\eta<5$ & $\left(72 \pm 0.3 \pm 6.8\right) \times 10^{3}$ & $4.1$ & $\left(295 \pm 1 \pm 28\right) \times 10^{3}$ \\

 LHCb (7 TeV pp)~\cite{LHCb:2013vjr} & $ \rm B^0, B^+, B_s$  & $\pT<$ 40 GeV, $2<y<4.5$ & $\left(48 \pm 6\right) \times 10^{3}$~(t.w.) & $5.1\pm 20\%$~(t.w.) & $\left(240 \pm 60\right) \times 10^{3}$~(t.w.) \\

 LHCb (7 TeV pp)~\cite{LHCb:2012sng} & $ \rm B^+$  & $\pT<$ 40 GeV, $2<y<4.5$ & $\left(51 \pm 5\right) \times 10^{3}$~(t.w.) & $5.1\pm 20\%$~(t.w.) & $\left(260 \pm 60\right) \times 10^{3}$~(t.w.) \\

 LHCb (7 TeV pp)~\cite{LHCb:2010wqx} & $ \rm D^0$ (n.p.) & all $\pT$, $2<\eta<6$ & $\left(75.3 \pm 5.4 \pm 13\right) \times 10^{3}$ & $3.77$ ($3.61$) & $\left(284 \pm 20 \pm 49\right) \times 10^{3}$  \\

 LHCb (7 TeV pp)~\cite{LHCb:2011zfl}  & $ \rm \jpsi$ (n.p.) & $\pT<$ 14 GeV, $2<y<5$ & $\left(49 \pm 7\right) \times 10^{3}$~(t.w.) & $5.88$ ($5.21$) & $\left(288 \pm 4 \pm 48\right) \times 10^{3}$  \\

 CMS (7 TeV pp)~\cite{CMS:2011oft} & $ \rm B^+$  & $\pT>$ 5 GeV, $|y|<2.4$ & $\left(69 \pm 11\right) \times 10^{3}$~(t.w.) & $3.45\pm 20\%$~(t.w.) & $\left(240 \pm 60\right) \times 10^{3}$~(t.w.) \\

 CMS (7 TeV pp)~\cite{CMS:2011pdu} & $ \rm B^0$  & $\pT>$ 5 GeV, $|y|<2.2$ & $\left(82 \pm 11\right) \times 10^{3}$~(t.w.) & $3.7\pm 20\%$~(t.w.) & $\left(300 \pm 70\right) \times 10^{3}$~(t.w.) \\

 CMS (7 TeV pp)~\cite{CMS:2011kew} & $ \rm B_s$  & $\pT$=8--50 GeV, $|y|<2.4$ & $\left(66 \pm 8\right) \times 10^{3}$~(t.w.) & $7.6\pm 20\%$~(t.w.) & $\left(500 \pm 120\right) \times 10^{3}$~(t.w.) \\

 CMS (7 TeV pp)~\cite{CMS:2012wje} & $ \rm \Lambda_b$ (n.p.) & $\pT>$ 10 GeV, $|y|<2.0$ & $\left(25 \pm 13\right) \times 10^{3}$~(t.w.) & $14.2\pm 20\%$~(t.w.) & $\left(350 \pm 200\right) \times 10^{3}$~(t.w.) \\

 CMS (7 TeV pp)~\cite{CMS:2010nis} & $ \rm \jpsi$ (n.p.) & $\pT^\mathrm{\jpsi}=6.5\mbox{--}30$ GeV, $|y|<2.4$ & $\left(19 \pm 3\right) \times 10^{3}$~(t.w.) & $15\pm 20\%$~(t.w.) & $\left(280 \pm 70\right) \times 10^{3}$~(t.w.) \\

 \multirow{ 2}{*}{CMS (7 TeV pp)~\cite{CMS:2011rxs}} & $ \rm \jpsi$ (n.p.) & $\pT^\mathrm{\jpsi}=8\mbox{--}30$ GeV, $|y|<2.4$ & $\left(10^{+1}_{-1}\right) \times 10^{3}$ & $26\pm 20\%$~(t.w.) & $\left(260 \pm 60\right) \times 10^{3}$~(t.w.) \\

  & $ \rm \psi(2S)$ (n.p.) & $\pT^\mathrm{\psi(2S)}=6.5\mbox{--}30$ GeV, $|y|<2.4$ & $\left(20^{+2}_{-2}\right) \times 10^{3}$ & $11.5\pm 20\%$~(t.w.) & $\left(230 \pm 60\right) \times 10^{3}$~(t.w.) \\

 \rowcolor{lightgray} CMS average (7 TeV pp) (t.w.) & (combined)  & (combined) & (combined) & -- & $\left(300 \pm 100\right) \times 10^{3}$~(t.w.) \\

 CMS (7 TeV pp)~\cite{CMS:2012xsp} & $ \rm \mu^\pm\mu^\pm$ (from b) & $\pT^{\mu_1,\mu_2} > 4$ GeV, $|\eta_\mu|<2.1$ & $2200 \pm 300$~(t.w.) & -- & -- \\

 CMS (7 TeV pp)~\cite{CMS:2011xhf} & $ \rm \mu^\pm$ (from b) & $\pT^{\mu} > 6$ GeV, $|\eta_\mu|<2.1$ & $3200 \pm 800$~(t.w.) & $60\pm 50\%$~(t.w.) & $\left(190 \pm 110\right) \times 10^{3}$~(t.w.) \\

 ATLAS (7 TeV pp)~\cite{ATLAS:2013cia} & $ \rm B^+$  & $\pT$=9--120 GeV, $|y|<2.25$ & $\left(26 \pm 3\right) \times 10^{3}$~(t.w.) & $10\pm 20\%$~(t.w.) & $\left(260 \pm 60\right) \times 10^{3}$~(t.w.) \\
\hline
 ALICE (5.02 TeV pp)~\cite{ALICE:2021mgk} & $ \rm D^0, D^+, D^{*+}$ (n.p.) & all $\pT$,\ $|y|<0.5$ & $\left(34.5 \pm 2.4 \pm 2.5 ^{+3.9}_{-1.4}\right) \times 10^{3}$  & $6.2\pm 20\%$~(t.w.) & $\left(210^{+60}_{-50}\right) \times 10^{3}$~(t.w.) \\

 LHCb (5.02 TeV pp)~\cite{LHCb:2021pyk} & $ \rm \jpsi$ (n.p.) & $\pT^\mathrm{\jpsi}<20$ GeV, $2<y<4.5$ & $\left(35 \pm 2\right) \times 10^{3}$~(t.w.) & $5.4\pm 20\%$~(t.w.) & $\left(190 \pm 40\right) \times 10^{3}$~(t.w.) \\

 \rowcolor{lightgray} ALICE+LHCb comb. (5.02 TeV pp)~\cite{Bai:2024pxk} & $ \rm D^0, \jpsi$ (n.p.) & all $\pT$,\ $|y|<0.5$ & $\left(29.6 \pm 2.4\right) \times 10^{3}$ & -- & $\left(191\pm 16\right) \times 10^{3}$ \\

 CMS (5.02 TeV pp)~\cite{CMS:2018bwt} & $ \rm D^0$  (n.p.) & $\pT$=2--100 GeV, $|y|<1.0$ & $\left(31 \pm 5\right) \times 10^{3}$~(t.w.) & $5.7\pm 20\%$~(t.w.) & $\left(180^{+50}_{-50}\right) \times 10^{3}$~(t.w.) \\

 CMS (5.02 TeV pp)~\cite{CMS:2017uoy} & $ \rm B^\pm$  & $\pT$=7--50 GeV, $|y|<2.4$ & $\left(16 \pm 2\right) \times 10^{3}$~(t.w.) & $7\pm 20\%$~(t.w.) & $\left(220 \pm 50\right) \times 10^{3}$~(t.w.) \\
\hline
 \rowcolor{lightgray} ALICE (2.76 TeV pp)~\cite{ALICE:2014aev} & $ \rm \rm e^\pm$ (from b) & all $\pT$, $|y|<0.5$ & $\left(29.1 \pm 2.6^{+9.8}_{-11.7}\right) \times 10^{3}$ & 5.57 & $\left(162 \pm 14^{+56}_{-65}\right) \times 10^{3}$ \\
\hline
 \rowcolor{lightgray} CDF (1.96 TeV p\=p)~\cite{CDF:2004jtw} & $ \rm \jpsi$ (n.p.) & all $\pT$, $|y|<0.6$ & $\left(17.6 \pm 0.4 \pm ^{+2.5}_{-2.3}\right) \times 10^{3}$ & $4.3\pm 20\%$~(t.w.) & $\left(76^{+19}_{-18}\right) \times 10^{3}$~(t.w.) \\
\hline
 UA1 (0.63 TeV p\=p)~\cite{UA1:1990vvp} & $ \rm \jpsi$, $\mu^\pm$ (from b) & all $\pT$, $|\eta|<1.5$ & $\left(12.8 \pm 4.7 \pm 6\right) \times 10^{3}$ & -- & $\left(19.3 \pm 7 \pm 9\right) \times 10^{3}$ \\

 \rowcolor{lightgray} UA1 (0.63 TeV p\=p)~\cite{UA1:1993jok} & $ \rm \mu^\pm$ (from b) & all $\pT$, $|\eta|<1.5$ & $\left(13 \pm 2.8 \pm ^{+6.9}_{-4.3}\right) \times 10^{3}$ & -- & $\left(19.7 \pm 4 ^{+10.4}_{-6.5}\right) \times 10^{3}$ \\
\hline
 PHENIX (0.51 TeV pp)~\cite{PHENIX:2017exy}+\cite{PHENIX:2011gyb} & $ \rm \jpsi$ (n.p.) & all $\pT$, $1.2<|y|<2.2$ & $3630^{+1920}_{-1700}$ & $3.3\pm 20\%$~(t.w.) & $\left(12^{+7}_{-6}\right) \times 10^{3}$~(t.w.) \\

 \rowcolor{lightgray} PHENIX (0.51 TeV pp)~\cite{PHENIX:2020iaj} & $ \rm \mu^\pm\mu^\pm$ (from b) & \Gape[0pt][2pt]{\makecell{$\pT^\mu>1$ GeV, $1.2<|y|<2.2$, \\$m_{\mu\mu}=5\mbox{--}10$ GeV}} & $26 \pm 5$~(t.w.) & ~500 & $\left(13.1 \pm 0.6 \pm 1.5 \pm 2.7\right) \times 10^{3}$ \\

 PHENIX (0.2 TeV pp)~\cite{PHENIX:2009dpd} & $ \rm \rm e^\pm$ (from b) & all $\pT$, $|y|<0.5$ & $920\ ^{+340}_{-310}\ ^{+390}_{-360}$ & $3.5\pm 8\%$ & $3200 ^{+1200}_{-1100}\ ^{+1400}_{-1300}$ \\

 PHENIX (0.2 TeV pp)~\cite{PHENIX:2008qav} & $ \rm \rm e^\pm$ (from b) & $\pT^\text{e}>0.2$ GeV, $|y|<0.35$ & -- & -- & $3900 \pm 2500^{+3000}_{-2000}$ \\

 PHENIX (0.2 TeV pp)~\cite{PHENIX:2017ztp} & $ \rm \rm e^\pm$ (from b) & $\pT^\text{e}>0.2$ GeV, $|y|<0.35$ & -- & -- & $3900 \pm 1300$ \\

 PHENIX (0.2 TeV dAu)~\cite{PHENIX:2014edx} & $ \rm \rm e^\pm$ (from b) & all $\pT$, $|y|<0.5$ & $1370\pm 280\pm 460$ & 2.5 & $3400 \pm 800 \pm 1100$ \\

 PHENIX (0.2 TeV dAu)~\cite{PHENIX:2017ztp} & $ \rm \rm e^\pm$ (from b) & $\pT>0.2$ GeV, $|y|<0.35$ & -- & -- & $2800 \pm 1300$ \\

 PHENIX (0.2 TeV pp)~\cite{PHENIX:2017caf}~\cite{PHENIX:2018dwt} & $ \rm \jpsi$ (n.p.) & all $\pT$, $1.2<|y|<2.2$ & $510 \pm 130 \pm 200$ & -- & $3800 \pm 900 \pm 1600$ \\

 \rowcolor{lightgray} PHENIX (0.2 TeV pp)~\cite{PHENIX:2018dwt} & $ \rm \mu^\pm\mu^\pm$ (from b) & all $\pT$, $1.2<|y|<2.2$ & $460 \pm 90$~(t.w.) & -- & $3750 \pm 240^{+350}_{-500} \pm 450$ \\

 \rowcolor{lightgray} STAR (0.2 TeV pp)~\cite{STAR:2011bqq} & $ \rm \rm e^\pm$ (from b) & $\pT^\text{e}=3\mbox{--}10$ GeV , $|\eta|<0.5$ & $9.8 \pm 2.8$~(t.w.) & -- & $4000 \pm 500 \pm 100 $ \\
\hline
 HERA-B (41.6~GeV  pA, A=C,Ti)~\cite{HERA-B:2002zbs} & $ \rm \jpsi$ (n.p.) & $x_{\rm F} \in [-0.25,0.15]$ & -- & -- & $32^{+14}_{-12}\ ^{+6}_{-7}$  \\

 HERA-B (41.6~GeV  pA, A=C,Ti)~\cite{HERA-B:2005tnp,Lourenco:2006vw} & $ \rm \jpsi$ (n.p.) & $x_{\rm F} \in [-0.35,0.15]$ & -- & -- & $14.9 \pm 2.2 \pm 2.3$  \\

 \rowcolor{lightgray} HERA-B (41.6~GeV  pA, A=C,Ti,W)~\cite{HERA-B:2006qsu}+\cite{HERA-B:2005tnp} & $ \rm \mu^\pm$ (from b) & $x_{\rm F} \in [-0.3,0.15]$ & -- & -- & $15.8 \pm 1.7 \pm 2.3$  \\

 \rowcolor{lightgray} E771 (38.7~GeV pSi)~\cite{E771:1997rbx,Lourenco:2006vw} & $ \rm \mu^\pm$ (from b) & $\pT^\mathrm{\mu}>6$ GeV, $x_{\rm F}^\mathrm{\mu} \in [-0.25,0.5]$ & -- & -- & $43\ ^{+27}_{-7} \pm 7$  \\

 \rowcolor{lightgray} E789 (38.7~GeV pAu)~\cite{Jansen:1994bz,Lourenco:2006vw} & $ \rm \jpsi$ (n.p.) & $\pT^\mathrm{\jpsi}<2$ GeV, $x_{\rm F}^\mathrm{\jpsi} \in [0,0.1]$ & -- & -- & $5.7 \pm 1.5 \pm 1.2$  \\
\bottomrule
\end{tabular}

}
\end{table}

\clearpage
\section{Theoretical predictions}
\label{sec:th}

The scattering process of interest in this work is the inclusive production of a heavy-quark pair ($\QQbar$) in \pp\ collisions, $\mathrm{p}(k_1) + \mathrm{p}(k_2) \to \mathrm{Q}\overline {\mathrm{Q}} + \rm X$,
where $k_i$ are the incoming momenta of the protons, leading to a collision with squared \cm\ energy $s=(k_1+k_2)^2$. The final state is required to have a massive on-shell quark-antiquark pair, implicitly assuming an integration over their momenta. 
We compare two different theoretical frameworks, based on the fixed flavour number (FFN) scheme and the general-mass variable flavour number scheme (GM-VFNS). In the FFN scheme, the number of active light-parton flavours is fixed ($N_f=3,\,4$ for charm and bottom, respectively), and heavy quarks such as charm and bottom are not included as proton partons, but are produced explicitly in the hard scattering with their full mass dependence retained. At NLO and beyond, the expressions for the total cross section involve logarithmic terms proportional to $\log(s/\mQ^2)$ which can potentially become large as the \cm\ energy grows. 
In the GM-VFN schemes, these logarithmic terms are summed up through the DGLAP renormalization group equations by introducing scale-dependent heavy-quark PDFs and FFs.

\subsection{FFN scheme: NNLO calculations}
\label{sec:th_MaunaKea}

First, we focus on fixed-order predictions within the FFN scheme with $N_f$ light (massless) quarks q and a single heavy (massive) quark Q.
The hadronic cross section $\sigma^{\pp\to \QQbar+\mathrm{X}}$ can be schematically expressed as a convolution of the PDFs $f(x,\muF^2)$ of the colliding hadrons, and the perturbative computable partonic cross section $\hat \sigma^{ij\to \QQbar+\mathrm{X}}$, as
\begin{equation}
    \sigma^{\pp\to \QQbar+\mathrm{X}}(s,\mQ^2,\muF^2,\muR^2) = \sum_{i,j} \int_{\xmin}^1\!\mathrm{d}x_1 \mathrm{d}x_2\, f_i(x_1,\muF^2) f_j(x_2,\muF^2) \, \hat \sigma^{ij\to \QQbar+\mathrm{X}}(\hat s = x_1x_2 s,\mQ^2,\muF^2,\muR^2) \,, \label{eq:fact}
\end{equation}
where the dependence on the factorization scale $\muF$, introduced by the collinear factorization, and the renormalization scale $\muR$, introduced by the $\alphaS$ running, are explicitly indicated.
Unless otherwise specified, we adopt $\muF = \muR = 2\mQ$ as the default scale choice, corresponding to the rest mass of the heavy-quark pair system. The sum of initial-state partons $i,j$ includes light quarks ($\rm q,\overline q$) and gluons ($g$), but not heavy quarks Q, 
with the full mass dependence $\mQ$ retained in the matrix elements, \ie we only consider the perturbative generation of heavy quarks, \eg\ $gg\to\QQbar$ and $\qqbar\to\QQbar$, and neglect any contributions from intrinsic heavy-quark PDFs~\cite{Ball:2022qks}. The cross section integration ranges in $x_{1,2}$ of \cref{eq:fact} extend up to the kinematic upper limit of unity, while the lower limit is given by
\begin{equation}
\xmin = \frac{4\mQ^2}{s},
\label{eq:xmin}
\end{equation}
which reflects the minimum partonic energy ($\hat s = 4\mQ^2$) required to produce a heavy-quark pair at rest.

The details of the NNLO PDFs employed in this work, NNPDF4.0~\cite{NNPDF:2021njg}, CT18~\cite{Hou:2019efy}, and  MSHT20~\cite{Bailey:2020ooq}, are listed in \cref{tab:PDFs}. For each PDF set, the following information is provided: number of effective quark flavours ($N_f=3,4$, relevant for $\ccbar$ and $\bbbar$ production, respectively), minimum parton momentum fraction ($\xgmin$) and virtuality ($\Qmin$),
charm and bottom masses ($\mc, \mb$), and type of error and number of replicas or eigenvectors used for the PDF uncertainty estimate. 
All PDFs adopt a reference value $\alphasmZ=0.118$, defined as the value obtained by evolving $\alphaS(Q_\mathrm{min})$ to the scale $m_\mathrm{Z}$ within the VFN scheme (up to $N_f = 5)$.
For consistency with the FFN scheme, all PDF sets used do not contain any intrinsic charm in the initial proton densities, \ie\ the charm distribution $f_\mathrm{c}(x,Q^2)$ is exactly zero at the initial scale and arises solely from pQCD evolution, \eg\ through $g\to\ccbar$ splitting.
For the NNPDF4.0 densities, this latter condition is achieved using the perturbative-charm (`pch') set, as the default set uses intrinsic charm~\cite{Ball:2022qks}.
The \texttt{NNPDF40\_nnlo\_pch\_as\_01180\_nf\_3} and \texttt{NNPDF40\_nnlo\_as\_01180\_nf\_4} parton densities are used as our default PDF choices for $\ccbar$ and $\bbbar$ production, respectively. 

\begin{table}[htbp!]
\caption{List of the NNLO PDFs used in the calculations, with their characteristics: number of effective quark flavours ($N_f=3,4$), minimum parton momentum fraction ($\xgmin$) and virtuality ($\Qmin$), charm and bottom masses ($\mc, \mb$), and type of error and number of replicas or eigenvectors used.
\label{tab:PDFs}}
\centering
\tabcolsep=1.5mm
\resizebox{\textwidth}{!}{%
\begin{tabular}{llcccccc}\hline
PDF set family & \lhapdf\ set name & $N_f$ & $\xgmin $ & $\Qmin$ [GeV] & $\mc$ [GeV] & $\mb$ [GeV] & \# replicas/eigenvectors \\ \hline
NNPDF4.0 & \texttt{NNPDF40\_nnlo\_pch\_as\_01180\_nf\_3} & 3 & $10^{-9}$ & 1.0 & 1.51 & -- & 100 (68\% CL) \\ 
              & \texttt{NNPDF40\_nnlo\_as\_01180\_nf\_4}      & 4 & $10^{-9}$ & 1.65 & -- & 4.92 & 100 (68\% CL)  \\ 
CT18    & \texttt{CT18NNLO\_NF3}                        & 3 & $10^{-9}$ & 1.295 & 1.30 & -- & 58 (Hessian 90\% CL) \\ 
              & \texttt{CT18NNLO\_NF4}                        & 4 & $10^{-9}$ & 1.295 & -- & 4.75 & 58 (Hessian 90\% CL) \\ 
MSHT20  & \texttt{MSHT20nnlo\_nf3}                      & 3 & $10^{-6}$ & 1.0 &  1.40 & -- & 64 (Hessian 68\% CL) \\ 
              & \texttt{MSHT20nnlo\_nf4}                      & 4 & $10^{-6}$ & 1.0 & -- & 4.75 & 64 (Hessian 68\% CL) \\ 
\hline
\end{tabular}
} 
\end{table}

The currently highest available fixed-order perturbative accuracy for heavy-quark production is NNLO, with the corresponding coefficient functions computed in Refs.~\cite{Aliev:2010zk,Czakon:2011xx,Czakon:2012pz,Czakon:2012zr,Czakon:2013goa,Barnreuther:2012wtj}.
Hence, we further decompose the partonic cross section as
\begin{equation}
    \hat \sigma^{ij\to \QQbar+\mathrm{X}}(\hat s,\mQ^2,\muF^2,\muR^2) = \sum_{k=0}^2 \left(\frac{\alphaS(\muR^2)}{4\pi}\right)^{2+k}\hat \sigma_k^{ ij\to \QQbar+\mathrm{X}}(\hat s,\mQ^2,\muF^2,\muR^2)\,, \label{eq:FFNSNNLO}
\end{equation}
with these coefficient functions implemented into the new \MaunaKea code.
In practice, we have taken them from \toppp~\cite{Czakon:2011xx}, and \MaunaKea differs from the latter in three relevant aspects:
(i) the number of light flavours $(N_f)$ is dynamic and not fixed, so it can be used to predict any heavy-quark (and not just $\ttbar$) production, (ii) \MaunaKea is interfaced to \pineappl\ to produce fast interpolation grids that are PDF-independent and allow setting arbitrary factorization and renormalization scales (effectively, this turns \cref{eq:fact} into a fast matrix operation), and (iii) \MaunaKea currently does not support soft gluon resummation, \ie all predictions are at fixed order. The \MaunaKea code has been extensively benchmarked against predictions by both \toppp~\cite{Czakon:2013goa} and \hathor~\cite{Aliev:2010zk}.
The code can be downloaded online~\cite{MaunaKea,felix_hekhorn_2024_14185847}.

We note that the FFN scheme coefficient functions in \cref{eq:FFNSNNLO} are obtained in the pole-mass (on-shell) renormalization scheme \ie\ the mass that appears in the coefficient functions is the same as the one that sets the kinematic boundaries, \eg\ in \cref{eq:xmin}.
The perturbative convergence of the FFN scheme cross section calculation is known to be improved when using the $\MSbar$ running-mass scheme compared with the pole-mass (on-shell) renormalization scheme~\cite{Bigi:1994em}.
Furthermore, the pole-mass definition carries an intrinsic uncertainty of order $\LQCD$ due to renormalon contributions~\cite{Bigi:1994em,Beneke:1994sw}.
This ambiguity is not included in the uncertainties of the pole-mass values quoted by the PDG~\cite{PDG}, which are obtained perturbatively from the best determinations of the running masses.
However, in this work we still adopt the pole-mass definition for the charm and bottom quarks, as it is the convention commonly used in PDF fits, and we will nevertheless need to specify the pole masses to set the kinematic boundaries, \eg\ in \cref{eq:xmin}.

\subsubsection*{Missing higher-order uncertainties}

\begin{figure}[htpb!]
    \centering
    \includegraphics[width=0.5\linewidth]{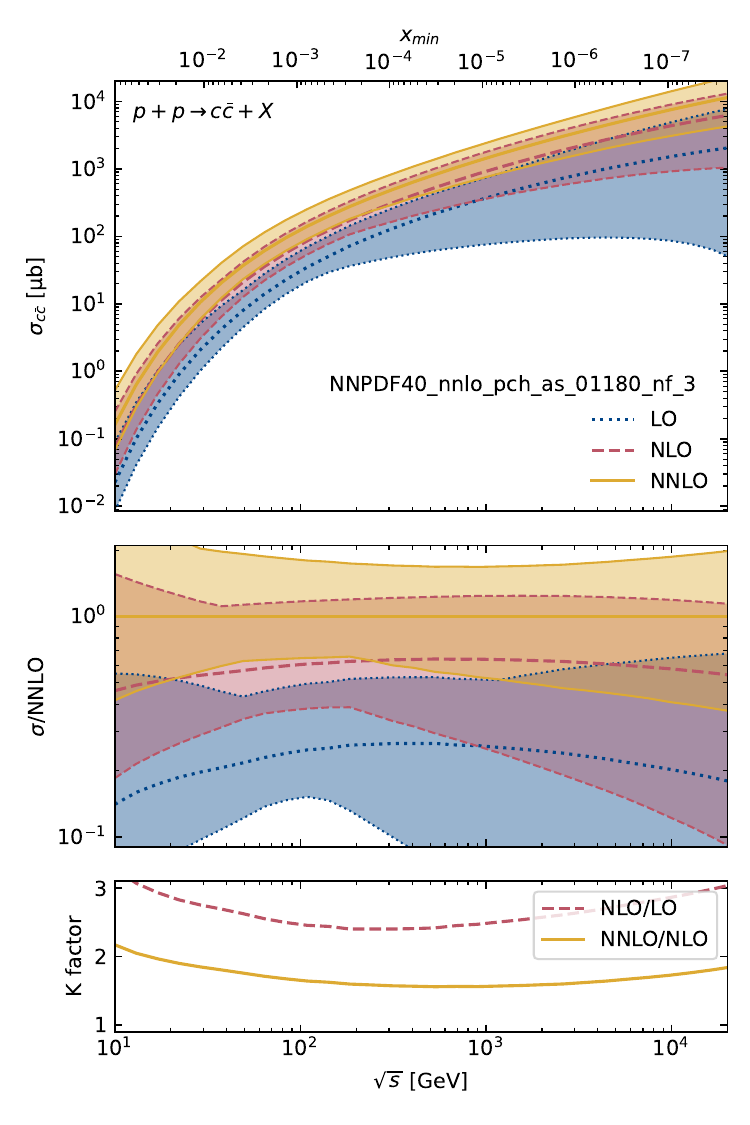}%
    \includegraphics[width=0.5\linewidth]{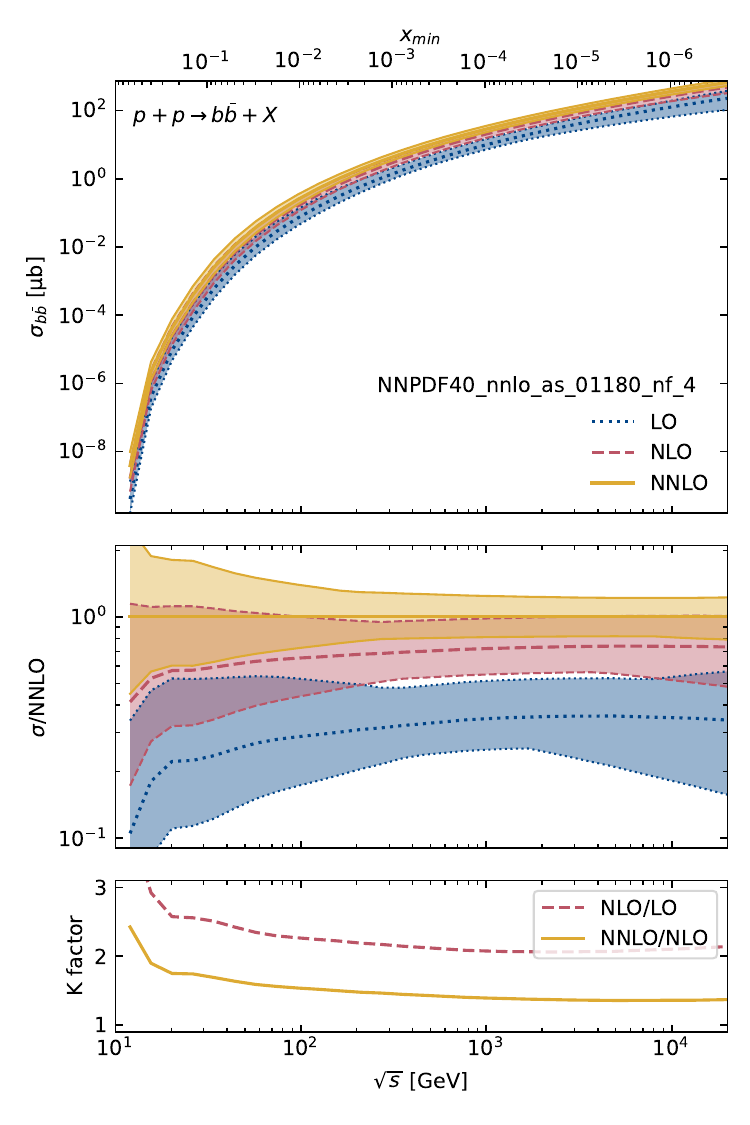}
    \caption{
    Inclusive cross sections for charm (left) and bottom (right) production in \pp\ collisions as a function of the \cm\ energy, over $\sqrts = 10$~GeV--20~TeV, computed at LO (blue), NLO (red), and NNLO (yellow) accuracy. All predictions are calculated with the NNLO PDFs. The upper panels display the absolute cross sections, with bands indicating the scale uncertainties. The middle panels show the relative scale uncertainty with respect to the central predictions, while the lower panels present the $K$-factors, \ie\ the ratios of the cross sections to those at the preceding perturbative order.
    The $\xmin$ values indicated in the upper $x$ axis correspond to \cref{eq:xmin}.
    \label{fig:pto}}
\end{figure}

We start by estimating the size of the theoretical uncertainties associated with missing higher-order corrections for increasing pQCD accuracies, and examining the perturbative convergence of the heavy-quark cross section calculations.
Figure~\ref{fig:pto} shows the inclusive cross sections for $\ccbar$ (left) and $\bbbar$ (right) production in \pp\ collisions as a function of $\sqrts$ predicted at LO, NLO, and NNLO accuracy.
All predictions have been calculated by using the same NNLO PDF, \ie\ the differences between the different-order calculations are due to the coefficient functions. The use of NNLO PDFs in conjunction with LO or NLO coefficient functions does not formally improve the perturbative accuracy of the calculation but, by keeping the PDFs fixed, it provides insight into how the full NNLO result is built up.
The error bands around the central curves indicate the missing higher-order theoretical uncertainty obtained through a 7-point scale variation, whereby the $\muF$ and $\muR$ scales are varied within factors of two (excluding variations in opposite directions) around our default scale choice of $\muF=\muR=2\mQ$.
The middle panels show the cross sections normalized to the NNLO predictions, and the bottom panels the so-called $K$-factors, namely, the ratios between cross sections obtained at a higher over the immediately lower pQCD accuracy. Despite the reduction in relative scale uncertainties when going from LO to NNLO, they remain substantial at NNLO, amounting on average to approximately $-40\%$ and $+100\%$ for charm and $\pm25\%$ for bottom, over the considered $\sqrts$ range.
We also observe that the central NNLO predictions are always outside the LO scale-variation bands and only barely within the NLO bands. 
In addition, the $K$-factors are quite large: for $\ccbar$ production the cross sections augment by up to a factor of three from LO to NLO and still by a factor of two from NLO to NNLO, while for $\bbbar$ production we find slightly smaller ratios, which indicates a better perturbative convergence with increasing heavy-quark mass.
All in all, the perturbative corrections are large.
This slow perturbative convergence can be linked to the relatively large values of the expansion parameter, $\alphaS(\muR)$, at the low scales of relevance for charm and bottom pair production, and possibly also to the size of the logarithmic terms $\sim \log(s/\mQ^2)$.
The large $K$-factors could also be due to other large contributions in the partonic cross sections that would require resummation, including the threshold and/or small-$x$ logarithms. 
In particular, in the high-energy (small-$x$) regime probed by charm production, resummation of $\log(x)$-type terms may be required to achieve more perturbatively stable
predictions~\cite{Ball:2017otu,xFitterDevelopersTeam:2018hym,Bonvini:2019wxf,Silvetti:2022hyc,Armesto:2022mxy}. A consistent resummation framework requires both resummed PDFs and resummed partonic cross sections, which is beyond the scope of this study.
Note that the high-energy resummation applicable for $\hat s \to \infty$ is opposite to the threshold resummation applicable for $\hat s\to 4\mQ^2$ (which \toppp{} can perform up to NNLL accuracy, but is not available in \MaunaKea{}), and is orthogonal to the resummation of $\log(s/\mQ^2)$ terms originating from collinear radiation discussed in \cref{subsec:resumm}. The calculation of beyond-NNLO perturbative corrections is mathematically challenging because of the increasingly large number of real and virtual gluon emission diagrams possible, and the presence of massive final-state particles.

\subsubsection*{Breakdown of partonic channels}

\begin{figure}[htpb!]
    \centering
    \includegraphics[width=0.5\linewidth]{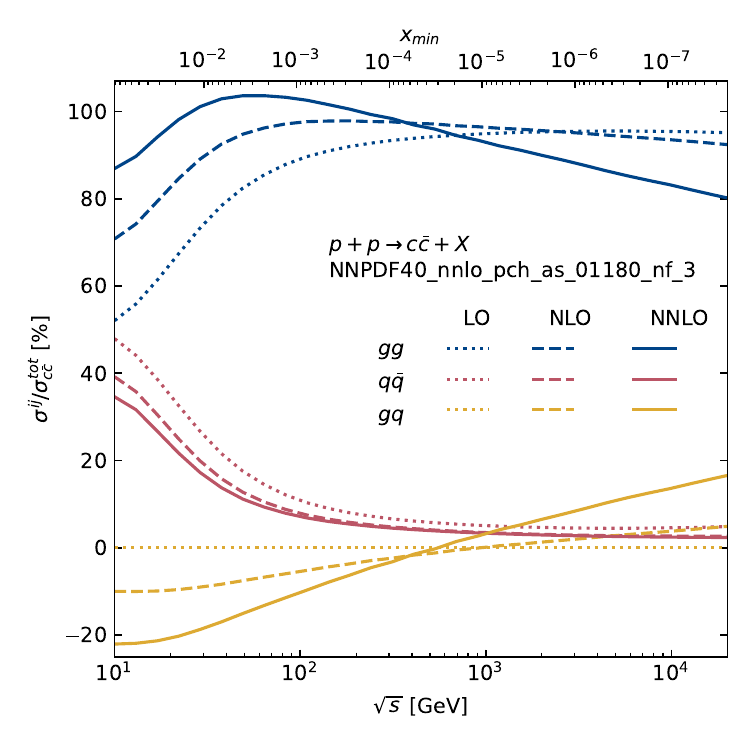}%
    \includegraphics[width=0.5\linewidth]{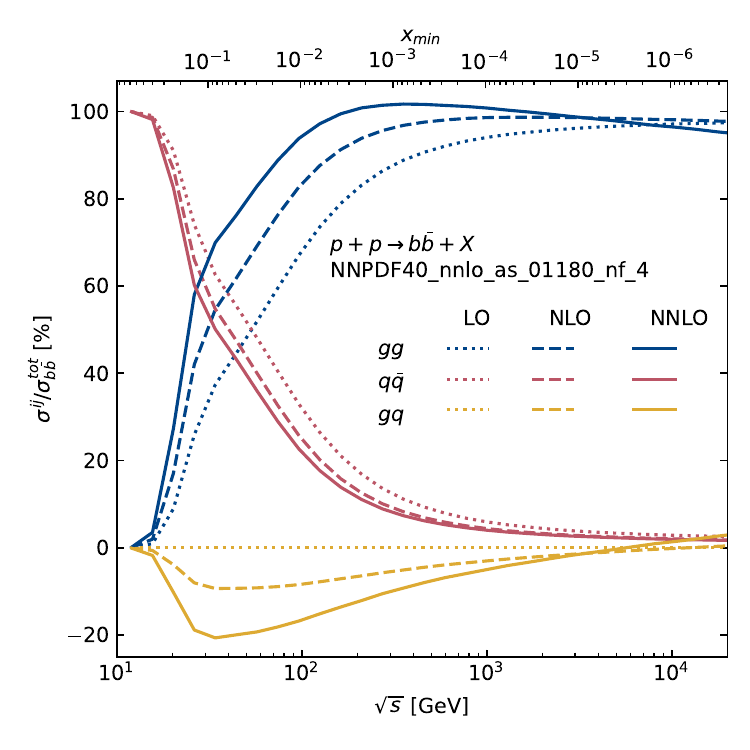}
    \caption{Breakdown of the relative contribution of different parton-parton luminosities to the total charm (left) and bottom (right) cross section in \pp\ collisions as a function of \cm\ energy, over $\sqrts = 10$~GeV--20~TeV: gluon-gluon (blue), quark-antiquark (red), and gluon-quark (yellow), at LO (dotted), NLO (dashed), and NNLO (solid) accuracy.
    The additional quark-quark luminosities that open up at NNLO are not plotted, as they provide only a minor contribution. The $\xmin$ values indicated in the upper $x$ axis correspond to \cref{eq:xmin}.
    \label{fig:lumi}}
\end{figure}

It is instructive to examine the relative contributions of different parton-parton collision combinations, indicated by the $i,j$ indices in \cref{eq:fact}, to the total heavy-quark pair cross section. The relative fractions of $gg, g \rm q$, and $\rm q\overline q$ processes are plotted as a function of $\sqrts$ in \cref{fig:lumi} for the different perturbative orders. At LO accuracy only the gluon-gluon ($gg$) and the quark-antiquark ($\rm q\overline q$) luminosities are possible, while at NLO also the gluon-quark ($g$q) luminosity starts to contribute. The additional quark-quark luminosities ($\rm qq$, $\rm qq'$ and $\rm q\overline q'$) that open up at NNLO and contribute to the small remaining fraction of the total cross section, are not shown in the plot. As the individual partonic channels do not represent physical cross sections, there can be contributions that are negative or bigger than 100\%. Figure~\ref{fig:lumi} clearly displays the well-known fact that the heavy-quark cross sections are clearly dominated by the gluon-gluon channel for increasing $\sqrts$ values. Towards low $\sqrts$, however, the $\qqbar$ channel begins to compete and eventually exceeds the size of the gluon-gluon contribution when approaching the threshold production. This can be explained by the increasingly important role of the valence quarks towards low values of $\sqrts$, which probe PDFs at higher $x$ as can be seen from Eq.~(\ref{eq:xmin}).

\subsubsection*{Heavy-quark mass dependence and uncertainties}

\begin{figure}[htpb!]
    \centering
    \includegraphics[width=0.5\linewidth]{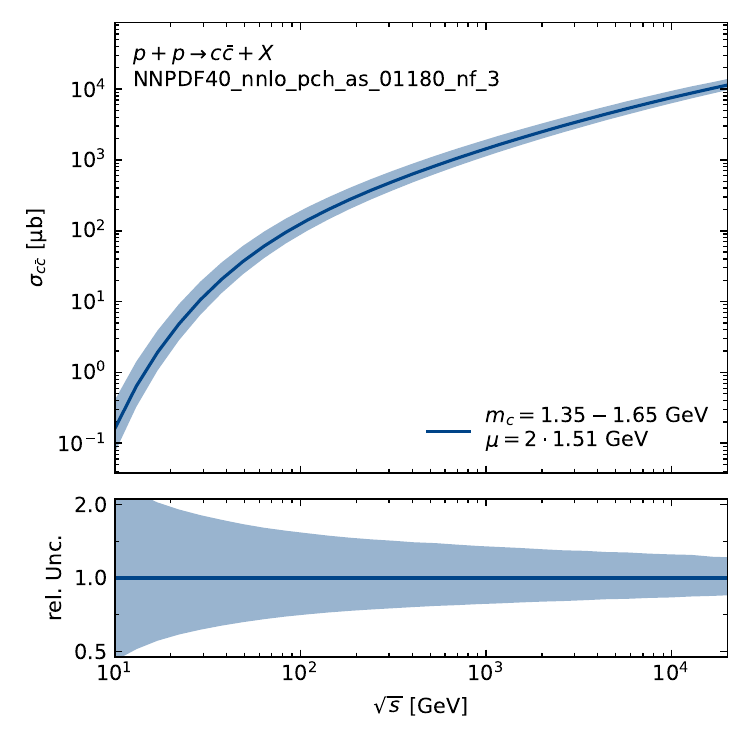}%
    \includegraphics[width=0.5\linewidth]{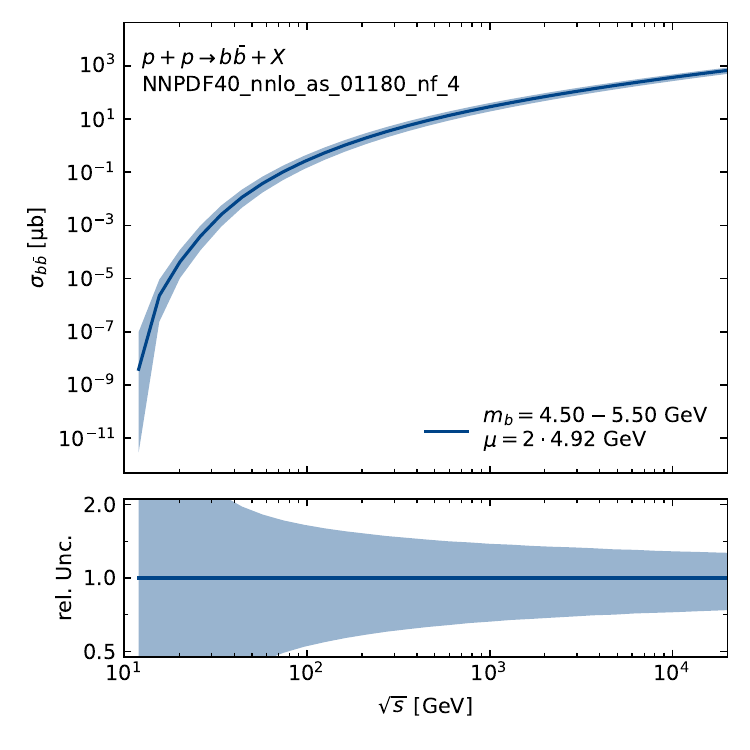}
    \caption{Inclusive charm (left) and bottom (right) NNLO production cross sections as a function of collision energy, over $\sqrts = 10$~GeV--20~TeV,     with the bands depicting the uncertainty associated with the heavy-quark mass value ($\mQ$) used in the calculations. The charm quark mass is varied over $\mc =1.35$--1.65~GeV 
    around the default choice of $\mc^0=1.51$~GeV. 
    The bottom quark mass is varied over $\mb=4.5$--5.5~GeV around the default choice of $\mb^0 = 4.92$~GeV.
    All scales are fixed to twice the default mass value $\muF = \muR = 2\mQ^0$.
    The upper panels indicate the total cross section, while the lower panels indicate the relative uncertainty with respect to the default mass value.
    \label{fig:mass-ccbar}}
\end{figure}

In \cref{fig:mass-ccbar}, we investigate the dependence of the inclusive NNLO $\QQbar$ cross sections on the value of the heavy-quark mass $\mQ$. The charm mass is varied over the range $\mc = 1.35$--1.65~GeV around the default choice of $\mc^{0}=\SI{1.51}{\GeV}$, and the bottom mass in the range $\mb=4.5$--$5.5$~ GeV around the default choice of $\mb^0 = 4.92$~GeV.
Such variations (shown in the figure as error bands around the central curves) represent relative uncertainties of about $\pm10\%$ in the charm and bottom masses. The variations are performed in the coefficient functions, but not also in the PDFs because only MSHT20 provides PDF sets for $N_f=3,4$ with $\mQ$ variants.
We fix the factorization scale and the renormalization scale to twice the default mass value $\muF = \muR = 2\mQ^0$, and vary only the value of the heavy quark mass in the hard matrix elements $\hat \sigma$ and the associated phase-space, \ie, effectively, in \cref{eq:xmin}. The default PDF sets from NNPDF4.0 are used, which do not provide alternative fits for varying $\mQ$ values, thus neglecting any correlations between quark masses and PDFs~\cite{Ball:2026qno}.
We have checked with the dedicated MSHT20 PDF sets for different $\mQ$ values, that the impact on the cross section of exactly matching $\mQ$ in the matrix elements and PDFs is small. 
The left plot of \cref{fig:mass-ccbar} indicates that the $\mc$ choice propagates as a maximum factor of two uncertainty in $\sigmaccbar$ at $\sqrts\approx 10$~GeV, decreasing to a $\pm(10$--20)\% uncertainty for $\sqrts\gtrsim 10$~TeV.
Namely, the uncertainty in the $\ccbar$ cross sections associated with the charm quark mass appears subleading with respect to the scale variation uncertainty (\cref{fig:pto}, left).
On the other hand, for $\bbbar$ production (\cref{fig:mass-ccbar}, right) the bottom-quark mass variations propagate as a relative uncertainty in the cross sections that is somewhat larger than for the $\ccbar$ case, and now dominates over the corresponding theoretical scale uncertainties (\cref{fig:pto} right).
This can be seen in particularly near the threshold production region, \ie\ for $\sqrts\to2\mb$, where the associated restriction on the available phase-space due to modifying the $\mb$ value has a stronger impact on the pair production cross sections.

\subsubsection*{PDF dependence and uncertainties}

\begin{figure}[htpb!]
    \centering
    \includegraphics[width=0.5\linewidth]{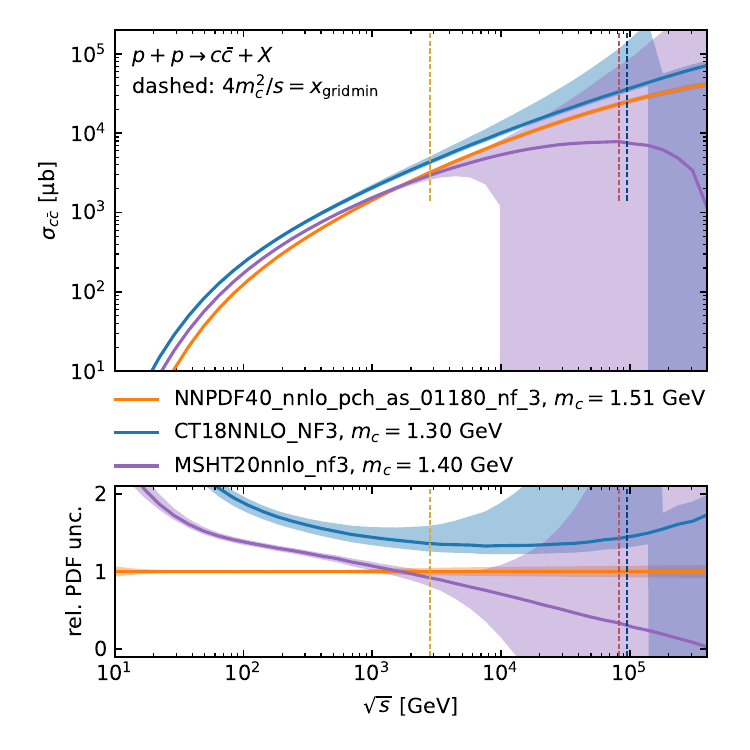}%
    \includegraphics[width=0.5\linewidth]{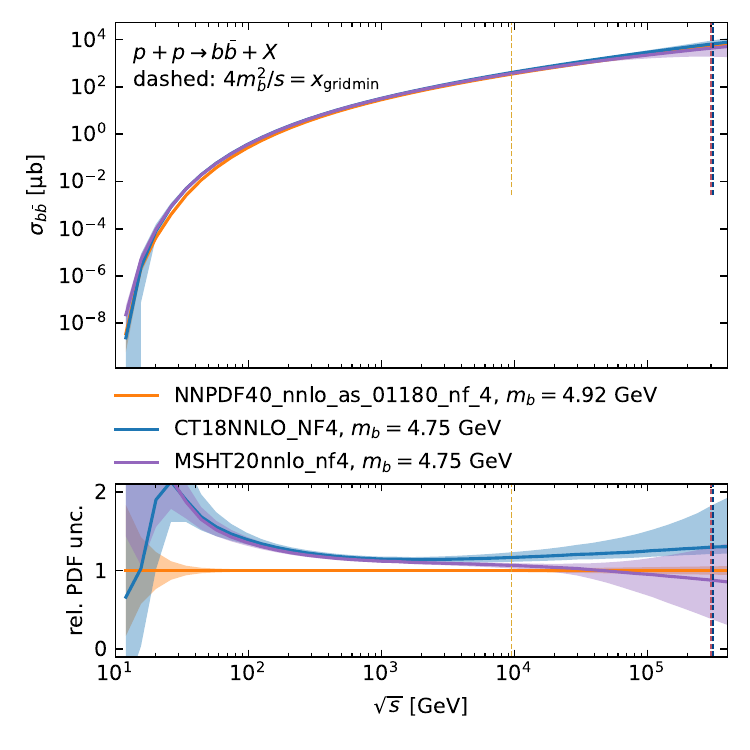}
    \caption{Total charm (left) and bottom (right) cross section at NNLO accuracy in \pp\ collisions as a function of \cm\ energy, over $\sqrts = 10$~GeV--400~TeV, predicted with different PDF sets: NNPDF4.0 (orange), MSHT20 (purple), and CT18 (blue), with bands indicating their corresponding PDF uncertainties. The upper panels show the total cross section, while the lower panels indicate their ratio to the NNPDF4.0 prediction. The vertical dashed lines indicate their associated $\xgmin$ values of the LHAPDF interpolation grids (Table~\ref{tab:PDFs}).
    \label{fig:pdf}}
\end{figure}

Figure~\ref{fig:pdf} shows the PDF dependence of the NNLO cross sections for charm (left) and bottom (right) production as a function of $\sqrts$, as obtained using three different PDF sets: NNPDF4.0 (orange), MSHT20 (purple), and CT18 (blue), with their associated PDF uncertainty bands. For the predictions based on the CT18 PDFs, we use the central values of the fixed-flavour variant \texttt{CT18NNLO\_NF3} and \texttt{CT18NNLO\_NF4} PDFs for charm and bottom, respectively, but the relative PDF error estimates are obtained from the 58 eigenvectors of the default (variable flavour) \texttt{CT18NNLO} set. In general, NNPDF4.0 predicts the lowest cross sections below $\sqrts \approx 2~(10)$~TeV for charm (bottom), whereas CT18 predicts larger values for all collision energies, and MSHT20 falls in between the other two PDFs except above $\sqrts \approx 2~(10)$~TeV where they appear to drop significantly. First of all, and beyond PDF differences discussed in detail below, a first major source of the spread between CT18, NNPDF4.0, and MSHT20 seen in Fig.~\ref{fig:pdf} can be attributed to the values of the heavy-quark masses adopted by each PDF set (Table~\ref{tab:PDFs}), to which the theoretical predictions exhibit strong sensitivity, as illustrated in \cref{fig:mass-ccbar}. The pronounced decline of the MSHT20 PDF predictions at high $\sqrts$ reflects the limitations associated with the very low momentum fractions probed at LHC energies, which, according to \cref{eq:xmin}, can reach values below $x=\num{1e-7}$ for either charm or bottom production. 
The fast decreasing and even negative cross sections for rising $\sqrts$ values occur simply because the MSHT20 PDF interpolation tables are not provided to as small momentum fractions as needed, and the calculation becomes sensitive to the technical extrapolation approach implemented in the LHAPDF interface~\cite{Buckley:2014ana}.
Indeed, the interpolation tables for this PDF are only provided down to $\xgmin=10^{-6}$ (indicated by the leftmost dashed vertical lines in \cref{fig:pdf}) and since certain error set members exhibit a negative slope in this low-$x$ regime, such a trend is exacerbated by the LHAPDF interface (which extrapolates the latest PDF points with a simple polynomial function) leading to vanishing predictions. As a result, the theoretical $\sigmaccbar$ cross sections obtained with MSHT20 have no meaningful physical interpretation above $\sqrts \approx \SI{3}{\TeV}$, as also reflected by the rapidly increasing uncertainty band, whose magnitude is not physically informative. 

Compared to MSHT20, the interpolation tables of CT18 and NNPDF extend to smaller values, down to $\xgmin=10^{-9}$ (indicated by the respective dashed vertical lines in \cref{fig:pdf}). Even so, the NNLO $\sigmaccbar$ predictions for those PDF sets become increasingly sensitive to the assumptions made about the behaviour of the parton densities in a region not directly constrained by the data. The datasets used to constrain PDFs in the small-$x$ region in modern global fits typically include DIS and gauge boson production at colliders, but neither provides directs constraints below $x\approx 10^{-5}$. Instead, PDF groups often impose a small-$x$ power-law behaviour of the form $f\propto x^\alpha$, based on Regge-theory arguments~\cite{Abarbanel:1969eh,Roberts:1990ww}, constrained only by the overall momentum sum rule. The question on how to approach this limit and what the exact scaling is, or at which scale it holds, remain open questions~\cite{Carrazza:2021yrg,Stegeman:2022bvt,Armesto:2022mxy}. In the case of bottom production, the higher factorization scale mitigates the issue, as the PDF behaviour at small $x$ is quickly governed by DGLAP evolution effects~\cite{Ball:1994du}, thereby reducing the dependence on the assumptions imposed on the PDFs at the fitting scale, usually taken near the charm mass threshold.

\begin{figure}[htpb!]
    \centering
    \includegraphics[width=0.5\linewidth]{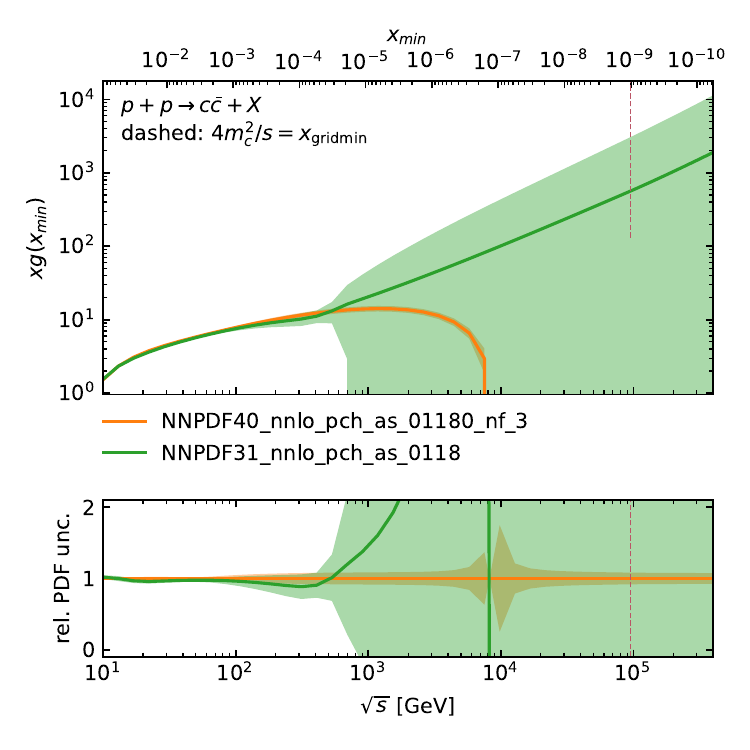}\\
    \includegraphics[width=0.5\linewidth]{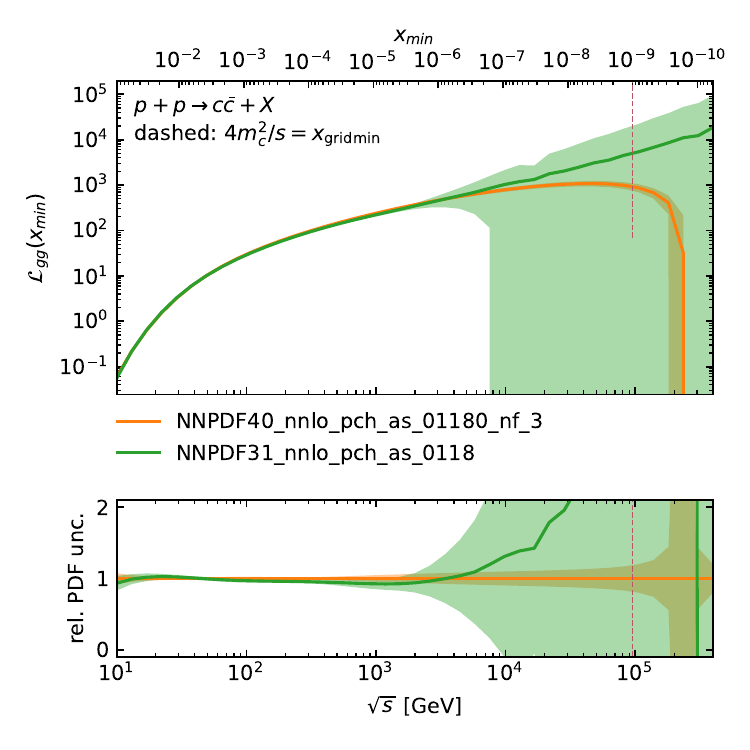}%
    \includegraphics[width=0.5\linewidth]{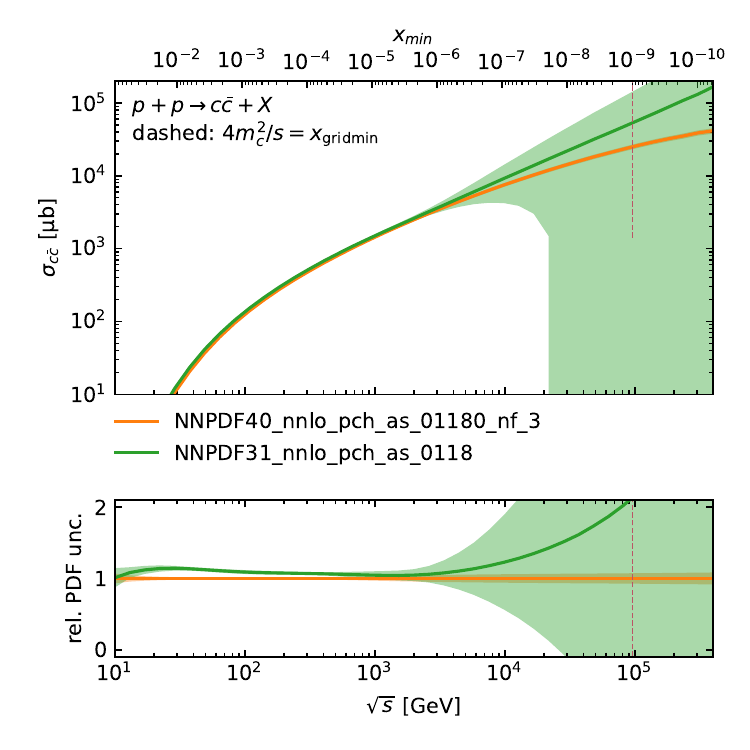}
    \caption{Comparison of the NNPDF3.1 (green) and NNPDF4.0 (orange) NNLO parton densities in \pp\ collisions as a function of $\sqrts$ (lower $x$ axis) and $\xmin$ (upper $x$ axis), for the following quantities:
    Gluon distribution $xg(x,\mu_0^2)$ (upper), gluon-gluon luminosity $\mathcal L_{gg}(\xmin, \mu_0^2)$, \cref{eq:L_gg}, (lower left), and total charm cross section (lower right).
    The lower panels show the relative PDF uncertainties with respect to the central NNPDF4.0 results as a function of $\sqrts$.
    The vertical dashed lines indicate the lower $\xgmin=10^{-9}$ values of the LHAPDF interpolation tables.}
    \label{fig:NNPDF}
\end{figure}

Whereas for PDFs with a fixed polynomial parametrization, such as CT18 and MSHT20, the extrapolation behavior is exclusively driven by the assumed small-$x$ exponent, the behavior of NNPDF sets is more subtle. This can be seen clearly in the top panel of \cref{fig:NNPDF} where the gluon density of the NNPDF3.1~\cite{NNPDF:2017mvq} (green) and NNPDF4.0 (orange) sets are compared, as probed in $\pp\to\ccbar+\mathrm{X}$ production as a function of $\sqrts$. Differences are apparent starting at $\sqrts \approx 1$~TeV, with NNPDF3.1 (NNPDF4.0) rising (decreasing) rapidly with collision energy.
Large differences between the two PDFs are not unexpected as both fits use significantly different datasets and, in particular, NNPDF4.0 includes a large amount of new LHC data that constrain better the gluon density.
However, these LHC constraints do not extend to the small-$x$ regime where the large differences appear, and the most decisive difference appears to be the  employed optimization algorithm:
NNPDF3.1 relied on a genetic algorithm, whereas NNPDF4.0 uses a stochastic gradient descent method (see Ref.~\cite{NNPDF:2021njg} for a detailed discussion).
The former yields largely fluctuating PDFs in the extrapolation region, whilst the latter simply extrapolates the behaviour last seen in the data region as is expected from smooth functions.
Neither of these assumptions have a solid first-principles foundation, nor have they been actually validated with experimental data, and thus should be taken with a grain of salt. 
The fact that the NNPDF4.0 gluon density seems to decrease rapidly with $\sqrts$ is not problematic per se, as the physical cross section, \cref{eq:fact}, involves a double multiplicative convolution.
We can disentangle the double integral by defining the usual parton-parton luminosities $\mathcal L_{ij}$ (see also \cref{fig:lumi})
as the multiplicative convolution over two PDFs, \ie\
\begin{equation}\label{eq:L_gg}
    \mathcal L_{ij}(x,\muF^2) = x \left(f_i(\muF^2)\otimes f_j(\muF^2)\right)(x)\,,
\end{equation}
such that the hadronic cross section is in turn just an ordinary multiplicative convolution of the parton-parton luminosities and the partonic cross sections, \ie\
\begin{equation}
    \sigma^{\pp\to \QQbar+\mathrm{X}}(\rho = 4\mQ^2/s,\muF^2,\muR^2) = \sum_{i,j} \left(\mathcal L_{ij}(\muF^2) \otimes \hat \sigma^{ij\to \QQbar+\mathrm{X}}(\muF^2,\muR^2)\right)(\rho)
\end{equation}
where the partonic cross sections are now folded over $\hat \rho = 4\mQ^2/\hat s$.
As one can see from \cref{fig:NNPDF}, a negative gluon distribution (upper panel) does not immediately imply a negative gluon-gluon luminosity (lower left) as those contributions to the integral need first to become dominant (which eventually occurs for $\sqrts \approx 200$~TeV in the NNPDF4.0 case).
Likewise, a negative gluon-gluon luminosity does not immediately lead to ill-behaved (negative) cross sections (lower right). 
As the differences between NNPDF3.1 and NNPDF4.0 in \cref{fig:NNPDF} clearly indicate, it should be useful to incorporate the $\sigmaccbar$ measurements at the LHC to the global PDF fits, helping to obtain a data-driven constraint on the gluon density at very small momentum fractions $x \lesssim 10^{-5}$. While the theoretical uncertainties associated with scale and mass variations are sizable, they can be systematically propagated into the PDF fits by well-established procedures~\cite{NNPDF:2024dpb}. As a result, though these new constraints would not be very strong, they would nonetheless reduce the sensitivity of the small-$x$ PDFs to assumptions about their functional forms and to the details of the fitting methodology or optimization algorithms.


\subsubsection*{QCD coupling uncertainties}

\begin{figure}[htpb!]
    \centering
    \includegraphics[width=0.5\linewidth]{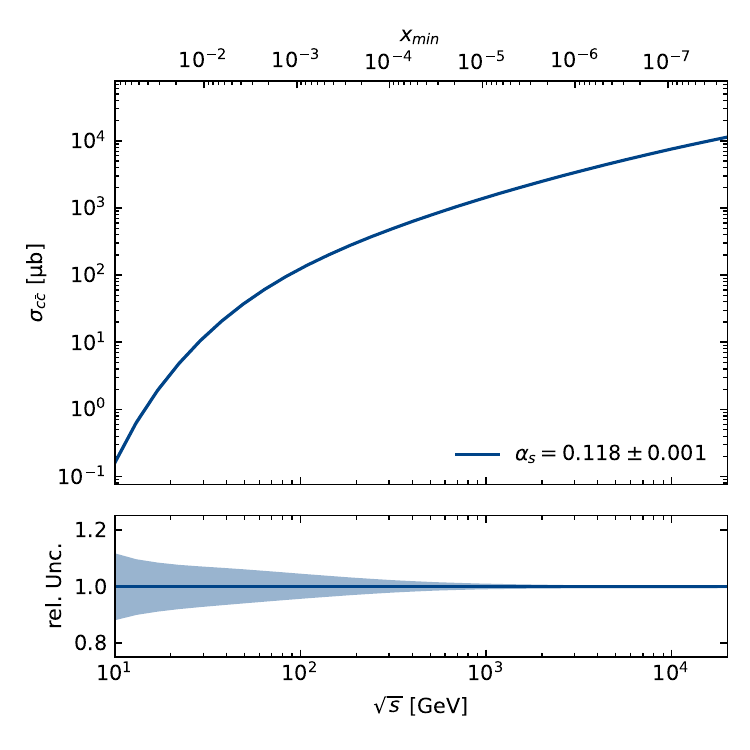}%
    \includegraphics[width=0.5\linewidth]{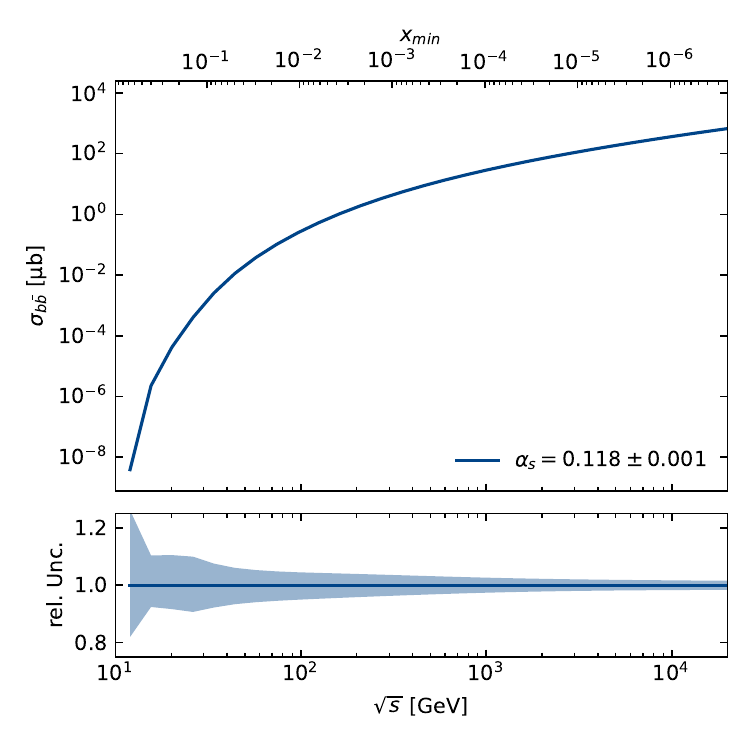}
    \caption{Inclusive charm (left) and bottom (right) NNLO production cross sections as a function of collision energy, over $\sqrts = 10$~GeV--20~TeV, with the bands showing the uncertainty associated with variations of the strong coupling constant over $\alphasmZ = 0.117$--0.119. The lower panels show the relative $\alphasmZ$ uncertainties as a function of $\sqrts$.
    \label{fig:alphas}}
\end{figure}

In \cref{fig:alphas} we evaluate the impact of the strong coupling uncertainty, $\alphasmZ = 0.118\pm0.001$~\cite{ParticleDataGroup:2008zun}, in the charm (left) and bottom (right) cross sections. In our calculations, the strong coupling is varied over $\alphasmZ = 0.117$--0.119 in the hard matrix elements as well as, consistently, in the PDFs. For the latter, we use the standard (VFN scheme) NNPDF4.0 sets, which are provided with $\alphasmZ$ variations, to account for the corresponding change in the PDFs~\cite{Forte:2020pyp,Forte:2025pvf}. For both $\ccbar$ and $\bbbar$ production the propagated $\alphaS$ uncertainty is small, of about $\SI{10}{\percent}$ at the lowest collision energies, further decreasing with $\sqrts$ and becoming negligible at the LHC. The larger impact on the cross section with respect to the actual change in its value is due to the running of the strong coupling, which enhances differences at the small renormalization scales used here ($\muR = 2\mQ$). We neglect any correlation that might appear by considering simultaneous variations of both the value of $\alphasmZ$ and the heavy quark mass $\mQ$, which, in principle, could appear through the gluon distribution. Note that apart from the purposes of this exercise, we always consider the correct FFN-scheme running of the strong coupling for the rest of the results presented in this work. In any case, the uncertainties associated with $\alphasmZ$ variations are clearly negligible compared with all other sources of theoretical uncertainties (scales, PDFs, quark masses) estimated in this work.

\subsection{VFN scheme: NLO calculations}
\label{subsec:resumm}

\begin{figure}[htbp!]
    \centering
    \includegraphics[width=0.495\linewidth]{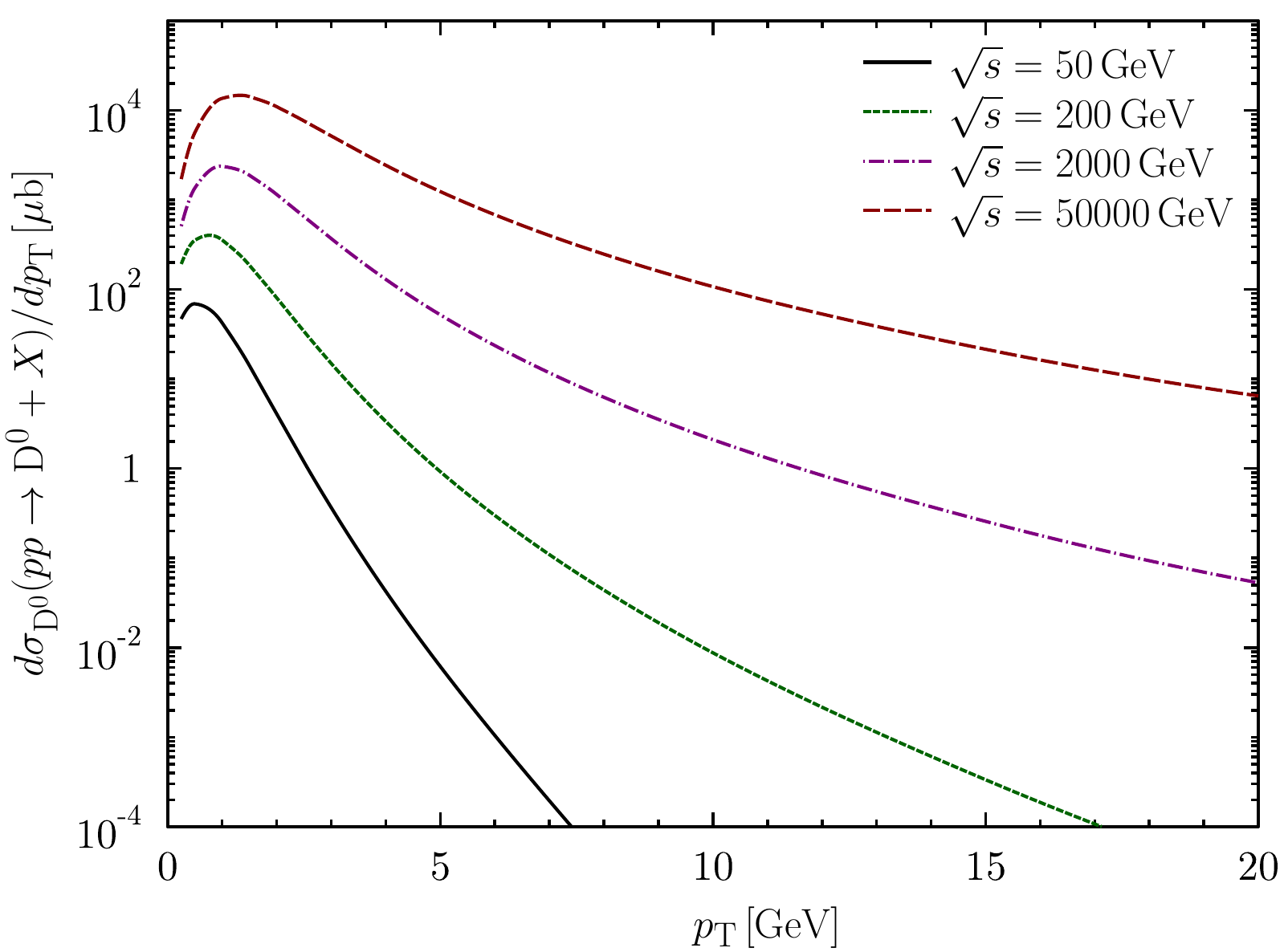}
    \includegraphics[width=0.495\linewidth]{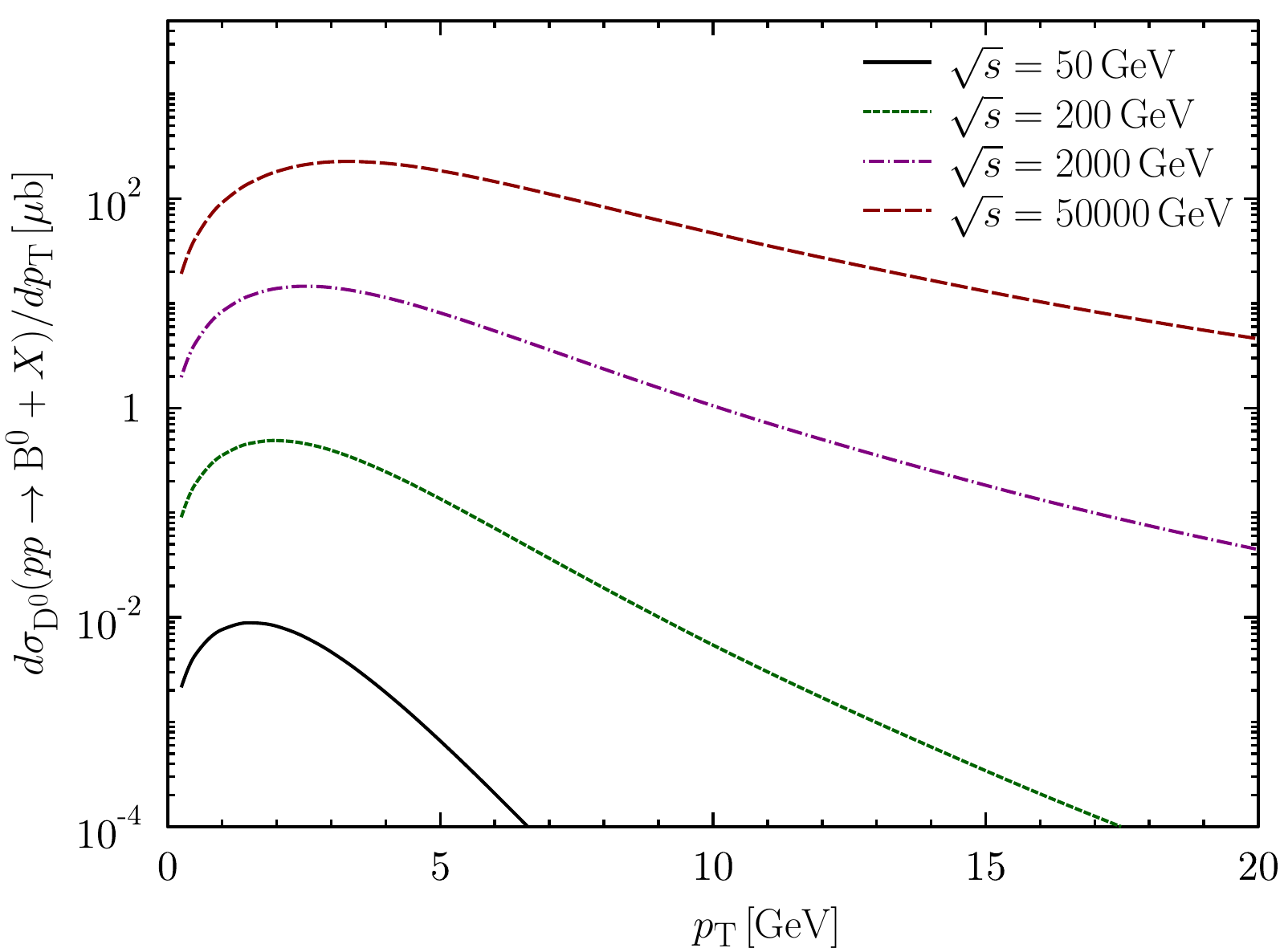}
    \caption{Inclusive transverse momentum $\pT$ spectra of D$^0$ (left) and B$^0$ (right) mesons produced in \pp\ collisions at various \cm\ energies as computed in the SACOT-$\mT$ framework.}
    \label{fig:nlo_sacot1}
\end{figure}

As discussed in the previous section, the perturbative convergence of the FFN-scheme calculations appears rather slow, indicating that higher-order contributions are still presumably significant. A certain class of higher-order contributions can be accounted for in the 
GM-VFN scheme, which resums contributions that originate from collinear radiation and splitting from the initial- and final-state partons. For example, a collinear splitting of an initial-state gluon into a heavy quark-antiquark pair gives rise to a logarithmic term of the form $\alphaS \log(\mQ^2/m^2_\mathrm{T})$, where $\mT$ is the transverse mass of the heavy quark. Logarithmic terms of this sort can be resummed in GM-VFNS by introducing a DGLAP-evolving heavy-quark PDF. In the same spirit, a final-state gluon can split into a heavy quark-antiquark pair giving rise to a similar logarithmic term, which in this case can be resummed by introducing a scale-dependent gluon-to-heavy-quark FF. The numerical significance of the resummation, however, is not so easy to assess: it is not only the value of $\alphaS \log(\mQ/\mT)$ that matters, but \eg the resummation always involves convolutions between partonic splitting functions and PDFs/FFs that obscure the simple power counting. In addition, and more importantly, the resummation suffers from scheme dependence, and this dependence is particularly severe at low values of transverse momentum which is precisely where most of the inclusive cross section accumulates. On the one hand, this leads to increased theoretical uncertainties and, in this sense, the FFN-scheme calculations for fully inclusive cross sections are typically considered more reliable. On the other hand, towards higher values of transverse momentum, the resummation becomes eventually necessary as the $\alphaS \log(\mQ^2/m^2_\mathrm{T})$ term grows. Whereas at low \cm energies the transverse-momentum spectra for D- and B-mesons are very narrow, peaked at low $\pT$, they become increasingly wider in the range of LHC and FCC collision energies and the average transverse momentum grows larger (\cref{fig:nlo_sacot1}). As a result, the importance of the GM-VFNS approach should increase towards large \cm energies.

The general structure of a GM-VFNS $\pT$-differential cross section for the production of a heavy-flavoured hadron $\mathrm{h}_3$ in a collision between hadrons $\mathrm{h}_1$ and $\mathrm{h}_2$,
\begin{equation}\label{eq:scattGM}
    \mathrm{h}_1(k_1) + \mathrm{h}_2(k_2) \to \mathrm{h}_3(p) + \mathrm{X} \,,
\end{equation}
is of the form (summed over incoming and outgoing partons $i, j$, and $l$),
\begin{align}
\frac{\mathrm{d}^3\sigma^{\mathrm{h}_1 + \mathrm{h}_2 \to \mathrm{h}_3 + \mathrm{X}}}{\mathrm{d}^3p/p^0}
& =
 \sum _{i,j,l} 
 \int_{z^\mathrm{min}}^1 \frac{\mathrm{d}z}{z^2} \int_{x_1^\mathrm{min}}^1 \mathrm{d}x_1 \int _{x_2^\mathrm{min}}^1 \mathrm{d}x_2 \,
 f_i^{\mathrm{h}_1}(x_1,\muF^2) \, f_j^{\mathrm{h}_2}(x_2,\muF^2) \, D_{l \to \mathrm{h}_3}(z,\muFF^2)
  \nonumber \\[3pt]
& \times  J(\vec {k}, \vec p) \times
\frac{\mathrm{d}^3\hat{\sigma}^{ij\to l+\mathrm{X}} (x_1, x_2, \mQ , \sqrts, \muR^2, \muF^2, \muFF^2)}{\mathrm{d}^3 k/k^0} 
\label{eq:master} 
- \text{subtraction \ terms.}
\end{align}
Here $p$ denotes the four-momentum of the outgoing hadron, $k$ is the four-momentum of the parton that fragments into $\mathrm{h}_3$, and $D_{l \to \mathrm{h}_3}(z,\muFF^2)$ denotes the corresponding parton-to-hadron fragmentation function at the fragmentation scale $\muFF$. The hadronic and partonic momenta are related by the fragmentation variable $z$, whose definition is not unique. A reasonable choice is
\begin{align}
z \equiv \frac{p \cdot \left(k_1 - k_2\right)}{k \cdot \left(k_1 - k_2\right)} \,,
\label{eq:fragvar}
\end{align}
as argued in Ref.~\cite{Helenius:2023wkn}. In this case the factor $J(\vec k, \vec p)$ in Eq.~(\ref{eq:master}) is
\begin{align}
J(\vec {k}, \vec p) = 
\sqrt{ \frac{ {\vec p}^{\,2} + \mhQ^2}{{\vec p}^{\,2}}  
\frac{{\vec {k}}^{\,2}}{{\vec k}^{\,2} + \mQ^2}  } \,,
\end{align}
where $\mhQ$ is the mass of the heavy-flavoured hadron. The partonic coefficient functions $\mathrm{d}^3\hat{\sigma}^{ij\to l+\mathrm{X}}$ as well as the integration limits $z^\mathrm{min}$ and $x_i^\mathrm{min}$ are scheme-dependent. The GM-VFNS results presented here are obtained in the so-called SACOT-$\mT$ scheme whose underlying idea is to retain the implicit (``kinematic") mass dependence also, \eg, in those partonic channels that do not explicitly involve a heavy quark~\cite{Helenius:2018uul}. The subtraction terms in Eq.~(\ref{eq:master}) remove the double counting of logarithmic terms that have been resummed into the scale-dependent heavy-quark PDFs and FFs. 

The integrated charm production cross section is calculated by evaluating the integrated cross section for D$^0$ production using the Kneesch--Kniehl--Kramer--Schienbein~\cite{Kneesch:2007ey} FFs and the \texttt{NNPDF40\_nlo\_pch\_as\_01180} proton PDFs, and then dividing the result by the appropriate $f(\rm c \to h_c)$ branching fraction. Since the FFs were fitted to $\epem$ data, we use here $f(\rm c \to {\rm D}^0) = 0.57$ consistently with the LEP data (Fig.~\ref{fig:fragfractions_charm}, upper plots).
For the integrated bottom production, we follow the same procedure but this time using the Kniehl--Kramer--Schienbein--Spiesberger FFs~\cite{Kniehl:2007erq} for B mesons, the \texttt{NNPDF40\_nlo\_as\_01180} proton PDFs,
and $f(\rm c \to {\rm B}^0) = 0.4$ (Fig.~\ref{fig:fragfractions_bottom}, upper left).
The renormalization, factorization, and fragmentation scales have been chosen as
\begin{align}
\mu_i = c_i\sqrt{\pT^2 + \mQ^2} \geq \mQ^2 \,,
\end{align}
with the default $c_i=1$ value. The sensitivity of our result to missing higher-order corrections is quantified by varying $c_i$ by a factor of two with the constraints
$1/2\leq \muF/\muR \leq 2$, $1/2\leq \muFF/\muR \leq 2$, and $\mu_i \geq \mQ^2$. 
This leads to 17 different combinations of scales, and the corresponding ``scale uncertainty" is then defined to be the maximum/minimum envelope found by evaluating the cross sections with these 17 combinations. The scale uncertainty obtained in this way is rather "pessimistic" in a sense that the extreme scale variations are known~\cite{Helenius:2018uul,Helenius:2023wkn} to clearly over-/under-estimate the LHC measurements at low $\pT$. In other GM-VFNS variants, such as FONLL, the scale uncertainties are smaller due to choices made in defining the scheme which forces the low-$\pT$ regime to be more FFN-like.

\subsection{Corrections for \pp-equivalent cross sections}

The \MaunaKea\ and \sacotmt\ cross section calculations are performed using the proton PDFs as expected for \pp\ collisions, but a fraction of the heavy-quark measurements have been performed in \ppbar\ (at  CERN Sp\=pS and Fermilab Tevatron colliders) and \pA\ (in fixed target, RHIC, or LHC) collisions. 
The \pA\ results collected in our world-data systematics (Tables~\ref{tab:charmmeasurements} and~\ref{tab:beautymeasurements}) are already `\pp-equivalent', namely they are quoted after normalizing the experimental cross section by the corresponding mass number $A$. 
If the actual collision system differs from \pp\ at a given $\sqrts$ value, a (small) correction factor needs to be further applied on the total predicted inclusive cross section to account for potential differences between proton and antiproton or nuclear PDFs, of the form
\begin{equation}\label{eq:pX}
    R_{\pX}(\sqrts) = \frac{\sigma^{\pX}(\sqrts)}{\sigma^{\pp}(\sqrts)}, \mbox{ with X = \=p,\,A.}
\end{equation}
For the \ppbar\ case, this correction takes into account the charge conjugation transformation of the PDFs, whereas for \pA\ collisions we use the corresponding nuclear PDF (nPDF) from the NLO EPPS21 set~\cite{Eskola:2021nhw}.
The partonic cross-sections remain of course the same, but we restrict them to NLO to match the perturbative accuracy of EPPS21.
We find very similar results when using NNLO partonic cross-section since $R_{\pX}(\sqrts)$ is a ratio of cross sections.
Moreover, it is worth recalling that the EPPS21 nPDFs are extracted from data by considering ratios of cross sections in the first place, which make them perturbatively more stable. The proton baseline in EPPS21 is CT18ANLO~\cite{Hou:2019efy}, which we thus use to evaluate \cref{eq:pX}.

\cref{tab:pA_factors} lists the different correction factors applied to transform the calculated \pp\ predictions for charm and bottom cross section into the corresponding values for $\ppbar$ or \pA\ collisions, accounting for the inclusive effect of the antiproton or nuclear PDFs on the total yields. For each colliding system we list the proton and nuclear PDFs used to derive the value of the correction factor given by \cref{eq:pX}. The correction factor for \pPb\ collision at the LHC is $R_{\pX}(\sqrts=5~\text{TeV}) = 0.77$ indicating a ${\sim}25\%$ gluon shadowing effect in the lead ion.
At \ppbar\ colliders, the differences between proton and antiproton PDFs are negligible for heavy-quark production because this process is dominated by gluon-gluon (rather than valence quark) scatterings, so the factor is $R_{\pX}(\sqrts) = 1$ for them. The same is true for proton-light-ion collisions, such as p-He and p-Ne, for which nPDFs effects are expected to be negligible within the current theoretical uncertainties. For fixed-target results, we use the gold nPDF for both p-Au collisions, as well as for the carbon (C), titanium (Ti), and tungsten (W)  targets.
The correction factors for such collisions are $R_{\pX}(\sqrts\approx40~\text{GeV}) = 1.05$, indicating a mild ${\sim}5\%$ gluon antishadowing in this kinematic regime.
Of course, these correction factors come with their own (few percent) uncertainty, which is nonetheless negligible compared with other sources of theoretical uncertainty (scales, PDFs, and heavy-quark masses).

\begin{table}[htbp!]
\caption{Overall correction factors applied to the theoretical inclusive $\sigma(\pp\to\QQbar)$ cross sections for charm and bottom, to transform them into the corresponding pp-equivalent $\sigma(\ppbar\to\QQbar)$ or $\sigma(\pA\to\QQbar)$ values, accounting for antiproton or nuclear PDF effects. For each colliding system we provide the PDFs used to derive the value of the correction factor through \cref{eq:pX}.
\label{tab:pA_factors}}
\centering
\tabcolsep=3.5mm
\begin{tabular}{llc}\hline
Measurement & PDFs used & Correction factor $R_{\pX}$ \\\hline
$\sigmaccbar$ cross sections: & & \\
LHC (5 TeV pPb)~\cite{LHCb:2017yua,ALICE:2016yta} & CT18ANLO (p), EPPS21 (Pb) & 0.77 \\ 
CDF (1.96 TeV p\=p)~\cite{CDF:2016lht} & CT18ANLO (p, \=p) & 1.00 \\ 
LHCb (86.6 GeV pHe)~\cite{LHCb:2018jry} & CT18ANLO (p, p) & 1.00 \\
LHCb (68.5~GeV pNe)~\cite{LHCb:2022cul} & CT18ANLO (p, p) & 1.00 \\
HERA-B (41.6~GeV pA, A=C,Ti,W)~\cite{HERA-B:2007rfd} & CT18ANLO (p), EPPS21 (Au) & 1.05 \\ 
E789 (38.8 GeV pAu)~\cite{E789:1994nhc} & CT18ANLO (p), EPPS21 (Au) & 1.05 \\ \hline 
$\sigmabbbar$ cross sections: & & \\
UA1 (630 TeV p\=p)~\cite{UA1:1993jok}  & CT18ANLO (p, \=p) & 1.00 \\ 
E771 (38.7~GeV pSi)~\cite{E771:1997rbx,Lourenco:2006vw} & CT18ANLO (p), EPPS21 (Ca) & 0.99 \\ 
E789 (38.7~GeV pAu)~\cite{Jansen:1994bz,Lourenco:2006vw} & CT18ANLO (p), EPPS21 (Au) & 0.98 \\ 
\hline
\end{tabular}
\end{table}

\section{Data versus pQCD}
\label{sec:compare}

We now turn to a comparison between the experimental data, compiled in \cref{sec:exp}, and the theoretical predictions, including all sources of uncertainties, discussed in \cref{sec:th}. We remind the reader that the presented predictions here do not take a stand concerning the heavy-quark baryon-versus-meson enhancement observed at the LHC, \ie\ they adopt a purely factorized leading-twist picture, while at least a fraction of the experimental data compared with the theory predictions is sensitive to the region where signals of HT effects have been observed.\\

In this section, we also present theoretical cross-section predictions extending up to the highest-energy hadronic collisions accessible at the FCC ($\sqrts\approx 100$~TeV), as well as in cosmic-ray interactions at the GZK cutoff ($\sqrts\approx 400$~TeV). Whereas the renormalization and factorization scales are formally unphysical, as their dependence would cancel in an all-orders prediction of the cross section, at fixed perturbative order a residual dependence remains due to the truncation of the perturbative series. For this reason, these scales must be chosen close to the characteristic hard scale of the process in order to avoid large logarithmic corrections and ensure good perturbative convergence. At very high collision energies, progressively smaller values of $x$ are probed and, in particular for the gluon PDF, the dependence on the factorization scale becomes especially pronounced near $\muF = \mc$, leading to a substantial amplification of the associated scale uncertainties. To circumvent this issue, we explore an alternative prescription for the default theoretical scales, $\muF$ and $\muR$, when evaluating the missing higher-order uncertainties of the \MaunaKea\ NNLO predictions. Instead of fixing the default scales to a static value $\mu_\text{static}=2\mQ$, corresponding to the production of a $\QQbar$ pair at rest, and following the \sacotmt\ results and discussion of Sec.~\ref{subsec:resumm}, we include in quadrature a kinematic dependence through the average $\pT$ of the outgoing quarks as follows,
\begin{align}
\mu_\text{dyn.} = c_i \sqrt{(2\mQ)^2 + \langle\pT\rangle^2} > 2\mQ \,,
\text{with } c_i=1\text{ for the default value and }c_i=\pm(2,1/2) \text{ for the variations.}
\label{eq:mu_dyn}
\end{align}
At low $\sqrts$, one has $\mu_\text{static} \simeq \mu_\text{dyn.}$, whereas at high $\sqrts$ the dynamical scale better captures the increasing effective energy probed by the underlying partonic scatterings, as reflected in the progressively extended high-$\pT$ tail of heavy-quark meson spectra in Fig.~\ref{fig:nlo_sacot1}. We have parametrized the approximately logarithmic dependence of $\langle\pT\rangle$ on $\sqrts$,
as predicted from the differential \sacotmt\ calculations, and combined it quadratically with $(2\mQ)$, which results in a modest $\approx$(20--30\%) enhancement of the $\muF$ and $\muR$ values at LHC and FCC/GZK-cutoff energies relative to the static scale choice. Using such a dynamic scale mitigates the large growth of the scale uncertainties that arises at very high \cm\ energies when using the fixed rest-mass of the heavy-quark pair system as the default theoretical scale. This points once more to the slow convergence of the perturbative expansion, which might be cured by resummation as discussed earlier.

\subsection{$\ccbar$ cross sections}

\begin{figure}[htbp!]
    \centering
    \resizebox{0.86\textwidth}{!}{%
        \includegraphics[width=\textwidth]{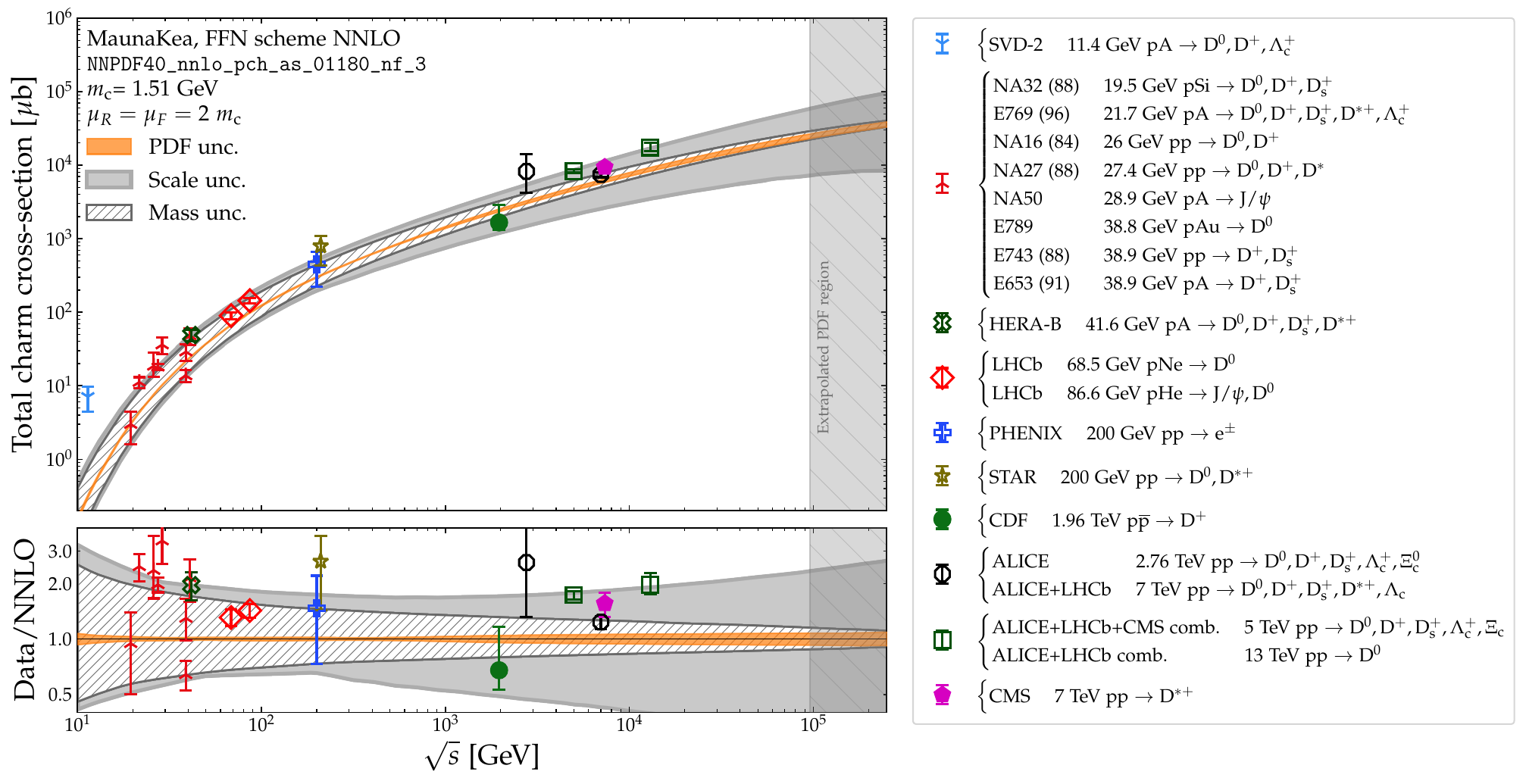}
    }
    \resizebox{0.86\textwidth}{!}{%
        \includegraphics[width=\textwidth]{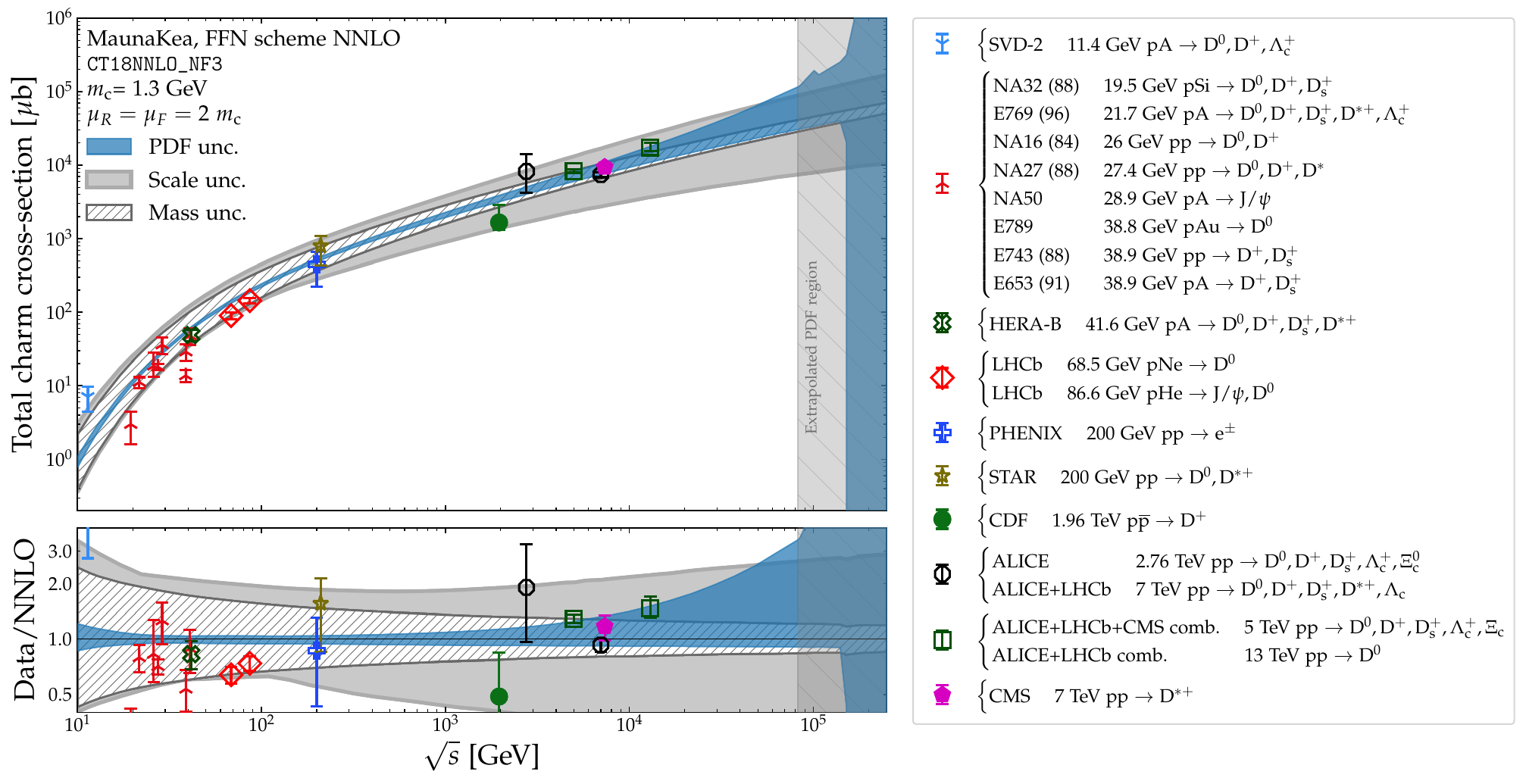}
    }
        \resizebox{0.86\textwidth}{!}{%
        \includegraphics[width=\textwidth]{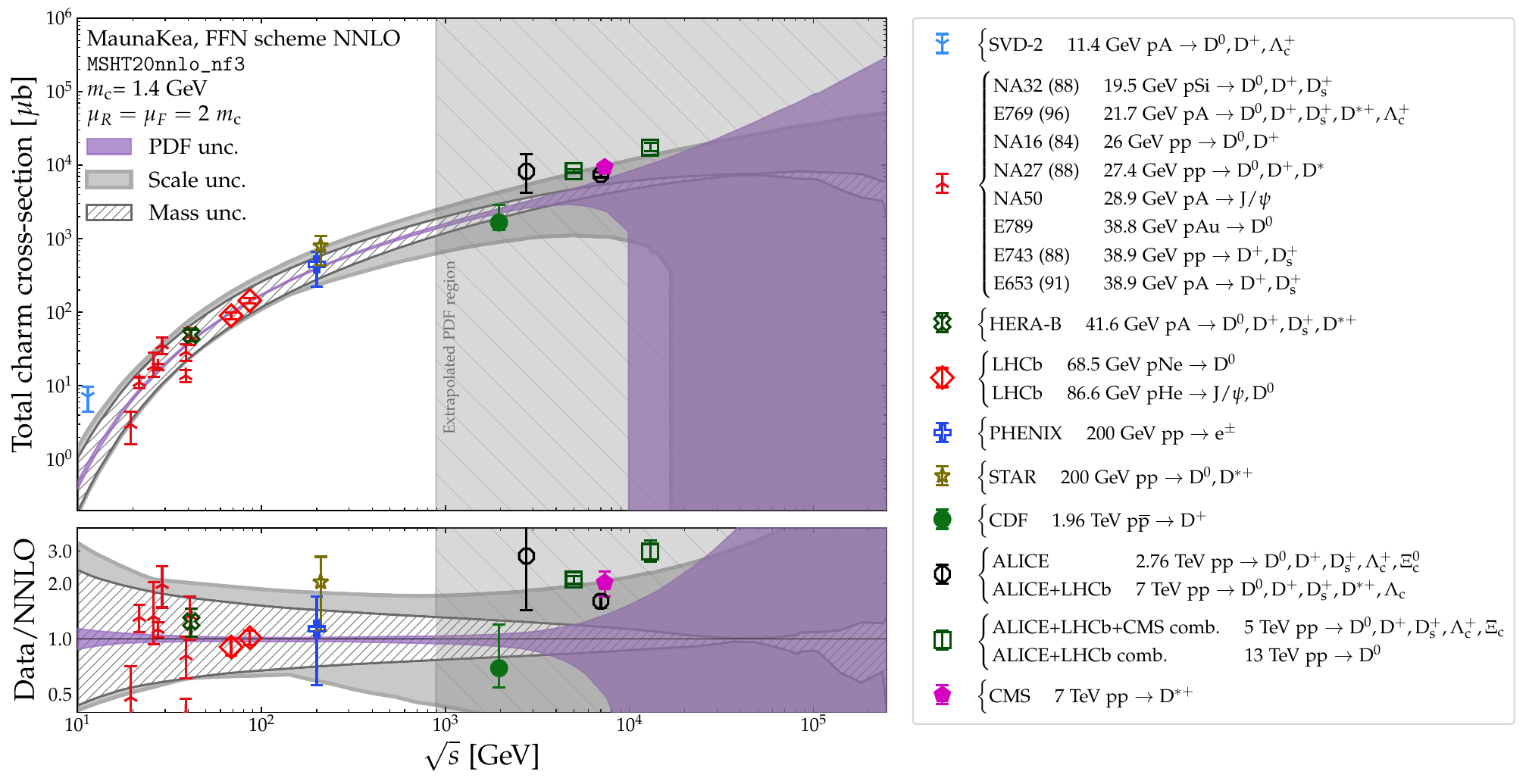}
    }
    \caption{Charm cross sections in \pp{} collisions as a function of \cm\ energy, over $\sqrts=10$~GeV--400~TeV: The experimental data (symbols, selected from \cref{tab:charmmeasurements}) are compared with theoretical \MaunaKea\ NNLO (curves) for the NNPDF4.0 (upper), CT18 (middle), and  MSHT20 (lower) PDFs. The PDF uncertainty (in orange, blue, and purple for NNPDF4.0, CT18, and MSHT20, respectively) (see also \cref{fig:pdf}), the scale variation uncertainties (gray) (\cref{fig:pto}), and the charm-quark mass uncertainty (striped white and dark gray) (\cref{fig:mass-ccbar}) are plotted as bands. The vertical shaded light-gray areas indicate the respective LHAPDF extrapolation regions (for which $\xmin\le\xgmin$, see text). The static scale $\muF=\muR = 2\mQ$ is used. The lower panels show the data over NNLO ratios.
    \label{fig:ccbar_exp_xsecs}}
\end{figure}

\begin{figure}[htbp!]
    \centering
    \resizebox{0.86\textwidth}{!}{%
        \includegraphics[width=\textwidth]{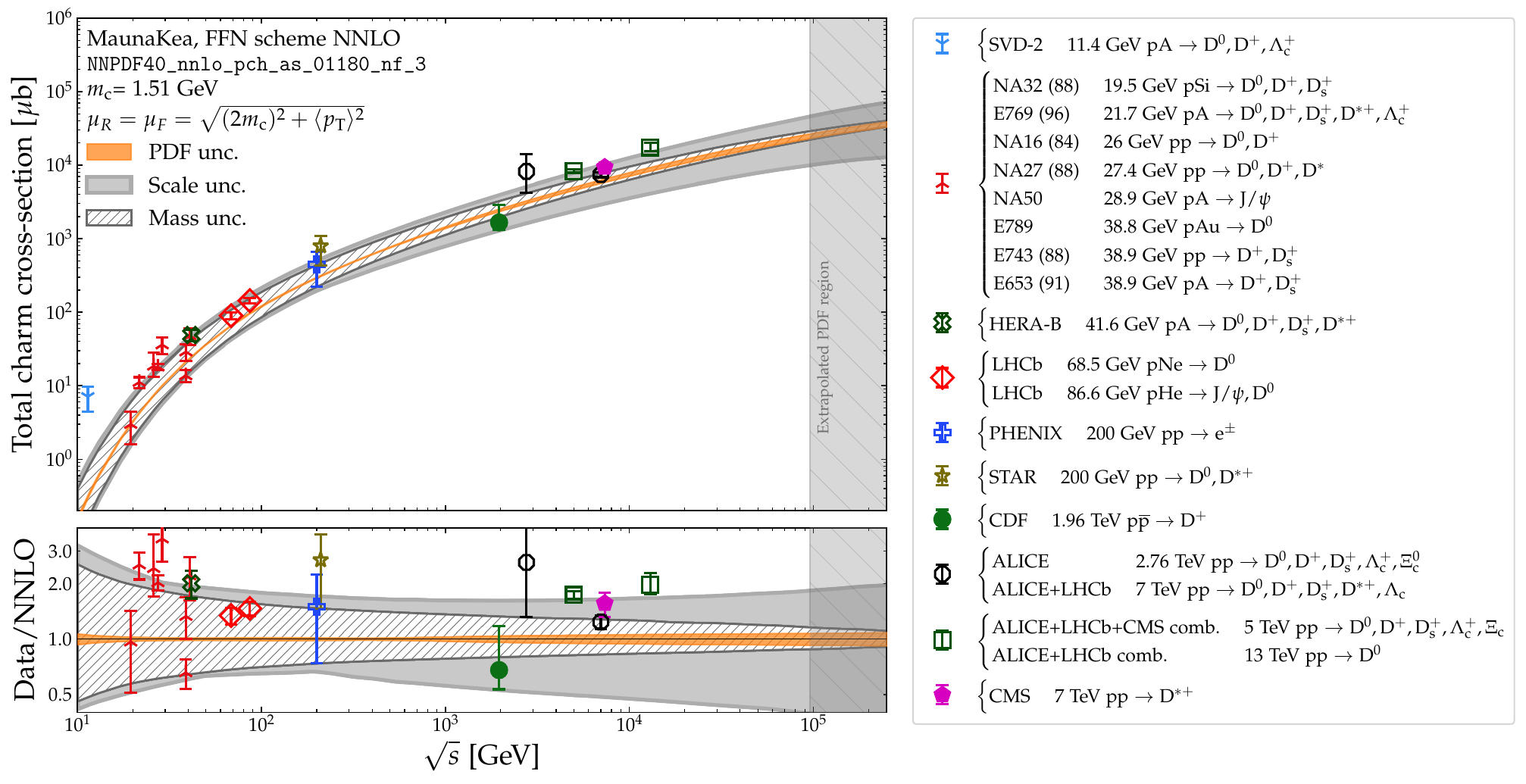}
    }
    \resizebox{0.86\textwidth}{!}{%
        \includegraphics[width=\textwidth]{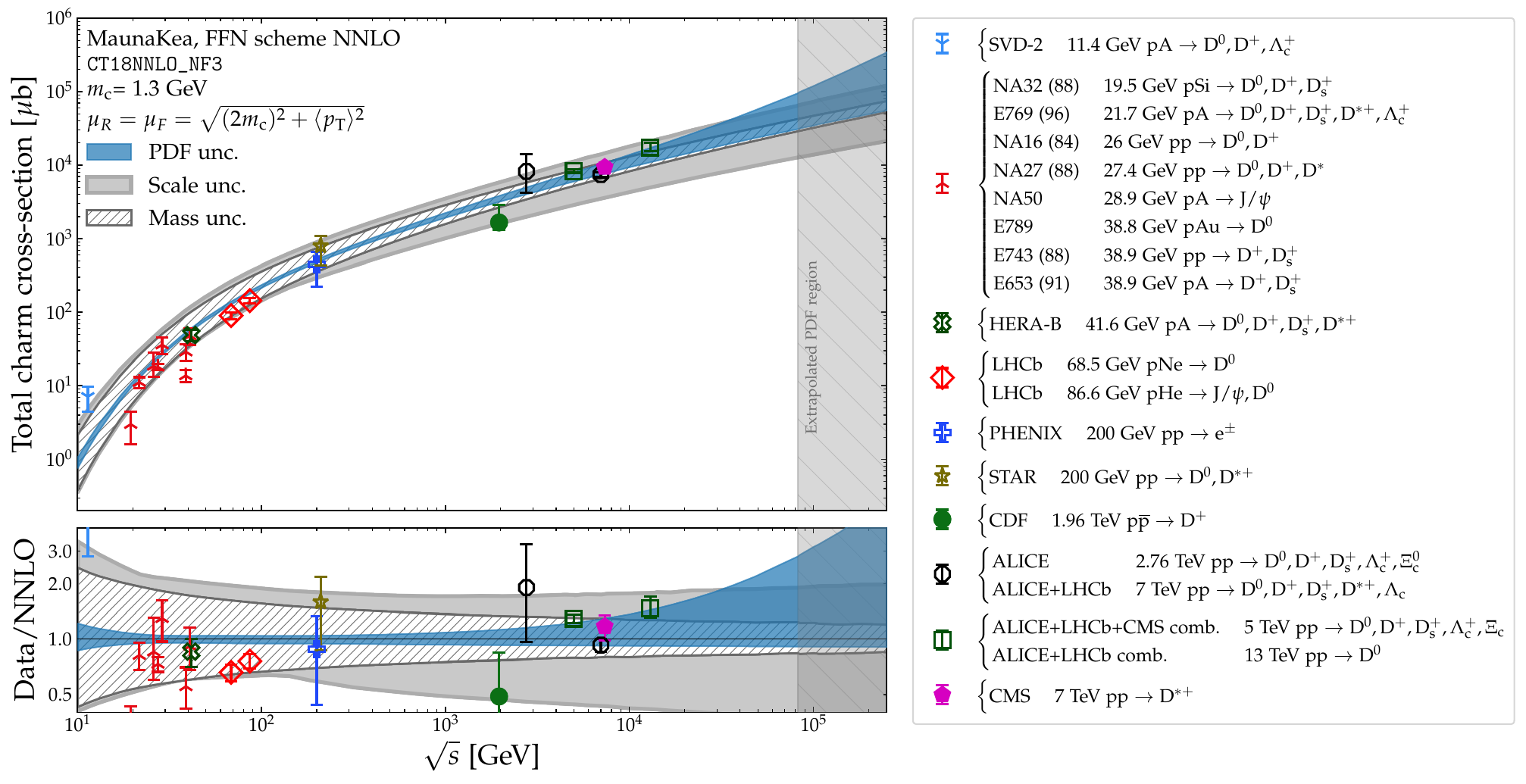}
    }
        \resizebox{0.86\textwidth}{!}{%
        \includegraphics[width=\textwidth]{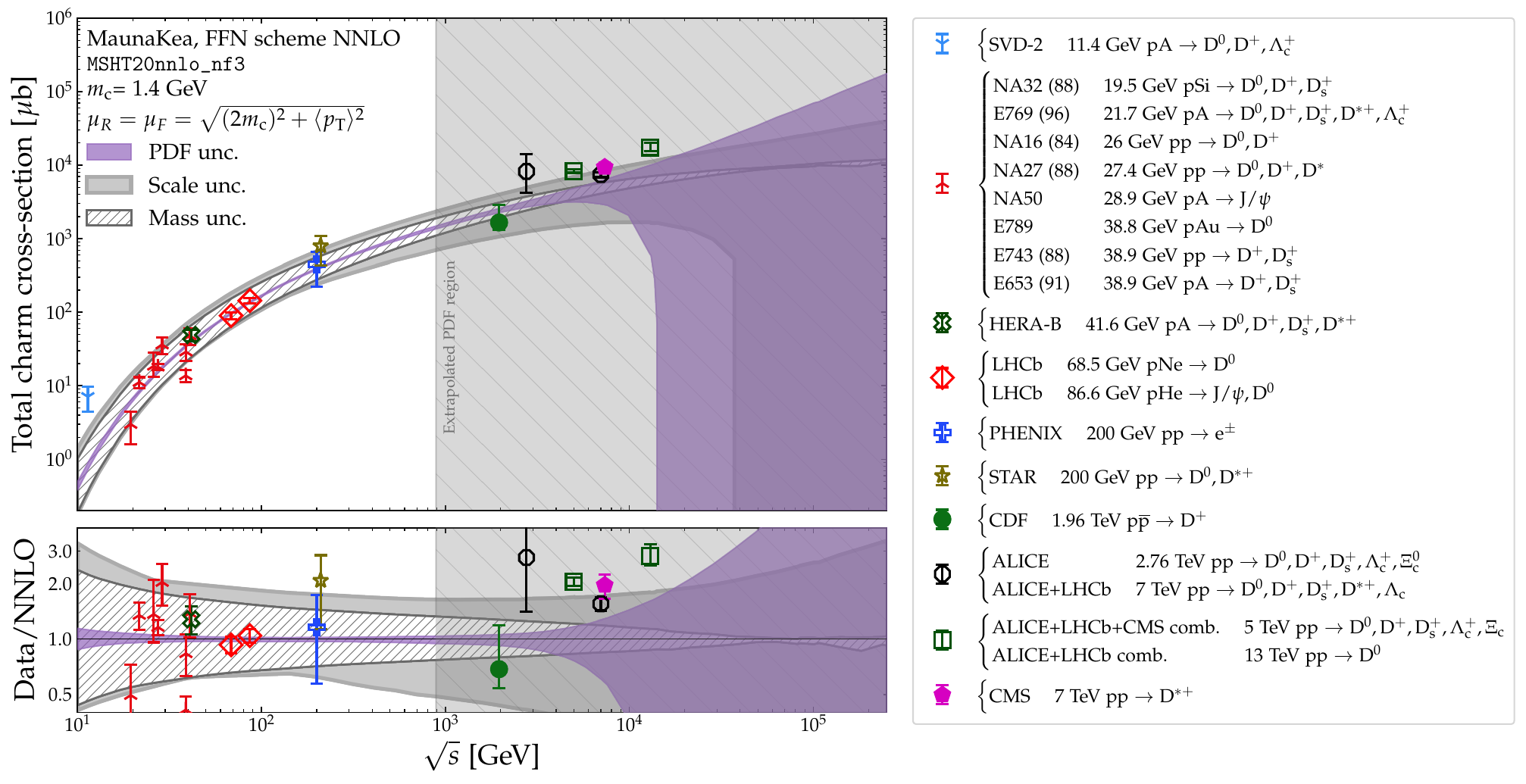}
    }
    \caption{
    The same as Fig.~\ref{fig:ccbar_exp_xsecs} but with the dynamic scale $\muF=\muR = \mu_\text{dyn.}$, \cref{eq:mu_dyn}.
    \label{fig:ccbar_exp_xsecs_dyn}}
\end{figure}

Figure~\ref{fig:ccbar_exp_xsecs} shows the selected set of experimental total charm production cross sections (gray entries in \cref{tab:charmmeasurements}) as a function of $\sqrts$ compared with the NNLO \MaunaKea\ calculations obtained with the NNPDF4.0, CT18, and MSHT20 proton PDFs. Within the considered sources of theoretical uncertainties (scales, PDFs, and charm quark mass), all three sets of PDFs lead to predictions consistent with the data over the \cm\ energies where the probed values of momentum fraction $x$ remain within the region covered by the LHAPDF grid files (\ie, for the range of $\sqrts$ to the left of the vertical shaded light-gray areas indicating the extrapolated PDF regions in the plots). At the LHC energies, the data appear to lie in the upper edge of the scale-uncertainty band of the NNPDF4.0 predictions, while being better in line with the CT18 results. The main reason for this difference is the value of the charm-quark mass, which is lower in CT18 than in NNPDF4.0 and leads to higher cross sections across all values of $\sqrts$. With the charm mass adopted by the NNPDF4.0 group an optimal agreement with the data would still require an N$^3$LO correction which increases the cross sections by a factor of two or so, \ie\ of the same size as the NNLO correction, see \cref{fig:pto}. At first sight, the PDF uncertainty of NNPDF4.0 would appear to be too small to bridge the factor of two difference with respect to the data. However, the NNPDF4.0 uncertainty is likely unrealistically small (due to significant parametrization bias), see \cref{fig:NNPDF}, as no data in the NNPDF4.0 global fit can directly constrain PDFs at the low values of $x$ and factorization scales probed by the LHC charm data. The CT18 uncertainty is larger but seems to only allow variations towards higher cross sections. This is due to the CT18 fit constraints, which force the PDFs to remain positive at the parametrization scale $\mc$. This is a rather strict requirement because the parametrization scale can, in principle, be chosen entirely freely. In other words, while being more realistic than the NNPDF4.0 one, also the CT18 PDF uncertainty is subject to rather restrictive assumptions at the global fit stage. Regarding the MSHT20 predictions, the LHC kinematic region already lies beyond the validity range of its associated PDF grids (indicated by the vertical gray shaded areas in the plot). From a practical perspective, it would therefore be beneficial to extend the $x$ grids coverage down to ${\sim}10^{-10}$ in future MSHT releases. Despite this problem, the central MSHT20 predictions are well in line with the measurements for collision energies below the LHC, whereas the central CT18 (NNPDF4.0) curve tends to somewhat overshoot (undershoot) the data. Also here, the hierarchy observed between the NNPDF4.0, CT18, and MSHT20 cross sections follows the different adopted charm-mass values of each PDF set.

Notwithstanding the sizable scale uncertainties in the theoretical calculations, we can conclude that $\sigmaccbar$ measurements at the LHC can provide valuable input in future PDF global fits to constrain the gluon density at very small momentum fractions $x\approx 10^{-6}$, ensuring that its behaviour is increasingly driven by experimental data rather than by assumptions underlying the small-$x$ extrapolation. Once the theoretical uncertainties are accounted for in the fits, the constraints will not be very strong, but can better ``anchor'' the low-$x$ gluon density and \eg\ exclude the clearly nonphysical cases with negative cross sections visible \eg\ for NNPDF3.1 in Fig.~\ref{fig:NNPDF}. Although (well) beyond the range of validity of the current PDF sets, Fig.~\ref{fig:ccbar_exp_xsecs} shows theoretical cross sections up to a few hundred TeV where charm (and bottom) production represents a substantial fraction of the total inelastic cross section. These are of relevance for future-hadron collider (FCC-hh) perspectives~\cite{Mangano:2016jyj,FCC:2018vvp}, and have an impact on muon and neutrino production in cosmic ray physics at ultrahigh energies, approaching the so-called ``GZK cutoff'' at $\sqrts\approx 400$~TeV~\cite{dEnterria:2016yhy,dEnterria:2018kcz}. A reliable understanding of this ultrahigh-energy regime underscores the need for improved control over low-$x$ parton dynamics.

Figure~\ref{fig:ccbar_exp_xsecs_dyn} presents the identical data as Fig.~\ref{fig:ccbar_exp_xsecs}, but now using the dynamical scale choice, \cref{eq:mu_dyn}, for the theoretical predictions.
We observe a slightly reduced scale sensitivity at high collision energies while the central cross section remains practically the same for all the considered PDF sets.
The impact of the dynamical scale choice is larger in the region not covered by the LHAPDF interpolation tables where for negative CT18 cross sections are avoided in the extrapolation region.
Similarly for MSHT20, the slightly increased factorization and renormalization scales delay the unphysical extrapolation features which yields better behaved theory predictions.
We stress again, that the extrapolation region is strongly driven by the last available points in the LHAPDF interpolation tables and, thus, is unreliable for physical interpretations.

\begin{figure}[htpb!]
    \centering
    \includegraphics[width=0.8\linewidth]{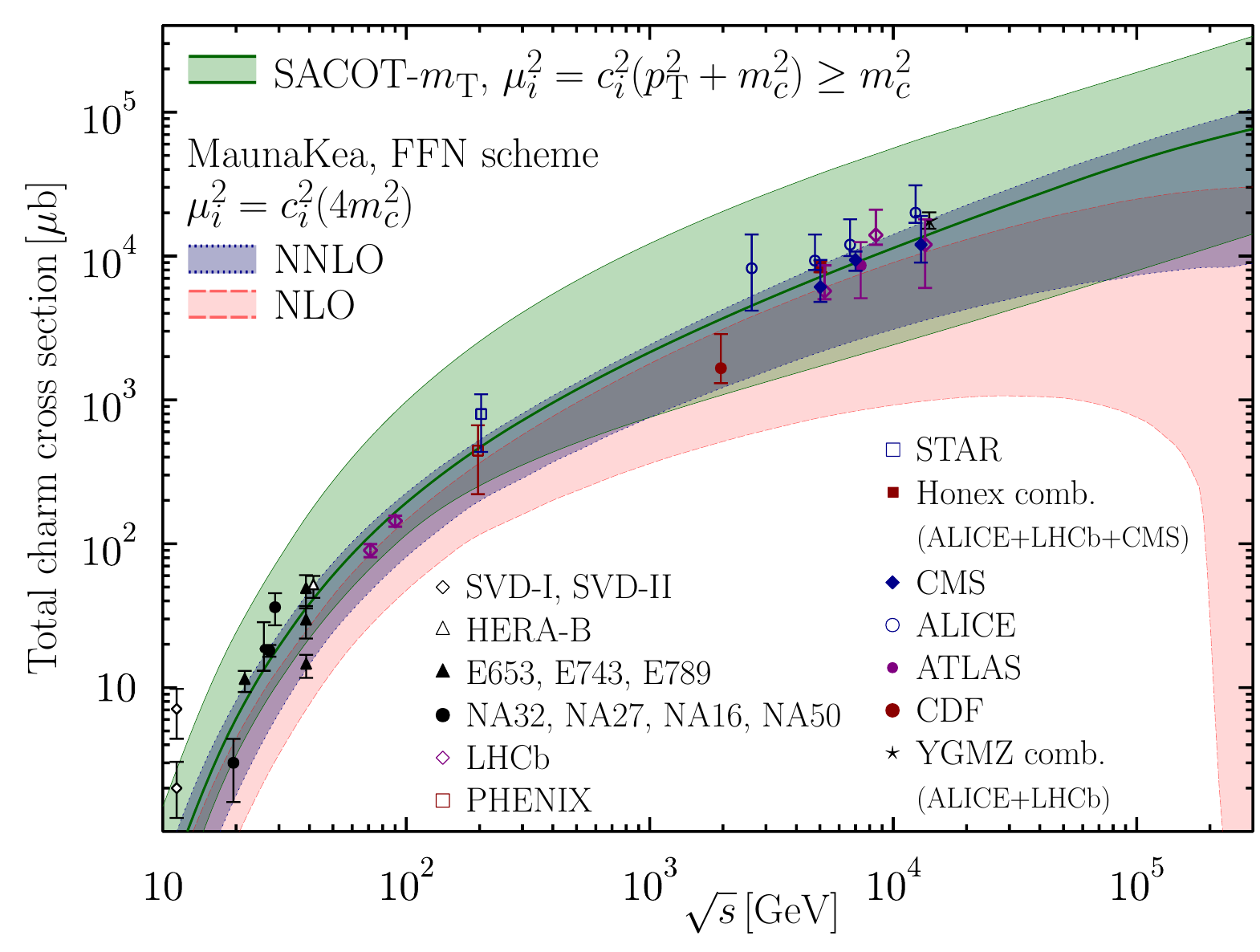}
    \caption{
    Compilation of representative experimental total $\ccbar$ cross sections as a function of $\sqrts$ (Honex and YGMZ indicate the data combinations of Refs.~\cite{Bierlich:2023ewv,Yang:2025pmq}, respectively) compared with GM-VFNS calculations within the SACOT-$\mT$ scheme, obtained using the corresponding NNPDF4.0 NLO PDF set. The corresponding \MaunaKea\ NLO- and NNLO-accurate FFNS calculations are shown for comparison. The coloured bands denote the corresponding theoretical scale uncertainties.
    \label{fig:master1}}
\end{figure}

Figure~\ref{fig:master1} presents a comparison between a somewhat larger selection of experimental charm cross section data than in Fig.~\ref{fig:ccbar_exp_xsecs} and the NLO SACOT-$\mT$ calculations obtained using the NNPDF4.0 PDFs. In addition, the FFNS \MaunaKea\ predictions at NLO and NNLO are included to better judge their differences. Here, we do not target a systematic charting of theoretical uncertainties, but only display the scale uncertainties, which were found to be the dominant ones for the FFNS calculations (Fig.~\ref{fig:pto}). The central SACOT-$\mT$ prediction describes the $\sqrts$ systematics surprisingly well, being generally a factor of two higher than the corresponding FFNS NNLO calculation with NNPDF4.0.
At large $\sqrts$, the SACOT-$\mT$ scale uncertainty clearly exceeds the FFNS one at NNLO, although it is actually comparable or even smaller than the FFNS NLO one, \ie\ the resummation of mass logarithms appears to have some relevance reducing higher-order uncertainties at very high $\sqrts$. At low $\sqrts$, the FFNS and SACOT-$\mT$ scale-uncertainty envelopes are comparable. However, we note that in the SACOT-$\mT$ case, scale variations below $\mc$ are not considered, which partially restricts the size of the downward uncertainty.
While the NNLO scale uncertainties appear smaller, potentially suggesting that the FFNS prediction are more reliable than the SACOT-$\mT$ calculation, we remind the reader that the perturbative FFNS series does not exhibit clear signals of convergence and that the scale variations at lower orders do not perfectly capture the size of the higher-order corrections (Fig.~\ref{fig:pto}). The SACOT-$\mT$ scale uncertainties are also expected to be somewhat overestimated, as comparisons with measurements~\cite{Helenius:2018uul} show that extreme scale variations fail to adequately describe the differential LHC D-meson cross section data. At $\sqrts \approx 100$~TeV, also the SACOT-$\mT$ calculation starts to integrate parton collisions below the NNPDF4.0 $\xgmin$ value, although the underlying parton distributions are generally effectively probed at larger $x$ values than in the FFNS calculations~\cite{Helenius:2018uul} and the predictions are, at the end, not as sensitive to the PDF-extrapolation region.

\subsection{$\bbbar$ cross sections}

\begin{figure}[!htbp]
    \centering
    \resizebox{0.77\textwidth}{!}{%
        \includegraphics[width=\textwidth]{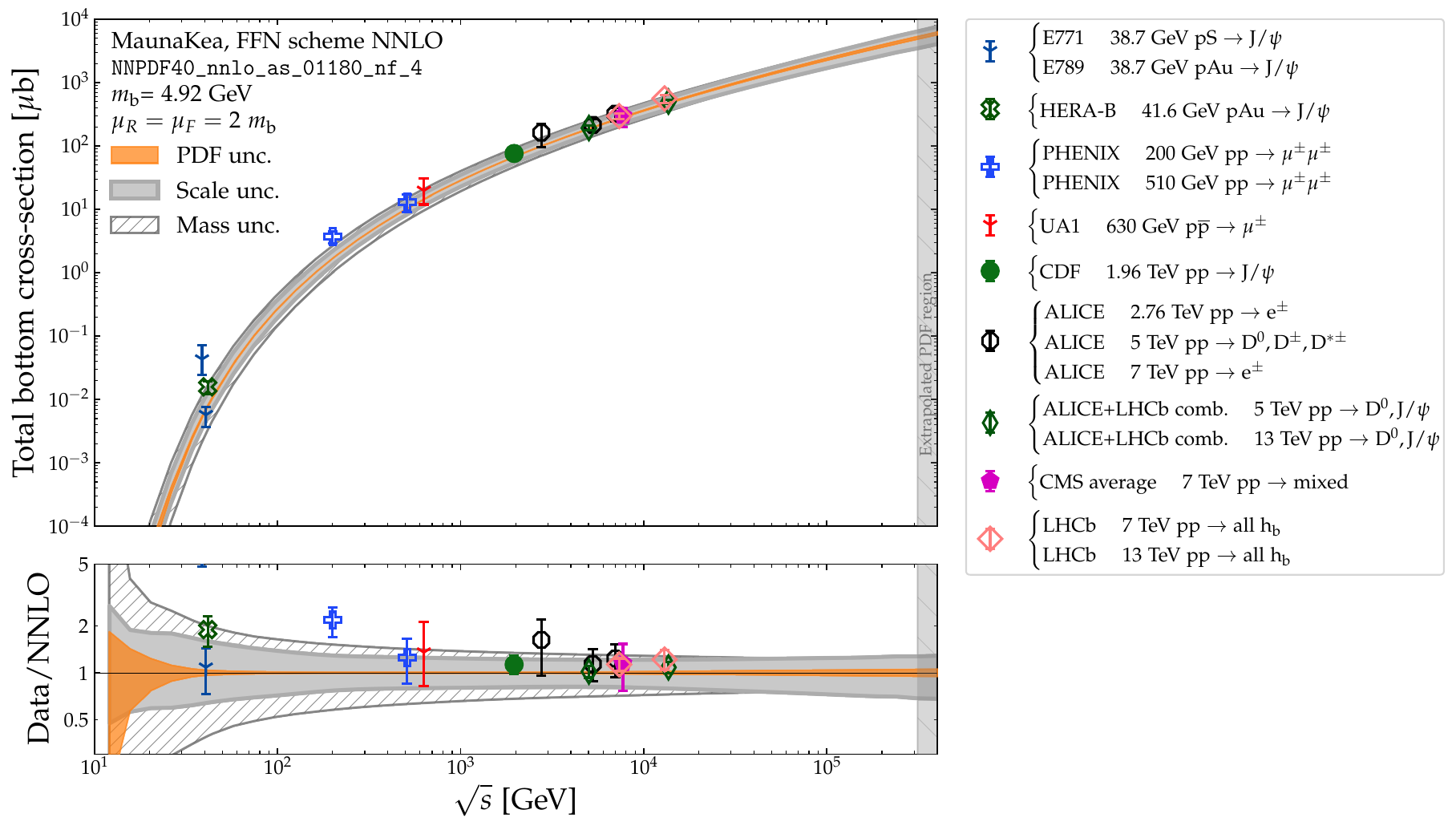}%
    }
    \resizebox{0.77\textwidth}{!}{%
        \includegraphics[width=\textwidth]{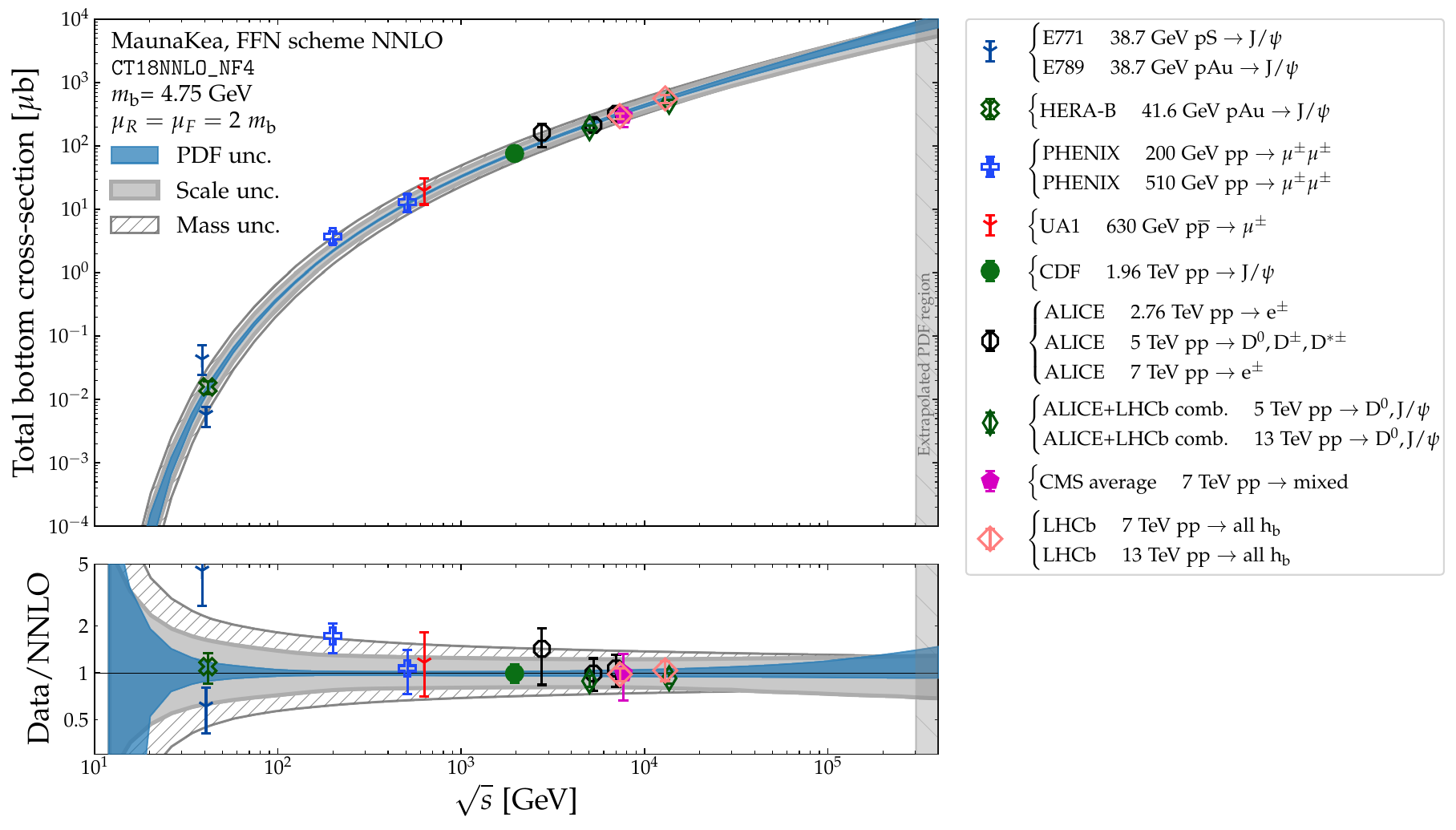}%
    }
    \resizebox{0.77\textwidth}{!}{%
        \includegraphics[width=\textwidth]{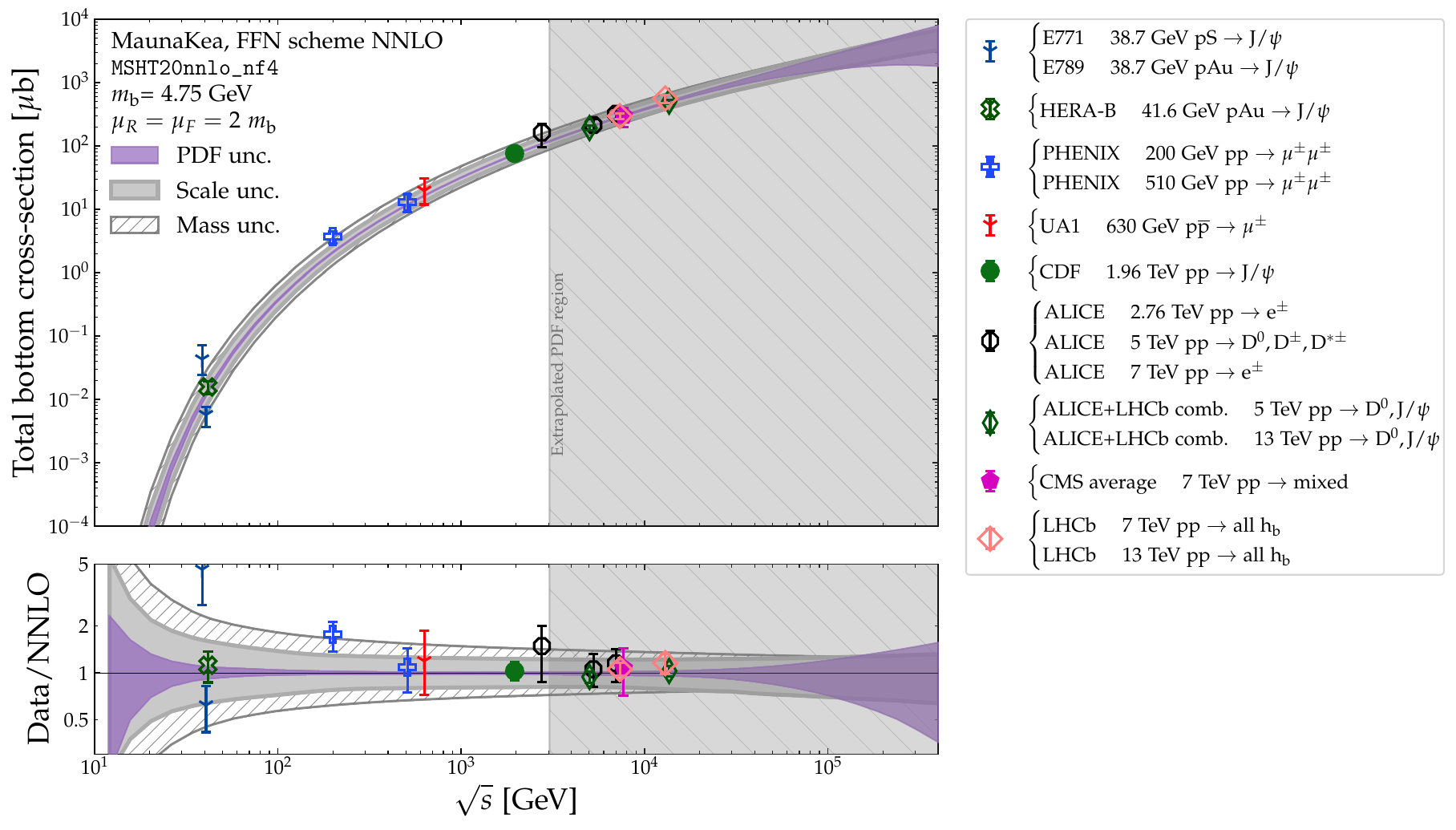}%
    }    
    \caption{Bottom cross sections in \pp{} collisions as a function of \cm\ energy, over $\sqrts=10$~GeV--400~TeV: The experimental data (symbols, selected from \cref{tab:beautymeasurements}) are compared with theoretical \MaunaKea\ NNLO (curves) for the NNPDF4.0 (upper), CT18 (middle), and  MSHT20 (lower) PDFs. The PDF uncertainty (in orange, blue, and purple for NNPDF4.0, CT18, and MSHT20, respectively) (see also \cref{fig:pdf}), the scale variation uncertainties (gray) (\cref{fig:pto}), and the bottom-quark mass uncertainty (striped white and dark gray) (\cref{fig:mass-ccbar}) are plotted as bands. The vertical shaded light-gray areas indicate the respective LHAPDF extrapolation regions (for which $\xmin\le\xgmin$, see text). The static scale $\muF=\muR = 2\mQ$ is used. The lower panels show the data over NNLO ratios.
    \label{fig:bbbar_exp_xsecs} }
\end{figure}

Figure~\ref{fig:bbbar_exp_xsecs} displays a selected set of experimental data for total bottom production cross sections (gray entries in Table~\ref{tab:beautymeasurements}) as a function of $\sqrts$ compared with the NNLO \MaunaKea\ calculations obtained using the NNPDF4.0, CT18, and MSHT20 proton PDFs. First, in general, all the data can be described within the theoretical uncertainties and do not leave room for very large remaining N$^3$LO QCD contributions. Compared to the charm case, the bottom-quark masses used in the three sets are relatively similar, and as a result all three PDF sets produce cross section predictions that are significantly more consistent with each other than observed for the former. The NNPDF4.0 set uses the largest $\mb$ value and, correspondingly, predicts the lowest $\sigmabbbar$ cross sections. At low $\sqrts$, the NNLO calculations become highly sensitive to the bottom-quark mass because  the mass variations modify the available phase space in the large-$x$ region where the PDFs are very steep functions of $x$. Unlike in the charm case, the effect of changing the bottom mass cannot be so easily compensated by the scale variations. As a result, improved measurements of bottom cross sections over $\sqrts\approx 10$--100~GeV would provide useful constraints on the bottom-quark pole mass. Due to the larger bottom compared to charm quark masses, the probed momentum fractions in the PDFs are clearly larger in $\bbbar$ production, see \cref{eq:xmin}, and the technical limitations of the LHAPDF grids are not as severe as for the $\sigmaccbar$ calculations. Nevertheless, the highest $\sqrts$ probed at the LHC cannot be fully consistently computed with the MSHT20 grids, since low-$x$ PDF extrapolations become necessary above $\sqrts\approx 10$~TeV for this PDF set. This observation further motivates an extension of the minimum $x$ coverage of the MSHT grids in future PDF releases.
The impact of choosing the dynamical scale, \cref{eq:mu_dyn}, is negligible in the case of bottom production due to the larger mass and thus already higher scales.
We list the corresponding values in \cref{sec:app2} for completeness.

\begin{figure}[htbp!]
    \centering
    \includegraphics[width=0.8\linewidth]{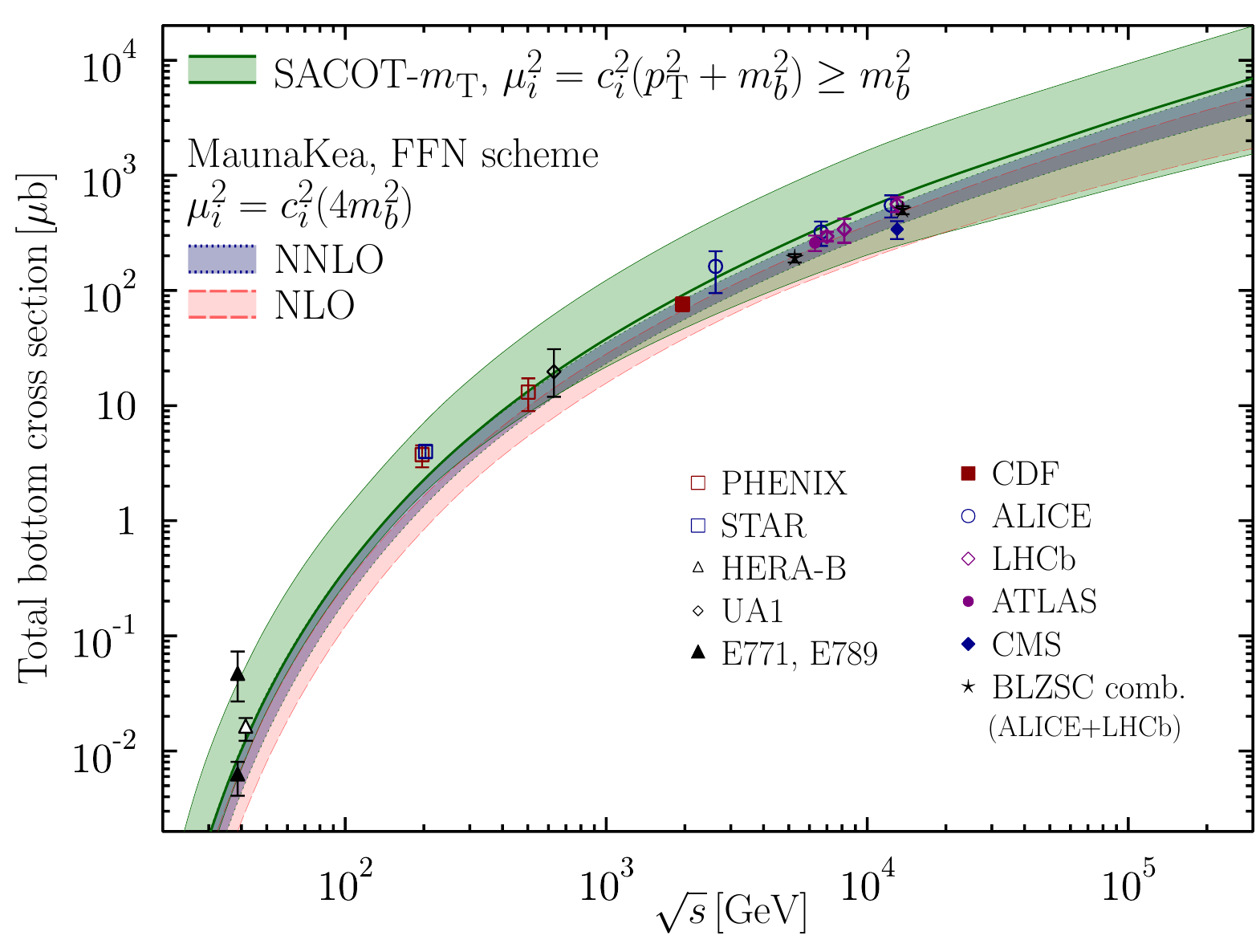}
    \caption{
     Compilation of representative experimental total $\bbbar$ cross sections as a function of $\sqrts$ (the BLZSC $\sigmabbbar$ combination corresponds to that reported in Ref.~\cite{Bai:2024pxk}) compared with GM-VFNS calculations within the SACOT-$\mT$ scheme, obtained using the corresponding NNPDF4.0 NLO PDF set. The corresponding \MaunaKea\ NNLO- and NLO-accuracy FFNS calculations are shown for comparison.  The coloured bands denote the corresponding theoretical scale uncertainties.
    \label{fig:master2}}
\end{figure}

A comparison of experimental $\sigmabbbar$ cross section measurements (including a few more data points than in Fig.~\ref{fig:bbbar_exp_xsecs}) as a function of $\sqrts$ with the SACOT-$\mT$ calculations, obtained with the NNPDF4.0 NLO set, is shown in Fig.~\ref{fig:master2}. Within the scale uncertainty bands, the calculation reproduces all data. However, the response of the theoretical cross sections to the scale variations is much larger than in the \MaunaKea\ FFNS calculations, even at NLO. Also here, it is known that the most extreme scale variation used in the SACOT-$\mT$ approach do not provide a good agreement with the differential B-meson data from the LHC, and therefore the scale-uncertainty band is most likely unrealistically wide.

\section{Summary}
\label{sec:summ}

A comprehensive collection of existing measurements of inclusive charm and bottom cross sections in hadronic collisions over three orders-of-magnitude in collision energy $\sqrts\approx 10$~GeV--13~TeV has been presented, and compared with calculations at next-to-next-to-leading order (NNLO) accuracy in perturbative QCD as implemented in the new \MaunaKea\ code, within the fixed flavour number (FFN) scheme. In addition, the data and NNLO calculations have been compared with the predictions obtained with the \sacotmt\ general-mass variable-flavour-number scheme (GM-VFNS) at next-to-leading-order (NLO) accuracy. The theoretical \MaunaKea\ cross sections are based on the matrix elements of the \toppp\ code, developed for the calculation of top-quark pair cross sections, modified to be applicable for fixed-order calculations of charm- and bottom-quark pair production and interfaced with the \pineappl\ fast interpolation grids. Theoretical predictions have been obtained for three different NNLO-accurate global-fit parton distribution functions (PDFs): NNPDF4.0, CT18, and  MSHT20. The uncertainties associated with missing higher-order corrections, PDFs, heavy-quark masses, and the strong coupling have been quantified, and detailed data-versus-NNLO comparisons have been presented for each PDF set. In addition, heavy-quark cross sections predictions are provided up to the highest \cm\ energies accessible at the CERN Future Circular Collider (FCC) and in ultrahigh-energy cosmic-ray interactions in the upper atmosphere, $\sqrts\approx 400$~TeV.

On the experimental side, a detailed discussion has been first provided about the determination of total charm and bottom cross sections from measurements of c- and b-hadrons (and/or their decay products), supplemented by extrapolations to heavy-quark-level cross sections based on parton-to-hadron fragmentation fractions. 
The determination of all relevant heavy-quark fragmentation fractions, and their dependence on the collision system where they have been measured ($\epem$, DIS, LHC), have been thoroughly discussed. A collection of approximately 50 measurements for charm hadrons and 50 measurements for bottom hadrons has been compiled, specifying the fiducial phase space and the extrapolation factors used to derive the corresponding experimental $\sigmaccbar$ and $\sigmabbbar$ cross sections over $\sqrts\approx 10$~GeV--13~TeV. About 10 (20) of the existing fiducial measurements of charm (bottom) hadrons production cross sections have been converted into inclusive $\sigmaccbar$ or $\sigmabbbar$ values, enabling an extended comparison of results. 
The impact of using different parton-to-hadron fragmentation fractions has been considered, in particular in the context of the enhanced heavy-quark baryon-to-meson ratios observed at the LHC.\\

On the theory side, the computed $K$-factors, defined as $K=\sigma($N$^{n+1}$LO)/$\sigma($N$^n$LO) (with $n=0,1$) are found to decrease when going from NLO/LO to NNLO/NLO for both charm and bottom cross sections, indicating that the perturbative expansion is converging. However, we also observe that the NNLO cross sections for both heavy quarks are enhanced by up to a factor of two relative to the NLO predictions, \ie\ the convergence of the perturbative series is slow. The theoretical uncertainties associated with the missing higher-order corrections have been estimated through scale variations, and found to be generally reduced when the perturbative accuracy of the calculation is increased. They still remain sizable, of the order of $\pm60\%$ ($\pm20\%$) for charm (bottom) production at NNLO, in the most favorable cases. We have also studied the sensitivity of the cross sections to the heavy-quark masses, by varying the values of $\mc$ and $\mb$ by $\pm 10\%$ in the matrix elements from their nominal values defined in the PDF fits. 
For charm, we observe that $\mc$ variations lead to cross-section uncertainties  of about $\pm100\%$ ($\pm20\%$) at low (LHC) energies. For bottom-quark production, $\mb$ variations remain the dominant source of uncertainty over the entire collision-energy range, leading to changes in the cross sections of about $\pm 100\%$ ($\pm 35\%$) at fixed-target (LHC) energies. Uncertainties associated with variations of $\alphasmZ$ within its current experimental precision are found to be subleading throughout the entire energy range, remaining (well) below 10\% for both charm and bottom production. 

As an alternative to the FFNS calculations, we also considered a specific GM-VFNS approach, \sacotmt, which resums terms proportional to $\log (\mT^2/\mQ^2$) that are significant in the large-$\pT$ limit. These contributions should become more pronounced at higher collision energies due to increasingly flatter $\pT$-spectra, and therefore could potentially improve the convergence of the perturbative expansion. Although the missing higher-order corrections of this approach are still too large to draw a strong conclusion at NLO, evidence for reduced scale uncertainties at $\sqrts \approx 100$~TeV for charm is visible with respect to the fixed-order \MaunaKea\ NLO result.\\

Detailed studies of the PDF dependence and propagated uncertainties of the heavy-quark cross sections have been presented. Except at the lowest collision energies in the bottom sector, heavy-quark pair production is dominated by gluon-gluon scattering, so the measured cross sections provide a particularly sensitive probe of the gluon PDF. In the current global-PDF fits, the gluon density is fairly well constrained down to $x \approx 10^{-5}$ where HERA data exist, but it is practically unconstrained at smaller parton fractional momenta. This small-$x$ region begins to contribute to heavy-quark cross section around $\sqrts \approx 1$~TeV in hadronic collisions and, accordingly, PDF uncertainties are found to increase in the multi-TeV regime. Technical issues appear for some of the employed PDFs when evaluated outside their nominal range of applicability in $x$. The \lhapdf\ grids of MSHT20 stop at $x=10^{-6}$, beyond which the densities rely on uncontrolled \lhapdf\ extrapolations leading to unphysical behaviour of the central PDFs and inflation of their uncertainty bands. The NNPDF4.0 and CT18 PDFs interpolation grids extend down to $x = 10^{-9}$, which provides some stability for the obtained cross sections beyond LHC energies, but one should keep in mind that there are no actual data constraints available, and the PDF uncertainties might be restricted due to parametrization biases.

Accounting for all, admittedly sizable, uncertainties, we find good agreement between the NNLO calculations and experimental data over the full range of collision energies explored, though the level of accord depends on the PDF set used. The theory-data agreement is driven mainly by the choice of heavy-quark masses over $\sqrts \approx 10$~GeV--1~TeV, and by intrinsic low-$x$ PDF differences in the multi-TeV regime. For $\sqrts>1$~TeV, the charm cross sections become sensitive to small-$x$ gluon distributions at low factorization scales where experimental constraints are absent, and in some cases depend on (arbitrary) LHAPDF extrapolations. In this context, three practical conclusions emerge. First, the existing $\sigmaccbar$ measurements at the LHC can serve as input to future global PDF fits, providing constraints on the gluon density at very small momentum fractions $x\approx 10^{-6}$, and reducing reliance on assumptions inherent to small-$x$ extrapolations. Second, extending the $x$-grid coverage down to $10^{-10}$ in future global-PDF releases would be advantageous, in particular in view of predictions for future hadron colliders. Third, in the case of $\bbbar$ production at low $\sqrts$, the NNLO calculations become highly sensitive to the bottom-quark mass, because it determines the available phase space, and thus precise measurements of $\sigmabbbar$ cross sections above threshold, over $\sqrts\approx 10$--100~GeV, would provide useful constraints on the bottom-quark pole mass.\\

Ultimately, high-precision charm and bottom cross sections will require improved experimental knowledge of heavy-flavour hadron production fractions over the largest possible phase space at the LHC, together with theoretical progress beyond NNLO fixed-order accuracy. Both developments are essential for a full understanding of heavy-quark production at future colliders such as the FCC-hh, as well as at the highest-energy hadronic collisions achieved in cosmic-ray interactions with the upper atmosphere. Further studies are therefore needed, and the present study provides a state-of-the-art baseline for such upcoming investigations.\\

\paragraph*{Acknowledgments --}

Useful discussions with A.~Verbytskyi on the experimental heavy-quark fragmentation fractions are acknowledged. A.~Geiser's contribution to collection of past experimental heavy-quark data is acknowledged. 
F.~H., I.~H., and H.~P. have been supported by the Academy of Finland projects 331545, 358090 and 361179, and were funded as a part of the Center of Excellence in Quark Matter of the Academy of Finland, project 346326.	

\clearpage

\appendix
\section{NNLO charm \pp\ cross section tables}
\label{sec:app}

\begin{table}[!htp]
\tabcolsep=2.0mm
   \centering
   \caption{Theoretical NNLO predictions for the inclusive $\ccbar$ cross sections in \pp\ collisions over $\sqrts\approx 10$~GeV--400~TeV obtained with \MaunaKea\ using the \texttt{NNPDF40\_nnlo\_pch\_as\_01180\_nf\_3} PDFs. For each system, we list the central cross section, $\sigmaccbar$, followed by the percent (positive and negative) uncertainties from missing higher-order corrections ($\Delta^{\pm}_\text{scale}$), parton densities ($\Delta^{\pm}_\text{PDF}$), charm quark mass ($\Delta^{\pm}_{\mc}$), and strong coupling constant ($\Delta^{\pm}_{\alphaS}$). The dotted line marks the lower $x$ limit of validity of the PDF set by $\xgmin$ (cross sections for c.m.\ energies above this threshold rely on extrapolations provided by LHAPDF). The static scale $\muF=\muR = 2\mQ$ is used as central value.
   \label{tab:table1a}}
\begin{tabular}{rcccccccccc}
\toprule
 & $\sigma(\ccbar)$~[\mub] & $\Delta^+_\mathrm{scale}~(\%)$ & $\Delta^-_\mathrm{scale}~(\%)$ & $\Delta^+_\mathrm{PDF}~(\%)$ & $\Delta^-_\mathrm{PDF}~(\%)$ & $\Delta^+_{\mc}~(\%)$ & $\Delta^-_{\mc}~(\%)$ & $\Delta^+_\mathrm{\alpha_s}~(\%)$ & $\Delta^-_\mathrm{\alpha_s}$~(\%) \\
$\sqrt{s}$~[GeV] &  &  &  &  &  &  &  &  &  \\
\midrule
11.6 & 0.36 & $+200$ & $-55$ & $+5$ & $-5$ & $+135$ & $-50$ & $+10.5$ & $-11$ \\
19.5 & 3.2 & $+135$ & $-50$ & $+3$ & $-3$ & $+95$ & $-45$ & $+7.5$ & $-9$ \\
21.7 & 4.5 & $+125$ & $-45$ & $+2$ & $-2$ & $+90$ & $-40$ & $+7.5$ & $-8$ \\
26 & 7.9 & $+110$ & $-45$ & $+2$ & $-2$ & $+85$ & $-40$ & $+7.5$ & $-7.5$ \\
27.4 & 9.2 & $+105$ & $-45$ & $+2$ & $-2$ & $+80$ & $-40$ & $+7$ & $-7.5$ \\
28.9 & 10.5 & $+105$ & $-45$ & $+2$ & $-2$ & $+80$ & $-40$ & $+7$ & $-7.5$ \\
38.7 & 22.0 & $+95$ & $-40$ & $+2$ & $-2$ & $+70$ & $-35$ & $+6.5$ & $-6.5$ \\
38.9 & 22.5 & $+95$ & $-40$ & $+2$ & $-2$ & $+70$ & $-35$ & $+6.5$ & $-6.5$ \\
41.6 & 26.0 & $+95$ & $-40$ & $+2$ & $-1$ & $+70$ & $-35$ & $+6.5$ & $-6.5$ \\
68.5 & 69.0 & $+85$ & $-35$ & $+2$ & $-2$ & $+60$ & $-30$ & $+5.5$ & $-5.5$ \\
86.6 & 100 & $+80$ & $-35$ & $+2$ & $-2$ & $+55$ & $-30$ & $+4.5$ & $-5$ \\
200 & 300 & $+75$ & $-35$ & $+2$ & $-2$ & $+45$ & $-25$ & $+3$ & $-3$ \\
1960 & 2450 & $+70$ & $-50$ & $+4$ & $-4$ & $+30$ & $-20$ & $+1$ & $-0.5$ \\
2760 & 3170 & $+75$ & $-55$ & $+5$ & $-5$ & $+30$ & $-20$ & $+0.5$ & $-1$ \\
5000 & 4820 & $+80$ & $-55$ & $+5$ & $-5$ & $+25$ & $-20$ & $+0$ & $-1$ \\
5020 & 4830 & $+80$ & $-55$ & $+5$ & $-5$ & $+25$ & $-20$ & $+0$ & $-1$ \\
7000 & 6030 & $+80$ & $-55$ & $+6$ & $-6$ & $+25$ & $-15$ & $+0$ & $-1$ \\
13000 & 8870 & $+90$ & $-60$ & $+6$ & $-6$ & $+25$ & $-15$ & $+0$ & $-0.5$ \\
13600 & 9120 & $+90$ & $-60$ & $+6$ & $-6$ & $+25$ & $-15$ & $+0$ & $-0.5$ \\
14000 & 9280 & $+90$ & $-60$ & $+6$ & $-6$ & $+25$ & $-15$ & $+0$ & $-0.5$ \\
84000 & 23600 & $+130$ & $-70$ & $+7$ & $-7$ & $+15$ & $-10$ & $+0$ & $-0$ \\ [-2ex ]
\multicolumn{11}{c}{\hdashrule{15cm}{1pt}{1pt}}\\
100000 & 25470 & $+135$ & $-70$ & $+7$ & $-7$ & $+15$ & $-10$ & $+0$ & $-0$ \\
400000 & 41500 & $+190$ & $-80$ & $+9$ & $-8$ & $+10$ & $-10$ & $+0$ & $-0$ \\
\bottomrule
\end{tabular}

\end{table}

\begin{table}[!htp]
\tabcolsep=2.0mm
   \centering
   \caption{Same as Table~\ref{tab:table1a}, but using the $\muF=\muR = \mu_\text{dyn.}$ dynamic scale, \cref{eq:mu_dyn}, as central value. 
   \label{tab:table1b}}
   \begin{tabular}{rcccccccccc}
\toprule
 & $\sigma(\ccbar)$~[\mub] & $\Delta^+_\mathrm{scale}~(\%)$ & $\Delta^-_\mathrm{scale}~(\%)$ & $\Delta^+_\mathrm{PDF}~(\%)$ & $\Delta^-_\mathrm{PDF}~(\%)$ & $\Delta^+_{\mc}~(\%)$ & $\Delta^-_{\mc}~(\%)$ & $\Delta^+_\mathrm{\alpha_s}~(\%)$ & $\Delta^-_\mathrm{\alpha_s}$~(\%) \\
$\sqrt{s}$~[GeV] &  &  &  &  &  &  &  &  &  \\
\midrule
11.6 & 0.36 & $+195$ & $-55$ & $+5$ & $-5$ & $+135$ & $-50$ & $+10$ & $-11$ \\
19.5 & 3.1 & $+135$ & $-50$ & $+3$ & $-3$ & $+95$ & $-45$ & $+8$ & $-8.5$ \\
21.7 & 4.4 & $+125$ & $-45$ & $+2$ & $-2$ & $+90$ & $-40$ & $+7.5$ & $-8$ \\
26 & 7.7 & $+110$ & $-45$ & $+2$ & $-2$ & $+85$ & $-40$ & $+7$ & $-7.5$ \\
27.4 & 9.0 & $+105$ & $-45$ & $+2$ & $-2$ & $+85$ & $-40$ & $+7$ & $-7.5$ \\
28.9 & 10.4 & $+105$ & $-45$ & $+2$ & $-2$ & $+80$ & $-40$ & $+7$ & $-7.5$ \\
38.7 & 21.6 & $+95$ & $-40$ & $+2$ & $-2$ & $+70$ & $-35$ & $+6.5$ & $-6.5$ \\
38.9 & 21.8 & $+95$ & $-40$ & $+2$ & $-2$ & $+70$ & $-35$ & $+6.5$ & $-6.5$ \\
41.6 & 25.4 & $+95$ & $-40$ & $+2$ & $-2$ & $+70$ & $-35$ & $+6$ & $-6.5$ \\
68.5 & 67.2 & $+85$ & $-35$ & $+2$ & $-2$ & $+60$ & $-30$ & $+5$ & $-5.5$ \\
86.6 & 98.5 & $+80$ & $-35$ & $+2$ & $-2$ & $+55$ & $-30$ & $+4.5$ & $-5$ \\
200 & 300 & $+70$ & $-35$ & $+2$ & $-2$ & $+45$ & $-25$ & $+3$ & $-3$ \\
1960 & 2450 & $+60$ & $-45$ & $+4$ & $-4$ & $+30$ & $-20$ & $+0.5$ & $-1$ \\
2760 & 3170 & $+60$ & $-50$ & $+5$ & $-5$ & $+30$ & $-20$ & $+0.5$ & $-1$ \\
5000 & 4820 & $+65$ & $-50$ & $+5$ & $-5$ & $+30$ & $-20$ & $+0$ & $-1$ \\
5020 & 4830 & $+65$ & $-50$ & $+5$ & $-5$ & $+30$ & $-20$ & $+0$ & $-1$ \\
7000 & 6020 & $+65$ & $-50$ & $+6$ & $-6$ & $+25$ & $-20$ & $+0$ & $-1$ \\
13000 & 8870 & $+65$ & $-55$ & $+6$ & $-6$ & $+25$ & $-15$ & $+0$ & $-0.5$ \\
13600 & 9110 & $+65$ & $-55$ & $+6$ & $-6$ & $+25$ & $-15$ & $+0$ & $-0.5$ \\
14000 & 9270 & $+65$ & $-55$ & $+6$ & $-6$ & $+25$ & $-15$ & $+0$ & $-0.5$ \\
84000 & 23580 & $+80$ & $-60$ & $+7$ & $-7$ & $+15$ & $-10$ & $+0$ & $-1$ \\[-2ex ]
\multicolumn{11}{c}{\hdashrule{15cm}{1pt}{1pt}}\\
100000 & 25500 & $+85$ & $-60$ & $+7$ & $-7$ & $+15$ & $-10$ & $+0$ & $-1.5$ \\
400000 & 41600 & $+110$ & $-65$ & $+8$ & $-8$ & $+5$ & $-10$ & $+0.5$ & $-2.5$ \\
\bottomrule
\end{tabular}

\end{table}


\begin{table}[!htp]
\tabcolsep=2.0mm
   \centering
   \caption{Theoretical NNLO predictions for the inclusive $\ccbar$ cross sections in \pp\ collisions over $\sqrts\approx 10$~GeV--400~TeV obtained with \MaunaKea\ using the \texttt{CT18NNLO\_NF3} PDFs. For each system, we list the central cross section, $\sigmaccbar$, followed by the percent (positive and negative) uncertainties from missing higher-order corrections ($\Delta^{\pm}_\text{scale}$), parton densities ($\Delta^{\pm}_\text{PDF}$), charm quark mass ($\Delta^{\pm}_{\mc}$), and strong coupling constant ($\Delta^{\pm}_{\alphaS}$). The dotted line marks the lower $x$ limit of validity of the PDF set by $\xgmin$ (cross sections for c.m.\ energies above this threshold rely on extrapolations provided by LHAPDF). The static scale $\muF=\muR = 2\mQ$ is used as central value.
   \label{tab:table1c}}
\begin{tabular}{rcccccccccc}
\toprule
 & $\sigma(\ccbar)$~[\mub] & $\Delta^+_\mathrm{scale}~(\%)$ & $\Delta^-_\mathrm{scale}~(\%)$ & $\Delta^+_\mathrm{PDF}~(\%)$ & $\Delta^-_\mathrm{PDF}~(\%)$ & $\Delta^+_{\mc}~(\%)$ & $\Delta^-_{\mc}~(\%)$ & $\Delta^+_\mathrm{\alpha_s}~(\%)$ & $\Delta^-_\mathrm{\alpha_s}$~(\%) \\
$\sqrt{s}$~[GeV] &  &  &  &  &  &  &  &  &  \\
\midrule
11.6 & 1.73 & $+210$ & $-55$ & $+17$ & $-11$ & $+125$ & $-55$ & $+13.5$ & $-11.5$ \\
19.5 & 10.5 & $+140$ & $-50$ & $+7$ & $-6$ & $+95$ & $-45$ & $+9.5$ & $-9$ \\
21.7 & 14.1 & $+125$ & $-45$ & $+6$ & $-5$ & $+90$ & $-45$ & $+9$ & $-8$ \\
26 & 22.5 & $+120$ & $-45$ & $+5$ & $-5$ & $+85$ & $-45$ & $+8$ & $-7$ \\
27.4 & 25.5 & $+120$ & $-45$ & $+5$ & $-5$ & $+85$ & $-45$ & $+8$ & $-7$ \\
28.9 & 29.0 & $+120$ & $-45$ & $+5$ & $-5$ & $+80$ & $-40$ & $+7.5$ & $-7$ \\
38.7 & 53.5 & $+110$ & $-40$ & $+4$ & $-5$ & $+70$ & $-40$ & $+6$ & $-6$ \\
38.9 & 54.0 & $+110$ & $-40$ & $+4$ & $-5$ & $+70$ & $-40$ & $+6$ & $-6$ \\
41.6 & 61.5 & $+110$ & $-40$ & $+4$ & $-5$ & $+70$ & $-40$ & $+5.5$ & $-6$ \\
68.5 & 140 & $+100$ & $-40$ & $+5$ & $-5$ & $+60$ & $-35$ & $+4.5$ & $-4$ \\
86.6 & 195 & $+95$ & $-40$ & $+4$ & $-5$ & $+55$ & $-35$ & $+4$ & $-3$ \\
200 & 510 & $+85$ & $-45$ & $+5$ & $-6$ & $+45$ & $-30$ & $+2$ & $-2.5$ \\
1960 & 3400 & $+90$ & $-60$ & $+14$ & $-7$ & $+30$ & $-25$ & $+2$ & $-1.5$ \\
2760 & 4330 & $+95$ & $-60$ & $+17$ & $-7$ & $+30$ & $-20$ & $+2$ & $-1.5$ \\
5000 & 6470 & $+105$ & $-65$ & $+25$ & $-8$ & $+30$ & $-20$ & $+1.5$ & $-2$ \\
5020 & 6480 & $+105$ & $-65$ & $+25$ & $-8$ & $+30$ & $-20$ & $+1.5$ & $-2$ \\
7000 & 8030 & $+110$ & $-65$ & $+30$ & $-8$ & $+30$ & $-20$ & $+2$ & $-1.5$ \\
13000 & 11850 & $+125$ & $-70$ & $+50$ & $-8$ & $+25$ & $-20$ & $+2$ & $-2$ \\
13600 & 12180 & $+125$ & $-70$ & $+55$ & $-8$ & $+25$ & $-20$ & $+2$ & $-1.5$ \\
14000 & 12400 & $+125$ & $-70$ & $+55$ & $-8$ & $+25$ & $-20$ & $+2$ & $-1.5$ \\[-2ex ]
\multicolumn{11}{c}{\hdashrule{15cm}{1pt}{1pt}}\\
84000 & 33780 & $+165$ & $-80$ & $+255$ & $-9$ & $+20$ & $-15$ & $+1.5$ & $-1.5$ \\
100000 & 36900 & $+170$ & $-80$ & $+420$ & $-9$ & $+20$ & $-15$ & $+1.5$ & $-1.5$ \\
400000 & 71850 & $+200$ & $-85$ & $+3835$ & $-100$ & $+20$ & $-15$ & $+1$ & $-1$ \\
\bottomrule
\end{tabular}

\end{table}

\begin{table}[!htp]
\tabcolsep=2.0mm
   \centering
   \caption{Same as Table~\ref{tab:table1c}, but using the $\muF=\muR = \mu_\text{dyn.}$ dynamic scale, \cref{eq:mu_dyn}, as central value.  
   \label{tab:table1d}}
   \begin{tabular}{rcccccccccc}
\toprule
 & $\sigma(\ccbar)$~[\mub] & $\Delta^+_\mathrm{scale}~(\%)$ & $\Delta^-_\mathrm{scale}~(\%)$ & $\Delta^+_\mathrm{PDF}~(\%)$ & $\Delta^-_\mathrm{PDF}~(\%)$ & $\Delta^+_{\mc}~(\%)$ & $\Delta^-_{\mc}~(\%)$ & $\Delta^+_\mathrm{\alpha_s}~(\%)$ & $\Delta^-_\mathrm{\alpha_s}$~(\%) \\
$\sqrt{s}$~[GeV] &  &  &  &  &  &  &  &  &  \\
\midrule
11.6 & 1.70 & $+210$ & $-55$ & $+17$ & $-11$ & $+125$ & $-55$ & $+13.5$ & $-11.5$ \\
19.5 & 10.2 & $+135$ & $-50$ & $+8$ & $-6$ & $+95$ & $-45$ & $+9.5$ & $-8.5$ \\
21.7 & 13.7 & $+125$ & $-45$ & $+6$ & $-5$ & $+90$ & $-45$ & $+9$ & $-8$ \\
26 & 21.8 & $+115$ & $-45$ & $+5$ & $-5$ & $+85$ & $-45$ & $+8$ & $-7.5$ \\
27.4 & 24.7 & $+115$ & $-45$ & $+5$ & $-5$ & $+85$ & $-45$ & $+7.5$ & $-7$ \\
28.9 & 28.0 & $+115$ & $-45$ & $+5$ & $-5$ & $+80$ & $-40$ & $+7.5$ & $-7$ \\
38.7 & 51.9 & $+105$ & $-40$ & $+4$ & $-5$ & $+70$ & $-40$ & $+6$ & $-6$ \\
38.9 & 52.4 & $+105$ & $-40$ & $+4$ & $-5$ & $+70$ & $-40$ & $+6$ & $-6$ \\
41.6 & 59.6 & $+105$ & $-40$ & $+4$ & $-5$ & $+70$ & $-40$ & $+6$ & $-5.5$ \\
68.5 & 137 & $+95$ & $-35$ & $+4$ & $-5$ & $+60$ & $-35$ & $+4$ & $-4$ \\
86.6 & 190 & $+90$ & $-35$ & $+4$ & $-5$ & $+55$ & $-35$ & $+3.5$ & $-3.5$ \\
200 & 500 & $+80$ & $-40$ & $+4$ & $-6$ & $+45$ & $-30$ & $+2$ & $-2$ \\
1960 & 3400 & $+70$ & $-55$ & $+12$ & $-7$ & $+35$ & $-25$ & $+1.5$ & $-1.5$ \\
2760 & 4330 & $+70$ & $-55$ & $+15$ & $-7$ & $+30$ & $-20$ & $+1.5$ & $-1.5$ \\
5000 & 6470 & $+75$ & $-55$ & $+25$ & $-8$ & $+30$ & $-20$ & $+2$ & $-1.5$ \\
5020 & 6480 & $+75$ & $-55$ & $+25$ & $-8$ & $+30$ & $-20$ & $+2$ & $-1.5$ \\
7000 & 8030 & $+75$ & $-55$ & $+30$ & $-8$ & $+30$ & $-20$ & $+2$ & $-2$ \\
13000 & 11880 & $+80$ & $-60$ & $+45$ & $-8$ & $+25$ & $-20$ & $+2$ & $-1.5$ \\
13600 & 12190 & $+80$ & $-60$ & $+45$ & $-8$ & $+25$ & $-20$ & $+2$ & $-1.5$ \\
14000 & 12400 & $+80$ & $-60$ & $+45$ & $-8$ & $+25$ & $-20$ & $+2$ & $-2$ \\[-2ex ]
\multicolumn{11}{c}{\hdashrule{15cm}{1pt}{1pt}}\\
84000 & 34550 & $+90$ & $-65$ & $+185$ & $-9$ & $+20$ & $-15$ & $+1.5$ & $-1.5$ \\
100000 & 37650 & $+95$ & $-65$ & $+215$ & $-9$ & $+20$ & $-15$ & $+1.5$ & $-1.5$ \\
400000 & 76650 & $+100$ & $-65$ & $+705$ & $-10$ & $+20$ & $-15$ & $+1.5$ & $-1.5$ \\
\bottomrule
\end{tabular}

\end{table}


\begin{table}[!htp]
\tabcolsep=2.0mm
   \centering
   \caption{Theoretical NNLO predictions for the inclusive $\ccbar$ cross sections in \pp\ collisions over $\sqrts\approx 10$~GeV--400~TeV obtained with \MaunaKea\ using the \texttt{MSHT20nnlo\_nf3} PDFs. For each system, we list the central cross section, $\sigmaccbar$, followed by the percent (positive and negative) uncertainties from missing higher-order corrections ($\Delta^{\pm}_\text{scale}$), parton densities ($\Delta^{\pm}_\text{PDF}$), charm quark mass ($\Delta^{\pm}_{\mc}$), and strong coupling constant ($\Delta^{\pm}_{\alphaS}$). The dotted line marks the lower $x$ limit of validity of the PDF set by $\xgmin$ (cross sections for c.m.\ energies above this threshold rely on extrapolations provided by LHAPDF). The static scale $\muF=\muR = 2\mQ$ is used as central value.
   \label{tab:table1e}}
\begin{tabular}{rcccccccccc}
\toprule
 & $\sigma(\ccbar)$~[\mub] & $\Delta^+_\mathrm{scale}~(\%)$ & $\Delta^-_\mathrm{scale}~(\%)$ & $\Delta^+_\mathrm{PDF}~(\%)$ & $\Delta^-_\mathrm{PDF}~(\%)$ & $\Delta^+_{\mc}~(\%)$ & $\Delta^-_{\mc}~(\%)$ & $\Delta^+_\mathrm{\alpha_s}~(\%)$ & $\Delta^-_\mathrm{\alpha_s}$~(\%) \\
$\sqrt{s}$~[GeV] &  &  &  &  &  &  &  &  &  \\
\midrule
11.6 & 0.90 & $+205$ & $-55$ & $+12$ & $-9$ & $+120$ & $-55$ & $+12$ & $-10.5$ \\
19.5 & 6.20 & $+135$ & $-50$ & $+8$ & $-4$ & $+90$ & $-45$ & $+8$ & $-8$ \\
21.7 & 8.50 & $+125$ & $-45$ & $+7$ & $-4$ & $+85$ & $-45$ & $+7.5$ & $-7.5$ \\
26 & 14.0 & $+110$ & $-45$ & $+6$ & $-4$ & $+80$ & $-40$ & $+7$ & $-6.5$ \\
27.4 & 16.0 & $+110$ & $-45$ & $+6$ & $-4$ & $+80$ & $-40$ & $+6.5$ & $-6.5$ \\
28.9 & 18.5 & $+110$ & $-45$ & $+5$ & $-3$ & $+75$ & $-40$ & $+6.5$ & $-6.5$ \\
38.7 & 35.5 & $+100$ & $-40$ & $+4$ & $-3$ & $+70$ & $-40$ & $+5.5$ & $-5.5$ \\
38.9 & 35.7 & $+100$ & $-40$ & $+4$ & $-3$ & $+70$ & $-40$ & $+5.5$ & $-5.5$ \\
41.6 & 41.0 & $+100$ & $-40$ & $+4$ & $-3$ & $+65$ & $-40$ & $+5$ & $-5.5$ \\
68.5 & 99.5 & $+90$ & $-35$ & $+3$ & $-3$ & $+60$ & $-35$ & $+4$ & $-3.5$ \\
86.6 & 140 & $+90$ & $-35$ & $+3$ & $-3$ & $+55$ & $-35$ & $+3.5$ & $-3$ \\
200 & 390 & $+80$ & $-40$ & $+3$ & $-3$ & $+45$ & $-30$ & $+1.5$ & $-2$ \\
1960 & 2390 & $+80$ & $-65$ & $+7$ & $-8$ & $+25$ & $-20$ & $+1$ & $-1$ \\[-2ex ]
\multicolumn{11}{c}{\hdashrule{15cm}{1pt}{1pt}}\\
2760 & 2925 & $+85$ & $-65$ & $+9$ & $-12$ & $+25$ & $-20$ & $+1$ & $-1$ \\
5000 & 3990 & $+95$ & $-70$ & $+20$ & $-28$ & $+20$ & $-15$ & $+1$ & $-1$ \\
5020 & 3995 & $+95$ & $-75$ & $+20$ & $-29$ & $+20$ & $-15$ & $+1$ & $-1$ \\
7000 & 4650 & $+105$ & $-75$ & $+35$ & $-48$ & $+15$ & $-15$ & $+1$ & $-1$ \\
13000 & 5875 & $+125$ & $-90$ & $+75$ & $-100$ & $+10$ & $-10$ & $+2$ & $-2$ \\
13600 & 5965 & $+130$ & $-90$ & $+80$ & $-100$ & $+10$ & $-10$ & $+2$ & $-2$ \\
14000 & 6020 & $+130$ & $-90$ & $+80$ & $-100$ & $+10$ & $-10$ & $+2$ & $-2$ \\
84000 & 7610 & $+340$ & $-100$ & $+860$ & $-100$ & $+10$ & $-15$ & $+2$ & $-2$ \\
100000 & 7425 & $+380$ & $-100$ & $+1130$ & $-100$ & $+10$ & $-15$ & $+2$ & $-2$ \\
400000 & 1140 & $+4900$ & $-100$ & $+49725$ & $-100$ & $+270$ & $-100$ & $+2$ & $-2$ \\
\bottomrule
\end{tabular}

\end{table}

\begin{table}[!htp]
\tabcolsep=2.0mm
   \centering
   \caption{Same as Table~\ref{tab:table1e}, but using the $\muF=\muR = \mu_\text{dyn.}$ dynamic scale, \cref{eq:mu_dyn}, as central value.  
   \label{tab:table1f}}
   \begin{tabular}{rcccccccccc}
\toprule
 & $\sigma(\ccbar)$~[\mub] & $\Delta^+_\mathrm{scale}~(\%)$ & $\Delta^-_\mathrm{scale}~(\%)$ & $\Delta^+_\mathrm{PDF}~(\%)$ & $\Delta^-_\mathrm{PDF}~(\%)$ & $\Delta^+_{\mc}~(\%)$ & $\Delta^-_{\mc}~(\%)$ & $\Delta^+_\mathrm{\alpha_s}~(\%)$ & $\Delta^-_\mathrm{\alpha_s}$~(\%) \\
$\sqrt{s}$~[GeV] &  &  &  &  &  &  &  &  &  \\
\midrule
11.6 & 0.88 & $+200$ & $-55$ & $+12$ & $-9$ & $+120$ & $-55$ & $+11.5$ & $-10.5$ \\
19.5 & 6.05 & $+135$ & $-50$ & $+8$ & $-4$ & $+90$ & $-45$ & $+8$ & $-8$ \\
21.7 & 8.32 & $+125$ & $-45$ & $+7$ & $-4$ & $+85$ & $-45$ & $+7.5$ & $-7.5$ \\
26 & 13.7 & $+110$ & $-45$ & $+6$ & $-4$ & $+80$ & $-40$ & $+7$ & $-6.5$ \\
27.4 & 15.6 & $+105$ & $-45$ & $+6$ & $-3$ & $+80$ & $-40$ & $+6.5$ & $-6.5$ \\
28.9 & 17.8 & $+105$ & $-45$ & $+5$ & $-3$ & $+75$ & $-40$ & $+6.5$ & $-6.5$ \\
38.7 & 34.4 & $+100$ & $-40$ & $+4$ & $-3$ & $+70$ & $-40$ & $+5.5$ & $-5.5$ \\
38.9 & 34.7 & $+100$ & $-40$ & $+4$ & $-3$ & $+70$ & $-40$ & $+5.5$ & $-5.5$ \\
41.6 & 39.8 & $+100$ & $-40$ & $+4$ & $-3$ & $+65$ & $-40$ & $+5$ & $-5$ \\
68.5 & 96.9 & $+90$ & $-35$ & $+3$ & $-3$ & $+60$ & $-35$ & $+4$ & $-3.5$ \\
86.6 & 138 & $+85$ & $-35$ & $+3$ & $-3$ & $+55$ & $-35$ & $+3$ & $-3$ \\
200 & 385 & $+75$ & $-40$ & $+2$ & $-3$ & $+45$ & $-30$ & $+2$ & $-2$ \\
1960 & 2420 & $+65$ & $-55$ & $+6$ & $-8$ & $+25$ & $-20$ & $+0.5$ & $-0.5$ \\[-2ex ]
\multicolumn{11}{c}{\hdashrule{15cm}{1pt}{1pt}}\\
2760 & 2980 & $+65$ & $-60$ & $+8$ & $-11$ & $+25$ & $-20$ & $+0$ & $0$ \\
5000 & 4095 & $+70$ & $-65$ & $+18$ & $-24$ & $+20$ & $-15$ & $+0.5$ & $-1$ \\
5020 & 4100 & $+70$ & $-65$ & $+18$ & $-24$ & $+20$ & $-15$ & $+0.5$ & $-1$ \\
7000 & 4800 & $+70$ & $-65$ & $+30$ & $-40$ & $+20$ & $-15$ & $+0.5$ & $-2$ \\
13000 & 6175 & $+80$ & $-75$ & $+60$ & $-91$ & $+15$ & $-15$ & $+1.5$ & $-5.5$ \\
13600 & 6265 & $+80$ & $-75$ & $+65$ & $-96$ & $+15$ & $-10$ & $+1.5$ & $-6$ \\
14000 & 6325 & $+80$ & $-75$ & $+65$ & $-99$ & $+15$ & $-10$ & $+2$ & $-6$ \\
84000 & 9830 & $+150$ & $-100$ & $+495$ & $-100$ & $+0$ & $-5$ & $+7.5$ & $-35$ \\
100000 & 10150 & $+160$ & $-100$ & $+585$ & $-100$ & $+0$ & $-5$ & $+8$ & $-39$ \\
400000 & 11900 & $+300$ & $-100$ & $+2295$ & $-100$ & $+10$ & $-15$ & $+125$ & $-92$ \\
\bottomrule
\end{tabular}

\end{table}

\clearpage
\section{NNLO bottom \pp\ cross section tables}
\label{sec:app2}

\begin{table}[!htp]
\tabcolsep=2.0mm
   \centering
   \caption{Theoretical NNLO predictions for the inclusive $\bbbar$ cross sections in \pp\ collisions over $\sqrts\approx 10$~GeV--400~TeV obtained with \MaunaKea\ using the \texttt{NNPDF40\_nnlo\_as\_01180\_nf\_4} PDFs. For each system, we list the central cross section, $\sigmabbbar$, followed by the percent (positive and negative) uncertainties from missing higher-order corrections ($\Delta^{\pm}_\text{scale}$), parton densities ($\Delta^{\pm}_\text{PDF}$), bottom quark mass ($\Delta^{\pm}_{\mb}$), and strong coupling constant ($\Delta^{\pm}_{\alphaS}$). The dotted line marks the lower $x$ limit of validity of the PDF set by $\xgmin$ (cross sections for c.m.\ energies above this threshold rely on extrapolations provided by LHAPDF). The static scale $\muF=\muR = 2\mQ$ is used as central value.
   \label{tab:table2a}}
\begin{tabular}{rcccccccccc}
\toprule
 & $\sigma(\bbbar)$~[\mub] & $\Delta^+_\mathrm{scale}~(\%)$ & $\Delta^-_\mathrm{scale}~(\%)$ & $\Delta^+_\mathrm{PDF}~(\%)$ & $\Delta^-_\mathrm{PDF}~(\%)$ & $\Delta^+_{\mb}~(\%)$ & $\Delta^-_{\mb}~(\%)$ & $\Delta^+_\mathrm{\alpha_s}~(\%)$ & $\Delta^-_\mathrm{\alpha_s}~(\%)$ \\
$\sqrt{s}$~[GeV] &  &  &  &  &  &  &  &  &  \\
\midrule
38.7 & 0.0056 & $+60$ & $-35$ & $+5$ & $-5$ & $+105$ & $-60$ & $+7$ & $-7$ \\
41.6 & 0.0083 & $+60$ & $-35$ & $+4$ & $-4$ & $+100$ & $-60$ & $+5.5$ & $-7.5$ \\
200 & 1.72 & $+30$ & $-20$ & $+1$ & $-1$ & $+50$ & $-40$ & $+4$ & $-4$ \\
510 & 10.4 & $+25$ & $-20$ & $+1$ & $-1$ & $+45$ & $-35$ & $+3$ & $-3$ \\
630 & 14.6 & $+25$ & $-20$ & $+1$ & $-1$ & $+40$ & $-35$ & $+3$ & $-3$ \\
1960 & 67.0 & $+25$ & $-20$ & $+1$ & $-1$ & $+35$ & $-30$ & $+2.5$ & $-2$ \\
2760 & 99.5 & $+25$ & $-20$ & $+1$ & $-1$ & $+35$ & $-30$ & $+2$ & $-2$ \\
7000 & 260 & $+20$ & $-20$ & $+2$ & $-2$ & $+30$ & $-30$ & $+1.5$ & $-2$ \\
13000 & 460 & $+20$ & $-20$ & $+2$ & $-2$ & $+30$ & $-25$ & $+1.5$ & $-1.5$ \\
13600 & 480 & $+20$ & $-20$ & $+2$ & $-2$ & $+30$ & $-25$ & $+1.5$ & $-2$ \\
14000 & 490 & $+20$ & $-20$ & $+2$ & $-2$ & $+30$ & $-25$ & $+1.5$ & $-2$ \\
84000 & 2050 & $+25$ & $-25$ & $+4$ & $-4$ & $+25$ & $-25$ & $+1.5$ & $-2$ \\
100000 & 2330 & $+25$ & $-25$ & $+4$ & $-4$ & $+25$ & $-25$ & $+1.5$ & $-1.5$ \\
[-2ex ]
\multicolumn{11}{c}{\hdashrule{15cm}{1pt}{1pt}}\\
400000 & 5950 & $+30$ & $-30$ & $+6$ & $-5$ & $+20$ & $-20$ & $+1$ & $-1$ \\
\bottomrule
\end{tabular}

\end{table}

\begin{table}[!htp]
\tabcolsep=2.0mm
   \centering
   \caption{Same as Table~\ref{tab:table2a}, but using the $\muF=\muR = \mu_\text{dyn.}$ dynamic scale, \cref{eq:mu_dyn}, as central value. 
   \label{tab:table2b}}
   \begin{tabular}{rcccccccccc}
\toprule
 & $\sigma(\bbbar)$[\mub] & $\Delta^+_\mathrm{scale}(\%)$ & $\Delta^-_\mathrm{scale}(\%)$ & $\Delta^+_\mathrm{PDF}(\%)$ & $\Delta^-_\mathrm{PDF}(\%)$ & $\Delta^+_{\mb}(\%)$ & $\Delta^-_{\mb}(\%)$ & $\Delta^+_\mathrm{\alpha_s}(\%)$ & $\Delta^-_\mathrm{\alpha_s}(\%)$ \\
$\sqrt{s}$~[GeV] &  &  &  &  &  &  &  &  &  \\
\midrule
38.7 & 0.0056 & $+60$ & $-35$ & $+5$ & $-5$ & $+105$ & $-60$ & $+6.5$ & $-7$ \\
41.6 & 0.0083 & $+60$ & $-35$ & $+4$ & $-4$ & $+100$ & $-60$ & $+6$ & $-7$ \\
200 & 1.7 & $+30$ & $-20$ & $+1$ & $-1$ & $+50$ & $-40$ & $+4$ & $-4$ \\
510 & 10.4 & $+25$ & $-20$ & $+1$ & $-1$ & $+45$ & $-35$ & $+3$ & $-3.5$ \\
630 & 14.5 & $+25$ & $-20$ & $+1$ & $-1$ & $+40$ & $-35$ & $+3$ & $-3$ \\
1960 & 66.8 & $+25$ & $-20$ & $+1$ & $-1$ & $+35$ & $-30$ & $+2$ & $-2$ \\
2760 & 98.9 & $+20$ & $-20$ & $+1$ & $-1$ & $+35$ & $-30$ & $+2$ & $-2$ \\
7000 & 260 & $+20$ & $-20$ & $+2$ & $-2$ & $+30$ & $-30$ & $+1.5$ & $-2$ \\
13000 & 460 & $+20$ & $-20$ & $+2$ & $-2$ & $+30$ & $-25$ & $+1.5$ & $-1.5$ \\
13600 & 480 & $+20$ & $-20$ & $+2$ & $-2$ & $+30$ & $-25$ & $+1.5$ & $-1.5$ \\
14000 & 490 & $+20$ & $-20$ & $+2$ & $-2$ & $+30$ & $-25$ & $+1.5$ & $-1.5$ \\
84000 & 2050 & $+25$ & $-25$ & $+4$ & $-4$ & $+25$ & $-25$ & $+1$ & $-1.5$ \\
100000 & 2340 & $+25$ & $-25$ & $+4$ & $-4$ & $+25$ & $-25$ & $+1$ & $-1.5$ \\[-2ex ]
\multicolumn{11}{c}{\hdashrule{15cm}{1pt}{1pt}}\\
400000 & 5950 & $+25$ & $-30$ & $+6$ & $-5$ & $+20$ & $-20$ & $+1$ & $-1.5$ \\
\bottomrule
\end{tabular}

\end{table}


\begin{table}[!htp]
\tabcolsep=2.0mm
   \centering
   \caption{Theoretical NNLO predictions for the inclusive $\bbbar$ cross sections in \pp\ collisions over $\sqrts\approx 10$~GeV--400~TeV obtained with \MaunaKea\ using the \texttt{CT18NNLO\_NF4} PDFs. For each system, we list the central cross section, $\sigmabbbar$, followed by the percent (positive and negative) uncertainties from missing higher-order corrections ($\Delta^{\pm}_\text{scale}$), parton densities ($\Delta^{\pm}_\text{PDF}$), bottom quark mass ($\Delta^{\pm}_{\mb}$), and strong coupling constant ($\Delta^{\pm}_{\alphaS}$). The dotted line marks the lower $x$ limit of validity of the PDF set by $\xgmin$ (cross sections for c.m.\ energies above this threshold rely on extrapolations provided by LHAPDF). The static scale $\muF=\muR = 2\mQ$ is used as central value.
   \label{tab:table2c}}
\begin{tabular}{rcccccccccc}
\toprule
 & $\sigma(\bbbar)$~[\mub] & $\Delta^+_\mathrm{scale}~(\%)$ & $\Delta^-_\mathrm{scale}~(\%)$ & $\Delta^+_\mathrm{PDF}~(\%)$ & $\Delta^-_\mathrm{PDF}~(\%)$ & $\Delta^+_{\mb}~(\%)$ & $\Delta^-_{\mb}~(\%)$ & $\Delta^+_\mathrm{\alpha_s}~(\%)$ & $\Delta^-_\mathrm{\alpha_s}~(\%)$ \\
$\sqrt{s}$~[GeV] &  &  &  &  &  &  &  &  &  \\
\midrule
38.7 & 0.010 & $+65$ & $-35$ & $+20$ & $-12$ & $+135$ & $-55$ & $+10.5$ & $-9$ \\
41.6 & 0.014 & $+60$ & $-35$ & $+18$ & $-11$ & $+125$ & $-55$ & $+8.5$ & $-9$ \\
200 & 2.17 & $+30$ & $-20$ & $+3$ & $-3$ & $+65$ & $-40$ & $+4.5$ & $-3.5$ \\
510 & 12.3 & $+25$ & $-20$ & $+3$ & $-3$ & $+55$ & $-35$ & $+2.5$ & $-2.5$ \\
630 & 17.0 & $+25$ & $-20$ & $+3$ & $-3$ & $+50$ & $-30$ & $+2.5$ & $-2.5$ \\
1960 & 76.5 & $+25$ & $-20$ & $+3$ & $-4$ & $+45$ & $-30$ & $+2$ & $-2$ \\
2760 & 115 & $+25$ & $-20$ & $+3$ & $-4$ & $+45$ & $-30$ & $+2$ & $-2$ \\
7000 & 300 & $+25$ & $-20$ & $+5$ & $-5$ & $+40$ & $-25$ & $+2$ & $-2$ \\
13000 & 540 & $+25$ & $-20$ & $+7$ & $-5$ & $+35$ & $-25$ & $+2$ & $-2$ \\
13600 & 565 & $+25$ & $-20$ & $+7$ & $-5$ & $+35$ & $-25$ & $+2$ & $-2$ \\
14000 & 580 & $+25$ & $-20$ & $+7$ & $-5$ & $+35$ & $-25$ & $+2$ & $-2$ \\
84000 & 2550 & $+25$ & $-25$ & $+18$ & $-6$ & $+30$ & $-20$ & $+2$ & $-2$ \\
100000 & 2910 & $+25$ & $-25$ & $+20$ & $-6$ & $+30$ & $-20$ & $+2$ & $-2$ \\[-2ex ]
\multicolumn{11}{c}{\hdashrule{15cm}{1pt}{1pt}}\\
400000 & 7760 & $+30$ & $-30$ & $+50$ & $-7$ & $+30$ & $-20$ & $+2$ & $-2$ \\
\bottomrule
\end{tabular}

\end{table}

\begin{table}[!htp]
\tabcolsep=2.0mm
   \centering
   \caption{Same as Table~\ref{tab:table2c}, but using the $\muF=\muR = \mu_\text{dyn.}$ dynamic scale, \cref{eq:mu_dyn}, as central value.  
   \label{tab:table2d}}
   \begin{tabular}{rcccccccccc}
\toprule
 & $\sigma(\bbbar)$[\mub] & $\Delta^+_\mathrm{scale}(\%)$ & $\Delta^-_\mathrm{scale}(\%)$ & $\Delta^+_\mathrm{PDF}(\%)$ & $\Delta^-_\mathrm{PDF}(\%)$ & $\Delta^+_{\mb}(\%)$ & $\Delta^-_{\mb}(\%)$ & $\Delta^+_\mathrm{\alpha_s}(\%)$ & $\Delta^-_\mathrm{\alpha_s}(\%)$ \\
$\sqrt{s}$~[GeV] &  &  &  &  &  &  &  &  &  \\
\midrule
38.7 & 0.010 & $+65$ & $-35$ & $+20$ & $-12$ & $+135$ & $-55$ & $+10$ & $-9$ \\
41.6 & 0.014 & $+60$ & $-35$ & $+18$ & $-11$ & $+125$ & $-55$ & $+9.5$ & $-8.5$ \\
200 & 2.15 & $+30$ & $-20$ & $+3$ & $-3$ & $+65$ & $-40$ & $+4$ & $-4$ \\
510 & 12.2 & $+25$ & $-20$ & $+3$ & $-3$ & $+55$ & $-35$ & $+2.5$ & $-2.5$ \\
630 & 16.9 & $+25$ & $-20$ & $+3$ & $-3$ & $+50$ & $-30$ & $+2.5$ & $-2.5$ \\
1960 & 76.1 & $+25$ & $-20$ & $+3$ & $-4$ & $+45$ & $-30$ & $+2$ & $-2$ \\
2760 & 113 & $+25$ & $-20$ & $+3$ & $-4$ & $+45$ & $-30$ & $+2$ & $-2$ \\
7000 & 300 & $+25$ & $-20$ & $+5$ & $-5$ & $+40$ & $-25$ & $+2$ & $-2$ \\
13000 & 540 & $+25$ & $-20$ & $+7$ & $-5$ & $+35$ & $-25$ & $+2$ & $-2$ \\
13600 & 560 & $+25$ & $-20$ & $+7$ & $-5$ & $+35$ & $-25$ & $+2$ & $-2$ \\
14000 & 575 & $+25$ & $-20$ & $+7$ & $-5$ & $+35$ & $-25$ & $+2$ & $-2$ \\
84000 & 2540 & $+25$ & $-25$ & $+18$ & $-6$ & $+30$ & $-20$ & $+2.5$ & $-2.5$ \\
100000 & 2900 & $+25$ & $-25$ & $+20$ & $-6$ & $+30$ & $-20$ & $+2.5$ & $-2.5$ \\[-2ex ]
\multicolumn{11}{c}{\hdashrule{15cm}{1pt}{1pt}}\\
400000 & 7760 & $+25$ & $-30$ & $+45$ & $-7$ & $+30$ & $-20$ & $+2.5$ & $-2.5$ \\
\bottomrule
\end{tabular}

\end{table}


\begin{table}[!htp]
\tabcolsep=2.0mm
   \centering
   \caption{Theoretical NNLO predictions for the inclusive $\bbbar$ cross sections in \pp\ collisions over $\sqrts\approx 10$~GeV--400~TeV obtained with \MaunaKea\ using the \texttt{MSHT20nnlo\_nf4} PDFs.  For each system, we list the central cross section, $\sigmabbbar$, followed by the percent (positive and negative) uncertainties from missing higher-order corrections ($\Delta^{\pm}_\text{scale}$), parton densities ($\Delta^{\pm}_\text{PDF}$), bottom quark mass ($\Delta^{\pm}_{\mb}$), and strong coupling constant ($\Delta^{\pm}_{\alphaS}$). The dotted line marks the lower $x$ limit of validity of the PDF set by $\xgmin$ (cross sections for c.m.\ energies above this threshold rely on extrapolations provided by LHAPDF). The static scale $\muF=\muR = 2\mQ$ is used as central value.
   \label{tab:table2e}}
\begin{tabular}{rcccccccccc}
\toprule
 & $\sigma(\bbbar)$~[\mub] & $\Delta^+_\mathrm{scale}~(\%)$ & $\Delta^-_\mathrm{scale}~(\%)$ & $\Delta^+_\mathrm{PDF}~(\%)$ & $\Delta^-_\mathrm{PDF}~(\%)$ & $\Delta^+_{\mb}~(\%)$ & $\Delta^-_{\mb}~(\%)$ & $\Delta^+_\mathrm{\alpha_s}~(\%)$ & $\Delta^-_\mathrm{\alpha_s}~(\%)$ \\
$\sqrt{s}$~[GeV] &  &  &  &  &  &  &  &  &  \\
\midrule
38.7 & 0.0098 & $+65$ & $-35$ & $+11$ & $-9$ & $+135$ & $-55$ & $+8.5$ & $-7.5$ \\
41.6 & 0.0140 & $+60$ & $-35$ & $+10$ & $-8$ & $+125$ & $-55$ & $+7$ & $-8$ \\
200 & 2.10 & $+30$ & $-20$ & $+3$ & $-2$ & $+65$ & $-40$ & $+3.5$ & $-3$ \\
510 & 12.0 & $+25$ & $-20$ & $+2$ & $-2$ & $+55$ & $-35$ & $+2.5$ & $-2.5$ \\
630 & 16.6 & $+25$ & $-20$ & $+2$ & $-2$ & $+50$ & $-30$ & $+2$ & $-2.5$ \\
1960 & 74 & $+25$ & $-20$ & $+2$ & $-2$ & $+45$ & $-30$ & $+2$ & $-1.5$ \\
2760 & 110 & $+25$ & $-20$ & $+2$ & $-2$ & $+40$ & $-30$ & $+2$ & $-1.5$ \\
7000 & 280 & $+20$ & $-20$ & $+2$ & $-3$ & $+40$ & $-25$ & $+1.5$ & $-1.5$ \\[-2ex ]
\multicolumn{11}{c}{\hdashrule{15cm}{1pt}{1pt}}\\
13000 & 485 & $+20$ & $-20$ & $+3$ & $-4$ & $+35$ & $-25$ & $+1.5$ & $-1.5$ \\
13600 & 505 & $+25$ & $-20$ & $+3$ & $-4$ & $+35$ & $-25$ & $+1.5$ & $-1.5$ \\
14000 & 520 & $+25$ & $-20$ & $+3$ & $-4$ & $+35$ & $-25$ & $+1.5$ & $-1.5$ \\
84000 & 1975 & $+25$ & $-30$ & $+18$ & $-25$ & $+30$ & $-20$ & $+0$ & $-1.5$ \\
100000 & 2215 & $+25$ & $-30$ & $+20$ & $-28$ & $+30$ & $-20$ & $+0$ & $-1$ \\
400000 & 5085 & $+30$ & $-35$ & $+55$ & $-64$ & $+25$ & $-15$ & $+0$ & $-1$ \\
\bottomrule
\end{tabular}

\end{table}

\begin{table}[!htp]
\tabcolsep=2.0mm
   \centering
   \caption{Same as Table~\ref{tab:table2e}, but using the $\muF=\muR = \mu_\text{dyn.}$ dynamic scale, \cref{eq:mu_dyn}, as central value.  
   \label{tab:table2f}}
   \begin{tabular}{rcccccccccc}
\toprule
 & $\sigma(\bbbar)$[\mub] & $\Delta^+_\mathrm{scale}(\%)$ & $\Delta^-_\mathrm{scale}(\%)$ & $\Delta^+_\mathrm{PDF}(\%)$ & $\Delta^-_\mathrm{PDF}(\%)$ & $\Delta^+_{\mb}(\%)$ & $\Delta^-_{\mb}(\%)$ & $\Delta^+_\mathrm{\alpha_s}(\%)$ & $\Delta^-_\mathrm{\alpha_s}(\%)$ \\
$\sqrt{s}$~[GeV] &  &  &  &  &  &  &  &  &  \\
\midrule
38.7 & 0.0097 & $+65$ & $-35$ & $+11$ & $-9$ & $+135$ & $-55$ & $+8$ & $-7.5$ \\
41.6 & 0.014 & $+60$ & $-35$ & $+10$ & $-8$ & $+125$ & $-55$ & $+7.5$ & $-7.5$ \\
200 & 2.1 & $+30$ & $-20$ & $+3$ & $-2$ & $+65$ & $-40$ & $+3.5$ & $-3.5$ \\
510 & 11.9 & $+25$ & $-20$ & $+2$ & $-2$ & $+55$ & $-35$ & $+2$ & $-2.5$ \\
630 & 16.5 & $+25$ & $-20$ & $+2$ & $-2$ & $+50$ & $-30$ & $+2$ & $-2.5$ \\
1960 & 73.5 & $+25$ & $-20$ & $+2$ & $-2$ & $+45$ & $-30$ & $+2$ & $-1.5$ \\
2760 & 108 & $+25$ & $-20$ & $+2$ & $-2$ & $+40$ & $-30$ & $+2$ & $-1.5$ \\
7000 & 278 & $+20$ & $-20$ & $+2$ & $-3$ & $+40$ & $-25$ & $+1.5$ & $-2$ \\[-2ex ]
\multicolumn{11}{c}{\hdashrule{15cm}{1pt}{1pt}}\\
13000 & 485 & $+20$ & $-20$ & $+3$ & $-4$ & $+35$ & $-25$ & $+1.5$ & $-1.5$ \\
13600 & 505 & $+20$ & $-20$ & $+3$ & $-4$ & $+35$ & $-25$ & $+1.5$ & $-1.5$ \\
14000 & 515 & $+20$ & $-20$ & $+3$ & $-4$ & $+35$ & $-25$ & $+1.5$ & $-1.5$ \\
84000 & 1980 & $+25$ & $-30$ & $+18$ & $-22$ & $+30$ & $-20$ & $+0$ & $-1.5$ \\
100000 & 2220 & $+25$ & $-30$ & $+20$ & $-26$ & $+30$ & $-20$ & $+0$ & $-1$ \\
400000 & 5100 & $+30$ & $-35$ & $+55$ & $-52$ & $+25$ & $-20$ & $+0$ & $-2.5$ \\
\bottomrule
\end{tabular}

\end{table}

\clearpage
\bibliographystyle{myutphys}
\bibliography{ccbar_bbbar_nnlo}

\end{document}